\documentclass[oneside,bibliography=totoc]{scrartcl}
\pdfoutput=1

\usepackage{geometry}
\usepackage[T1]{fontenc}
\usepackage[utf8]{inputenc}
\usepackage{lmodern}

\usepackage{amsmath,amsthm,amssymb}

\usepackage{xcolor} %

\definecolor{col1}{RGB}{254,97,0}
\definecolor{col2}{RGB}{100,143,255}
\definecolor{col3}{RGB}{120, 94, 240}
\definecolor{col4}{RGB}{220, 38, 127}
\definecolor{col5}{RGB}{255, 176, 0}

\usepackage{numprint}
\newcommand{\lfrac}[2]{\frac{\numprint{#1}}{\numprint{#2}}}
\newcommand{\Periods}{\mathcal{P}}
\newcommand{\motivic}{\mathfrak{m}}

\usepackage{tikz,tikz-3dplot}
\usetikzlibrary{positioning,calc,arrows.meta,decorations.pathreplacing,patterns}

\usepackage{longtable}

\makeatletter
 \tikzoption{canvas is plane}[]{\@setOxy#1}
 \def\@setOxy O(#1,#2,#3)x(#4,#5,#6)y(#7,#8,#9)%
   {\def\tikz@plane@origin{\pgfpointxyz{#1}{#2}{#3}}%
    \def\tikz@plane@x{\pgfpointxyz{#4}{#5}{#6}}%
    \def\tikz@plane@y{\pgfpointxyz{#7}{#8}{#9}}%
    \tikz@canvas@is@plane
   }
\makeatother  

\newcommand{\rada}{0.6}

\usepackage{hyperref}

\newsavebox{\solidedgebox}
\savebox{\solidedgebox}{%
\begin{tikzpicture}[baseline={([yshift=-0.7ex]0,0)}] \coordinate (v1) at (0,0); \coordinate (v2) at (.5,0); \draw (v1) -- (v2); \filldraw (v1) circle (1.3pt); \filldraw (v2) circle (1.3pt); \end{tikzpicture}%
}
\newcommand{\solidedge}{{\usebox{\solidedgebox}}}

\newsavebox{\dashededgebox}
\savebox{\dashededgebox}{%
\begin{tikzpicture}[baseline={([yshift=-0.7ex]0,0)}] \coordinate (v1) at (0,0); \coordinate (v2) at (.5,0); \draw[dashed] (v1) -- (v2); \filldraw (v1) circle (1.3pt); \filldraw (v2) circle (1.3pt); \end{tikzpicture}%
}
\newcommand{\dashededge}{{\usebox{\dashededgebox}}}

\newsavebox{\dottededgebox}
\savebox{\dottededgebox}{%
\begin{tikzpicture}[baseline={([yshift=-0.7ex]0,0)}] \coordinate (v1) at (0,0); \coordinate (v2) at (.5,0); \draw[dotted] (v1) -- (v2); \filldraw (v1) circle (1.3pt); \filldraw (v2) circle (1.3pt); \end{tikzpicture}%
}

\newcommand{\dottededge}{{\usebox{\dottededgebox}}}

\newcommand{\A}{{\mathbb A}}
\newcommand{\BB}{{\mathbb B}}
\newcommand{\CC}{{\mathbb C}}
\newcommand{\DD}{{\mathbb D}}
\newcommand{\EE}{{\mathbb E}}
\newcommand{\FF}{{\mathbb F}}
\newcommand{\GG}{{\mathbb G}}
\newcommand{\HH}{{\mathbb H}}
\newcommand{\II}{{\mathbb I}}
\newcommand{\JJ}{{\mathbb J}}
\newcommand{\KK}{{\mathbb K}}
\newcommand{\LL}{{\mathbb L}}
\newcommand{\MM}{{\mathbb M}}
\newcommand{\NN}{{\mathbb N}}
\newcommand{\OO}{{\mathbb O}}
\newcommand{\PP}{{\mathbb P}}
\newcommand{\QQ}{{\mathbb Q}}
\newcommand{\RR}{{\mathbb R}}
\renewcommand{\SS}{{\mathbb S}}
\newcommand{\TT}{{\mathbb T}}
\newcommand{\UU}{{\mathbb U}}
\newcommand{\VV}{{\mathbb V}}
\newcommand{\WW}{{\mathbb W}}
\newcommand{\XX}{{\mathbb X}}
\newcommand{\YY}{{\mathbb Y}}
\newcommand{\ZZ}{{\mathbb Z}}
\newcommand{\ee}{\mathrm{e}}
\newcommand{\ii}{\mathrm{i}}
\newcommand{\dR}{\mathfrak{dr}}
\newcommand{\sL}{\mathfrak{sl}}
\newcommand{\dd}{\mathrm{d}}
\newcommand{\Li}{\mathrm{Li}\,}
\newcommand{\aaa}{{\overline{a}}}
\newcommand{\ddd}{{\overline{d}}}
\newcommand{\ff}{{\overline{f}}}
\newcommand{\gggg}{{\overline{g}}}
\newcommand{\hh}{{\overline{h}}}
\newcommand{\kk}{{\overline{k}}}
\newcommand{\mm}{{\overline{m}}}
\newcommand{\pp}{{\overline{p}}}
\newcommand{\qq}{{\overline{q}}}
\newcommand{\rr}{{\overline{r}}}
\newcommand{\sss}{{\overline{s}}}
\newcommand{\xx}{{\overline{x}}}
\newcommand{\zz}{{\overline{z}}}
\newcommand{\Gg}{{\overline{G}}}
\newcommand{\sC}{\mathcal{C}}
\newcommand{\sD}{\mathcal{D}}
\newcommand{\sE}{\mathcal{E}}
\newcommand{\sF}{\mathcal{F}}
\newcommand{\sG}{\mathcal{G}}
\newcommand{\sI}{\mathcal{I}}
\newcommand{\sJ}{\mathcal{J}}
\newcommand{\sM}{\mathcal{M}}
\newcommand{\sS}{\mathcal{S}}
\newcommand{\sV}{\mathcal{V}}
\newcommand{\PE}{\mathcal{P}}

\newcommand{\td}[1]{\widetilde{#1}}
\newcommand{\bb}[1]{{\boldsymbol{#1}}}
\newcommand{\sVGint}{{\sV_G^\mathrm{int}}}
\newcommand{\VGint}{{|\sVGint|}}
\newcommand{\sVGext}{{\sV_G^\mathrm{ext}}}
\newcommand{\VGext}{{|\sVGext|}}
\newcommand{\intsv}{\int_\mathrm{sv}}
\renewcommand{\Im}{\mathop\mathrm{Im}}
\renewcommand{\Re}{\mathop\mathrm{Re}}

\theoremstyle{plain}
\newtheorem{thm}{Theorem}

\newtheorem{con}[thm]{Conjecture}

\hypersetup{
  pdfauthor={Michael Borinsky, Oliver Schnetz},
  pdftitle={Recursive computation of Feynman periods},
  pdfsubject={},
  pdfkeywords={},
  linkcolor  = col3,
  citecolor  = col4,
  urlcolor   = col5,
  colorlinks = true,
}

\begin{document}
\title{Recursive computation of Feynman periods}
\author{Michael Borinsky 
\\
\small \text{Institute for Theoretical Studies}\\[-1ex]
\small \text{ETH Z\"urich}\\[-1ex]
\small \text{8092 Z\"urich, Switzerland}
\and Oliver Schnetz
\\
\small \text{Department Mathematik}\\[-1ex]
\small \text{Friedrich Alexander Universität}\\[-1ex]
\small \text{91058 Erlangen, Germany}
}
\date{}

\maketitle
\begin{abstract}
Feynman periods are Feynman integrals that do not depend on external kinematics. Their computation, which is necessary for many applications of quantum field theory, is greatly facilitated by graphical functions or the equivalent conformal four-point integrals. We describe a set of transformation rules that act on such functions and allow their recursive computation in arbitrary even dimensions. As a concrete example we compute all subdivergence-free Feynman periods in $\phi^3$ theory up to six loops and 561 of 607 Feynman periods at seven loops analytically. Our results support the conjectured existence of a coaction structure in quantum field theory and suggest that $\phi^3$ and $\phi^4$ theory share the same number content.
\end{abstract}

\tableofcontents
\section{Introduction}
\subsection{Feynman periods}
Feynman graphs and the associated integrals are both fascinating mathematical objects and necessary ingredients for many computations in quantum field theory (QFT). %

The primary subject of this article are special Feynman integrals that only depend trivially on external kinematics: \emph{Feynman periods}. 
Feynman periods are equivalent to Feynman integrals of massless two-point functions also called \emph{single-scale} or \emph{p-integrals}.
We will give explicit definitions of Feynman periods in Sections~\ref{sec:momperiods}, \ref{sec:parametric} and \ref{sec:positionperiod} in momentum, parametric and in position-space.

Many types of computations in perturbative QFT can be performed by evaluating Feynman periods. Among them is the computation of renormalization group functions~\cite{tHooft:1973mfk}, the expansion by region method to evaluate Feynman integrals in kinematic limits~\cite{Semenova:2018cwy} and many instances of the general operator product expansion~\cite{Gorishnii:1983su}. Many further applications rely on Feynman period computations as an ingredient. For instance, the {differential equation method}~\cite{Kotikov:1990kg,Remiddi:1997ny} requires the computation of a boundary value that can be chosen to be a Feynman period. 

A new and surprising application of Feynman periods in pure mathematics is the detection of non-vanishing homology classes in the commutative graph complex \cite{Brown:2021umn} which itself allows the identification of classes in  the highest non-zero weight-graded piece of the cohomology of $\mathcal M_g$, the moduli space of curves of genus $g$~\cite{chan2021tropical}. 
We will comment on this new application in Section~\ref{sec:wheels}.

\subsection{Graphical functions and conformal four-point integrals}
The secondary subject of this article are \emph{graphical functions} or, equivalently,
\emph{conformal four-point integrals} (see Section~\ref{sec:completegf} for an explanation of this equivalence).
We will introduce these functions as a tool to perform Feynman period computations in arbitrary even dimensions $\geq4$. The necessary mathematical background has been developed by the authors in~\cite{gfe} and previously by the second author in~\cite{Schnetz:2013hqa,Schnetz:2016fhy}. 
Our methods also apply in the more general realm of dimensionally regularized Feynman integrals where also divergent Feynman periods can be computed (see~\cite{Schnetz:2016fhy,7loops} for details). 
In this article, however, we will focus on the finite and even dimensional case.

Graphical functions constitute a natural generalization of Feynman periods by allowing a \emph{minimal} dependence on external kinematics. 
These functions are massless three-point Feynman integrals up to a scale factor, which are parameterized to be functions on the complex plane. We describe this parameterization together with the basic theory of graphical functions in Section~\ref{sec:graphicalfunctions}.
The theory of graphical functions combines the rich structure of Feynman integrals with tools from complex analysis. The structure of Feynman periods strongly constrains the space of
graphical functions. The efficacy of the graphical function method comes from these constraints.

At the heart of the graphical function method is a set of \emph{transformation rules}. These rules modify the underlying graph while performing a corresponding analytic operation on the associated graphical function while keeping the correspondence between graph and function intact. These transformation rules allow one to increase the complexity of the original graph quite drastically and analytic expressions for a large number of graphs can be computed in this \emph{recursive} manner, in the spirit of~\cite{Britto:2005fq}. We will describe the most important transformation rules in Sections~\ref{sec:trafo} where we also provide many examples of their applications.

The transformation rules have been proved in~\cite{gfe} by exploiting the constraints that the graphical functions must fulfill.
These constraints also suggest a relation to the \emph{bootstrap and integrability approach} where Feynman integrals are computed using constraints from \emph{hidden symmetries}~(see,~e.g.~\cite{Zamolodchikov:1980mb,Isaev:2003tk,Drummond:2013nda,Basso:2017jwq,Loebbert:2019vcj,Basso:2021omx}).  %
Our methods, however, just rely on inherent structures of Feynman integrals with their properties as functions on $\CC$ 
and
do not exploit any hidden symmetries. 
Still, graphical functions \emph{complement} and \emph{leverage} results that have been obtained by integrability and other methods. For instance, \emph{fishnet graphs}, which all come with explicit analytic expressions thanks to an impressive achievement of the integrability approach~\cite{Basso:2017jwq,Basso:2021omx}, can serve as nontrivial starting points for the transformation rules (see Section~\ref{sec:external}). Each fishnet graph gives rise to an infinite family of conformal four-point integrals and, equivalently, graphical functions all of which can be computed explicitly.
The transformation rules also generalize known properties of conformal four-point integrals. For instance, they incorporate the magic identities from~\cite{Drummond:2006rz} and they directly generate known families of Feynman integrals such as the ladder graphs~\cite{Usyukina:1993ch}.

In the context of Feynman periods, these functions have been instrumental for the proof of the \emph{zig-zag conjecture}~\cite{Broadhurst:1995km,Brown:2015ztw}, for which recently an alternative, integrability based proof was found~\cite{Derkachov:2022ytx}. 

All this indicates that there is huge potential in further synthesis of the integrability and our complex analysis and single-valuedness based methods which still remains to be exploited for Feynman integral computations.

\subsection{Number-theoretical content of QFT}
The application of Feynman period computations that shall serve as the guiding motivation in this article 
is the  study of the \emph{number-theoretical} content of QFT.

Feynman periods are periods in the sense of Kontsevich and Zagier \cite{Kontsevich2001}.
They are a family of real numbers that include all rational and algebraic numbers, but only certain transcendental numbers.
For instance, $\pi$, $\log2$ and integer values of the Riemann-$\zeta$ function are periods.
Periods play a critical role in many parts of mathematics and from this perspective, it is natural to ask which subset of periods appears in QFTs.

By the number theoretical content of a QFT, we mean the set of numbers to which the renormalization group 
independent part of the Feynman graphs in the theory evaluate. Equivalently, we take this set to be the numbers that appear as the $\frac{1}{\varepsilon}$ residue of all logarithmically divergent
subdivergence-free Feynman integrals in the respective dimensionally regularized theory.

Questions on the number theoretical content of QFT 
have been raised already quite some time ago~(see e.g.~\cite{Broadhurst:1995km}). Since then many theoretical developments on Feynman integrals went hand in hand with better understanding of the underlying number theory. 
A recent example for this interaction is the conjectured \textit{coaction principle} which implies that Feynman integrals are heavily constrained by a rather mysterious number theoretical symmetry \cite{Brown:2015fyf,Schnetz:2016fhy}. A hypothetical \emph{coaction} defined on Feynman integrals should reduce to a coaction which is related to the Goncharov--Brown coaction \cite{Goncharov:2001iea,PanzerSchnetz:2017coact,brown2012mixed,P3P}
when restricted to polylogarithmically valued Feynman integrals \cite{brown2012mixed,Brown:2015fyf}. Even more recently this coaction has also been conjectured to be realizable diagrammatically on the level of Feynman graphs \cite{Abreu:2021vhb}.  %

In this article we gather further data which support the Feynman period version of the coaction conjecture. To do so, we tap a new and highly resourceful data-source: scalar $\phi^3$ theory in six dimensions. Preceding studies on the transcendental number content of quantum field theories at higher loop order have been conducted in scalar $\phi^4$ theory~\cite{Broadhurst:1995km,Bloch:2005bh,Schnetz:2008mp,PanzerSchnetz:2017coact} and to a limited extend in QED~\cite{Broadhurst:1995dq,Schnetz:2017bko}. Looking at the number content of $\phi^3$ theory is a natural extension of these studies. An advantage of $\phi^3$ theory is that it contains comparatively many Feynman graphs and therefore many more examples to test a coaction principle: 
The number of subdivergence-free 1PI $4$-point diagrams in $\phi^4$ theory is
$e^{-\frac{15}{4}}\left(2/3\right)^{n+2} (n+2)!/\sqrt{2} \pi$ 
for large loop order $n$ whereas $\phi^3$ theory has 
$e^{-\frac{10}{3}} \left(3/2\right)^{n+1} (n+1)!/\pi$ subdivergence-free 1PI $3$-point diagrams at $n$ loops for $n\rightarrow \infty$~\cite{Borinsky:2017hkb}. The dominating exponential factor of $\approx (9/4)^n$ between both asymptotic numbers illustrates the increased difficulty of $\phi^3$ theory computations.
For instance, there are only 11 different $\phi^4$ period graphs at 7 loops \cite{Schnetz:2008mp}, whereas there are 607 such $\phi^3$ graphs at the same loop order. Moreover, $\phi^3$ theory is an obvious intermediate step before the exploration of Gauge theories such as QCD or Yang--Mills theories.

Explicitly, we will compute all finite Feynman periods that contribute to $\phi^3$ theory in six-dimensional spacetime up to 6 loops
and 561 of 607 Feynman periods at 7 loops analytically.

\subsection{Comparison to traditional methods of Feynman period evaluation}
The traditional method of computing Feynman periods (i.e.~massless propagators) involves a \emph{IBP-reduction step} and the evaluation of the reduced \emph{master integrals}. This approach has been put forward in~\cite{Chetyrkin:1981qh}, made more systematic in~\cite{Laporta:2000dsw} and has been applied in similar style for numerous computations since then.  In the IBP-reduction step the Feynman periods of interest are expressed in terms of a basis of master integrals. Although this reduction is essentially a linear algebra problem, it is still computationally demanding. It constitutes the bottleneck for many desirable computations that involve Feynman integrals. 

Even though it has been noticed early in the 1980's that all Feynman periods up to three-loop-level can be evaluated in integer dimension~\cite{Kazakov:1983ns}, it was only proven in the late 2000's by introducing the concept of \emph{linear reducibility} that all Feynman periods up to four loops evaluate to Laurent expansions with multiple zeta values as coefficients and that all Feynman periods at five loops evaluate to Laurent expansions with multiple zeta values ramified at $6$-th roots of unity~\cite{Brown:2008um,Brown:2009ta}. The complete set of master integrals at four-loop level was given in~\cite{Baikov:2010hf,Lee:2011jt}. 
At least in principle all Feynman integrals with up to five loops became accessible with the implementation of Brown's algorithm~\cite{Brown:2008um,Brown:2009ta} as \texttt{HyperInt}~\cite{Panzer:2014caa}.
The five-loop master integrals have been computed in \cite{Georgoudis:2018olj,Georgoudis:2021onj}.

The main difference of the graphical function method from the traditional one is that it aims to work with combinatorial features of the underlying graphs as long as possible before resorting to any heavy analytic machinery. The reduction to master integrals is replaced by a \emph{combinatorial} reduction of the Feynman periods. This combinatorial reduction is more powerful than the traditional IBP-reduction, as it reduces to a set of small \emph{kernel} graphs whose number is usually much smaller than the number of master integrals. The reason for this is that, effectively, the combinatorial reduction makes use of many more identities than just the IBP identity. As a consequence the graphical function method can rather quickly deal with quite huge amounts of graphs. Moreover, the combinatorial reduction allows the computation of comparatively large graphs, where tools like \texttt{HyperInt} would hopelessly fail. Furthermore the graphical function methods unifies and extends many known techniques to evaluate and relate period integrals into one framework. Examples are the \emph{cut-and-glue} identity~\cite{Chetyrkin:1980pr,Chetyrkin:1981qh,Baikov:2010hf} and the magic identities from~\cite{Drummond:2006rz}.

Even though the graphical function method can deal with divergent (but properly dimensionally regularized) Feynman periods and higher orders in the associated $\varepsilon$ expansion (see, e.g., \cite{Schnetz:2016fhy,Borinsky:2021jdb,7loops}), we will focus on finite Feynman periods in this paper. All tools described in Sections~\ref{feynpergf} and \ref{sec:trafo} apply almost verbatim to (divergent) dimensionally regularized Feynman integrals. %

All methods described in this paper apply to Feynman periods in any even dimension $\geq 4$ and in any scalar QFT. In fact, they can be used to calculate the Feynman periods associated to graphs that are rather exotic from the perspective of QFT (e.g.\ for the previously mentioned application to topology~\cite{Brown:2021umn}). For concreteness, we will focus on periods that appear in six-dimensional $\varphi^3$ theory in the last part of the paper (Sections~\ref{phi3} and \ref{sectsyst}), where we will present some explicit computations.

\subsection{Six-dimensional \texorpdfstring{$\phi^3$}{phi3} theory}

\begin{table}
\begin{center}
\begin{tabular}{rr|ll}
$\ell$&$\!$wt&number&value\\\hline
1&0&$Q_0=1$&1\\\hline
3&3&$Q_3=\zeta(3)$&1.202~056~903~159\\\hline
4&5&$Q_5=\zeta(5)$&1.036~927~755~143\\\hline
5&7&$Q_7=\zeta(7)$&1.008~349~277~381\\\hline
6&8&$Q_8=N_{3,5}$&0.070~183~206~556\\
&9&$Q_9=\zeta(9)$&1.002~008~392~826\\\hline
7&10&$Q_{10}=N_{3,7}$&0.090~897~338~299\\
&11&$Q_{11,1}=\zeta(11)$&1.000~494~188~604\\
&&$Q_{11,2}=-\zeta(3,5,3)\!+\!\zeta(3)\zeta(5,3)$&0.042~696~696~025\\\hline
8&12&$Q_{12,1}=N_{3,9}$&0.096~506~102~637\\
&&$Q_{12,2}=N_{5,7}$&0.020~460~547~937\\
&&$Q_{12,3}=\pi^{12}/10!$&0.254~703~808~841\\
&&$Q_{12,4}$&-371~077.332~598\\
&&$Q_{12,5}$&194~862.842~178\\
&13&$Q_{13,1}=\zeta(13)$&1.000~122~713~347\\
&&$Q_{13,2}=-\zeta(5,3,5)\!+\!11\zeta(5)\zeta(5,3)\!+\!5\zeta(5)\zeta(8)$&5.635~097~688~692\\
&&$Q_{13,3}=-\zeta(3,7,3)\!+\!\zeta(3)\zeta(7,3)\!+\!12\zeta(5)\zeta(5,3)\!+\!6\zeta(5)\zeta(8)\hspace*{-5pt}$&6.725~631~947~085
\end{tabular}
\end{center}
\caption{List of known $\phi^3$ and $\phi^4$ transcendentals up to weight 13.}
\label{tabQ}
\end{table}

\begin{table}
\begin{center}
\begin{tabular}{rr|l}
$\ell$&$\!$wt&base\\[1ex]\hline&&\\[-1.5ex]
6&8&$N_{3,5}=\frac{27}{80}\zeta(5,3)+\frac{45}{64}\zeta(5)\zeta(3)-\frac{261}{320}\zeta(8)$\\[1ex]\hline&&\\[-1.5ex]
7&10&$N_{3,7}=\frac{423}{3584}\zeta(7,3)+\frac{189}{256}\zeta(7)\zeta(3)+\frac{639}{3584}\zeta(5)^2-\frac{7137}{7168}\zeta(10)$\\[1ex]\hline&&\\[-1.5ex]
8&12&$N_{3,9}=\frac{27}{512}\zeta(4,4,2,2)+\frac{55}{1024}\zeta(9,3)+\frac{231}{256}\zeta(9)\zeta(3)+\frac{447}{256}\zeta(7)\zeta(5)-\frac{9}{512}\zeta(3)^4$\\[1ex]\hline&&\\[-1.5ex]
&&\hspace{13mm}$-\frac{27}{448}\zeta(7,3)\zeta(2)-\frac{189}{128}\zeta(7)\zeta(3)\zeta(2)-\frac{1269}{1792}\zeta(5)^2\zeta(2)+\frac{189}{512}\zeta(5,3)\zeta(4)$\\[1ex]
&&\hspace{13mm}$+\frac{945}{512}\zeta(5)\zeta(3)\zeta(4)+\frac{9}{64}\zeta(3)^2\zeta(6)-\frac{7322453}{5660672}\zeta(12)$\\[1ex]
&&$N_{5,7}=-\frac{81}{512}\zeta(4,4,2,2)+\frac{19}{1024}\zeta(9,3)-\frac{477}{1024}\zeta(9)\zeta(3)-\frac{4449}{1024}\zeta(7)\zeta(5)+\frac{27}{512}\zeta(3)^4$\\[1ex]
&&\hspace{13mm}$+\frac{81}{448}\zeta(7,3)\zeta(2)+\frac{567}{128}\zeta(7)\zeta(3)\zeta(2)+\frac{3807}{1792}\zeta(5)^2\zeta(2)-\frac{567}{512}\zeta(5,3)\zeta(4)$\\[1ex]
&&\hspace{13mm}$-\frac{2835}{512}\zeta(5)\zeta(3)\zeta(4)-\frac{27}{64}\zeta(3)^2\zeta(6)+\frac{3155095}{5660672}\zeta(12)$\\[1ex]
&&$\pi^{12}/10!=\frac{45045}{176896}\zeta(12)$\\[1ex]\hline&&\\[-1.5ex]
9&12&$Q_{12,4}=\lfrac{8634368}{135}f^2_1f^2_5f^2_3f^2_3-\lfrac{3899392}{135}f^2_1f^2_3f^2_5f^2_3+\lfrac{458752}{27}f^2_1f^2_3f^2_3f^2_5$\\[1ex]
&&\qquad\qquad$+\lfrac{222208}{38475}f^2_1f^2_3\pi^8-\lfrac{71206701679851520}{59408350617}f^2_1f^2_{11}$\\[1ex]
&&$Q_{12,5}=\lfrac{777728}{45}f^2_1f^2_5f^2_3f^2_3+\lfrac{990976}{45}f^2_1f^2_3f^2_5f^2_3-\lfrac{163072}{9}f^2_1f^2_3f^2_3f^2_5$\\[1ex]
&&\qquad\qquad$-\lfrac{194432}{38475}f^2_1f^2_3\pi^8+\lfrac{9739832477359040}{25460721693}f^2_1f^2_{11}$
\end{tabular}
\end{center}
\caption{Conversions of the expressions in Table~\ref{tabQ} into MZVs or Euler sums, see \cite{Schnetz:2008mp}. The numbers $Q_{12,4}$ and $Q_{12,5}$ were defined in \cite{PanzerSchnetz:2017coact}.
They are given in terms of the motivic $f$-alphabet for Euler sums that is used in \texttt{HyperlogProcedures}~\cite{Shlog}.}
\label{tabN}
\end{table}

Our calculations of Feynman periods in $\phi^3$ theory were performed using a tailor-made \texttt{C++} program by the first author  that feeds information
into \texttt{HyperlogProcedures}~\cite{Shlog} by the second author. The details of our approach are described in Section~\ref{sectsyst}. The results comprise Feynman periods in $\phi^3$ theory up to nine loops which have been included into the new release of the \texttt{HyperlogProcedures} \texttt{Maple}\footnote{\texttt{Maple} is a trademark of Waterloo Maple Inc.} package. A list of known Feynman periods in $\phi^3$ theory up to seven loops can be found in Appendix~\ref{app}. This list is also available in machine readable form in the ancillary files to the \texttt{arXiv} version of this article.
Up to eight loops all periods which could be calculated can be expressed in terms of the numbers $Q_\bullet$ which were identified in $\phi^4$ theory
\cite{Schnetz:2008mp,PanzerSchnetz:2017coact}. These numbers are listed in Tables~\ref{tabQ} and \ref{tabN}. We found the number $Q_{12,4}$ in various 8 loop $\phi^3$ periods, see \cite{Shlog}
(note that periods of $\leq7$ loops conjecturally have weights $\leq11$, see Conjecture~\ref{con:coaction}).
The number $Q_{12,5}$, which is conjectured in $\phi^4$ theory at 10 loops,
is absent in our (incomplete) $\phi^3$ theory data. At nine loops only a few $\phi^3$ periods were calculated. All of these are multiple zeta values (MZVs), i.e.\ rational linear
combinations of
\begin{equation*}
\zeta(n_d,\ldots,n_2,n_1)=\sum_{k_d>\cdots>k_1\geq1}\frac{1}{k_d^{n_d}\!\cdots k_2^{n_2}k_1^{n_1}}\quad\text{with}\quad n_d\geq 2.
\end{equation*}
The weight of $\zeta(n_d,\ldots,n_2,n_1)$ is $n_d+\ldots+n_2+n_1$. The weight is conjectured (proved in the motivic setup) to be a grading on MZVs.

Our results are compatible with the following conjecture. 

\begin{con}\label{con:coaction}\mbox{}
\begin{enumerate}
\item The number content of six-dimensional $\phi^3$ periods is a subset of, and possible equal to, the number content of four-dimensional $\phi^4$ periods.
\item The motivic Galois coaction closes on $\phi^3$ periods,
\begin{equation*}
\Delta\colon\Periods_{\phi^3}^{\motivic}\longrightarrow\Periods^\dR\otimes\Periods_{\phi^3}^{\motivic}.
\end{equation*}
\item The motivic Galois coaction is consistent with the loop-grading, i.e.\ the non-trivial part of the Galois coaction $\Delta'$ is provided by periods of lower loops.
\begin{equation*}
\Delta'\colon\Periods_{\phi^3,\leq n}^{\motivic}\longrightarrow\Periods^\dR \otimes\Periods_{\phi^3,\leq n-1}^{\motivic},
\end{equation*}
where $\Periods_{\phi^3,\leq n}$ is the $\QQ$ vector space of motivic $\phi^3$ periods at $\leq n$ loops.
\item The maximum weight in $\Periods_{\phi^3,\leq n}^{\motivic}$ is $2n-3$.
\end{enumerate}
\end{con}
The first statement of the conjecture supports the picture of the existence of a universal number content in all QFTs associated to the \emph{cosmic Galois group}.
The second statement is the coaction conjecture \cite{PanzerSchnetz:2017coact}.
The third statement implies the second statement. It is stronger than the respective conjecture in four-dimensional $\phi^4$ theory where one has to generalize the right hand side of the tensor product to all (not only $\phi^4$) log-divergent periods in four dimensions \cite{PanzerSchnetz:2017coact}.
Our data is compatible with the stronger statement. But the analogy to $\phi^4$ theory indicates that a period might appear at higher loop order that only fulfills the weaker statement.

The $\phi^4$ theory analogue of the fourth statement is easy to prove. The $\phi^3$ theory result is much more obscure, as one might expect a much higher weight from a naive analysis of the associated integrals (for a more detailed discussion see \cite[Conjecture 46]{gfe}).

The main difference between $\phi^3$ and $\phi^4$ periods is weight mixing (i.e.\ numbers of different weights appear in a $\QQ$-linear combination within the same period).
The mixing of weights in $\phi^4$ periods is severely constrained \cite{PanzerSchnetz:2017coact}, whereas in $\phi^3$ periods the weights mix much more freely. Within the period
of a single $n$-loop $\phi^3$ graph all weights from 0 to $2n-3$ may mix. In this context $\phi^3$ theory behaves more generically than $\phi^4$ theory. 

To facilitate the search for a \emph{diagrammatic coaction formula} for Feynman integrals and periods, 
i.e.~an explicit formula for $\Delta$ in Conjecture~\ref{con:coaction},
we included diagrammatic representations for the graphs in the period list in Appendix~\ref{app}. Such a coaction formula would likely 
provide the key to the proof of Conjecture~\ref{con:coaction} and related conjectures. Moreover, an explicit coaction 
formula might give rise to new powerful analytic evaluation techniques for Feynman periods and integrals --- for instance along the lines of~\cite{Brown:2011ik}.

\subsection{The \texorpdfstring{$\phi^3$}{phi3} theory Hepp bound is not a perfect invariant}
\label{sec:hepp}
In integer dimensions, the Hepp bound is a rational number associated to a graph~\cite{Panzer:2019yxl}. It can be seen as a certain tropicalization of the Feynman period. The Hepp bound fulfills similar combinatorial identities as the period and in $\phi^4$ theory Panzer conjectured that the Hepp bound is a \emph{complete invariant} with respect to the period of the associated graph~\cite[Conjecture~1.2]{Panzer:2019yxl}. This means that two $\phi^4$ graphs evaluate to the same period if and only if they have the same Hepp bound.

In $\phi^3$ theory we observe that equality of the Hepp bound does not imply equality of periods. The smallest counter example appears at 
$5$ loops. The period completed graphs (in the numbering of the table in Appendix~\ref{app})
\begin{align*} G_{5,5}^\star &= \begin{tikzpicture}[x=1.5em,y=1.5em,baseline={([yshift=-.7ex]current bounding box.center)}] \coordinate (v0) at (-1.000000,0.004067); \coordinate (v1) at (-0.196236,-0.420209); \coordinate (v2) at (-0.714069,-0.854259); \coordinate (v3) at (-0.439579,0.800952); \coordinate (v4) at (0.934325,0.635561); \coordinate (v5) at (0.393879,-0.356861); \coordinate (v6) at (1.000000,-0.414090); \coordinate (v7) at (0.114463,1.000000); \coordinate (v8) at (0.271573,0.423585); \coordinate (v9) at (0.643406,0.224135); \coordinate (v10) at (0.133537,-1.000000); \coordinate (v11) at (-0.348499,0.232895); \draw[preaction={draw, white, line width=.9pt, -},line width=.3pt] (v0) -- (v1); \draw[preaction={draw, white, line width=.9pt, -},line width=.3pt] (v0) -- (v2); \draw[preaction={draw, white, line width=.9pt, -},line width=.3pt] (v0) -- (v3); \draw[preaction={draw, white, line width=.9pt, -},line width=.3pt] (v1) -- (v9); \draw[preaction={draw, white, line width=.9pt, -},line width=.3pt] (v1) -- (v11); \draw[preaction={draw, white, line width=.9pt, -},line width=.3pt] (v2) -- (v10); \draw[preaction={draw, white, line width=.9pt, -},line width=.3pt] (v2) -- (v11); \draw[preaction={draw, white, line width=.9pt, -},line width=.3pt] (v3) -- (v7); \draw[preaction={draw, white, line width=.9pt, -},line width=.3pt] (v3) -- (v8); \draw[preaction={draw, white, line width=.9pt, -},line width=.3pt] (v4) -- (v6); \draw[preaction={draw, white, line width=.9pt, -},line width=.3pt] (v4) -- (v7); \draw[preaction={draw, white, line width=.9pt, -},line width=.3pt] (v4) -- (v8); \draw[preaction={draw, white, line width=.9pt, -},line width=.3pt] (v5) -- (v8); \draw[preaction={draw, white, line width=.9pt, -},line width=.3pt] (v5) -- (v9); \draw[preaction={draw, white, line width=.9pt, -},line width=.3pt] (v5) -- (v10); \draw[preaction={draw, white, line width=.9pt, -},line width=.3pt] (v6) -- (v9); \draw[preaction={draw, white, line width=.9pt, -},line width=.3pt] (v6) -- (v10); \draw[preaction={draw, white, line width=.9pt, -},line width=.3pt] (v7) -- (v11); \draw[line width=.3pt] (v0) -- ($(v0)!1pt!(v1)$); \draw[line width=.3pt] (v1) -- ($(v1)!1pt!(v0)$); \draw[line width=.3pt] (v0) -- ($(v0)!1pt!(v2)$); \draw[line width=.3pt] (v2) -- ($(v2)!1pt!(v0)$); \draw[line width=.3pt] (v0) -- ($(v0)!1pt!(v3)$); \draw[line width=.3pt] (v3) -- ($(v3)!1pt!(v0)$); \draw[line width=.3pt] (v1) -- ($(v1)!1pt!(v9)$); \draw[line width=.3pt] (v9) -- ($(v9)!1pt!(v1)$); \draw[line width=.3pt] (v1) -- ($(v1)!1pt!(v11)$); \draw[line width=.3pt] (v11) -- ($(v11)!1pt!(v1)$); \draw[line width=.3pt] (v2) -- ($(v2)!1pt!(v10)$); \draw[line width=.3pt] (v10) -- ($(v10)!1pt!(v2)$); \draw[line width=.3pt] (v2) -- ($(v2)!1pt!(v11)$); \draw[line width=.3pt] (v11) -- ($(v11)!1pt!(v2)$); \draw[line width=.3pt] (v3) -- ($(v3)!1pt!(v7)$); \draw[line width=.3pt] (v7) -- ($(v7)!1pt!(v3)$); \draw[line width=.3pt] (v3) -- ($(v3)!1pt!(v8)$); \draw[line width=.3pt] (v8) -- ($(v8)!1pt!(v3)$); \draw[line width=.3pt] (v4) -- ($(v4)!1pt!(v6)$); \draw[line width=.3pt] (v6) -- ($(v6)!1pt!(v4)$); \draw[line width=.3pt] (v4) -- ($(v4)!1pt!(v7)$); \draw[line width=.3pt] (v7) -- ($(v7)!1pt!(v4)$); \draw[line width=.3pt] (v4) -- ($(v4)!1pt!(v8)$); \draw[line width=.3pt] (v8) -- ($(v8)!1pt!(v4)$); \draw[line width=.3pt] (v5) -- ($(v5)!1pt!(v8)$); \draw[line width=.3pt] (v8) -- ($(v8)!1pt!(v5)$); \draw[line width=.3pt] (v5) -- ($(v5)!1pt!(v9)$); \draw[line width=.3pt] (v9) -- ($(v9)!1pt!(v5)$); \draw[line width=.3pt] (v5) -- ($(v5)!1pt!(v10)$); \draw[line width=.3pt] (v10) -- ($(v10)!1pt!(v5)$); \draw[line width=.3pt] (v6) -- ($(v6)!1pt!(v9)$); \draw[line width=.3pt] (v9) -- ($(v9)!1pt!(v6)$); \draw[line width=.3pt] (v6) -- ($(v6)!1pt!(v10)$); \draw[line width=.3pt] (v10) -- ($(v10)!1pt!(v6)$); \draw[line width=.3pt] (v7) -- ($(v7)!1pt!(v11)$); \draw[line width=.3pt] (v11) -- ($(v11)!1pt!(v7)$); \filldraw (v0) circle (.3pt); \filldraw (v1) circle (.3pt); \filldraw (v2) circle (.3pt); \filldraw (v3) circle (.3pt); \filldraw (v4) circle (.3pt); \filldraw (v5) circle (.3pt); \filldraw (v6) circle (.3pt); \filldraw (v7) circle (.3pt); \filldraw (v8) circle (.3pt); \filldraw (v9) circle (.3pt); \filldraw (v10) circle (.3pt); \filldraw (v11) circle (.3pt); \end{tikzpicture} & G_{5,9} ^\star &= \begin{tikzpicture}[x=1.5em,y=1.5em,baseline={([yshift=-.7ex]current bounding box.center)}] \coordinate (v0) at (-0.428401,0.550063); \coordinate (v1) at (0.363962,0.930476); \coordinate (v2) at (-1.000000,0.311604); \coordinate (v3) at (-0.243397,-0.190103); \coordinate (v4) at (0.626905,0.103757); \coordinate (v5) at (0.836549,-0.621373); \coordinate (v6) at (0.091171,-0.536065); \coordinate (v7) at (0.134825,0.483444); \coordinate (v8) at (-0.041976,-1.000000); \coordinate (v9) at (-0.636055,-0.617077); \coordinate (v10) at (1.000000,0.301642); \coordinate (v11) at (-0.403284,1.000000); \draw[preaction={draw, white, line width=.9pt, -},line width=.3pt] (v0) -- (v1); \draw[preaction={draw, white, line width=.9pt, -},line width=.3pt] (v0) -- (v2); \draw[preaction={draw, white, line width=.9pt, -},line width=.3pt] (v0) -- (v3); \draw[preaction={draw, white, line width=.9pt, -},line width=.3pt] (v1) -- (v10); \draw[preaction={draw, white, line width=.9pt, -},line width=.3pt] (v1) -- (v11); \draw[preaction={draw, white, line width=.9pt, -},line width=.3pt] (v2) -- (v9); \draw[preaction={draw, white, line width=.9pt, -},line width=.3pt] (v2) -- (v11); \draw[preaction={draw, white, line width=.9pt, -},line width=.3pt] (v3) -- (v7); \draw[preaction={draw, white, line width=.9pt, -},line width=.3pt] (v3) -- (v8); \draw[preaction={draw, white, line width=.9pt, -},line width=.3pt] (v4) -- (v6); \draw[preaction={draw, white, line width=.9pt, -},line width=.3pt] (v4) -- (v7); \draw[preaction={draw, white, line width=.9pt, -},line width=.3pt] (v4) -- (v10); \draw[preaction={draw, white, line width=.9pt, -},line width=.3pt] (v5) -- (v6); \draw[preaction={draw, white, line width=.9pt, -},line width=.3pt] (v5) -- (v8); \draw[preaction={draw, white, line width=.9pt, -},line width=.3pt] (v5) -- (v10); \draw[preaction={draw, white, line width=.9pt, -},line width=.3pt] (v6) -- (v9); \draw[preaction={draw, white, line width=.9pt, -},line width=.3pt] (v7) -- (v11); \draw[preaction={draw, white, line width=.9pt, -},line width=.3pt] (v8) -- (v9); \draw[line width=.3pt] (v0) -- ($(v0)!1pt!(v1)$); \draw[line width=.3pt] (v1) -- ($(v1)!1pt!(v0)$); \draw[line width=.3pt] (v0) -- ($(v0)!1pt!(v2)$); \draw[line width=.3pt] (v2) -- ($(v2)!1pt!(v0)$); \draw[line width=.3pt] (v0) -- ($(v0)!1pt!(v3)$); \draw[line width=.3pt] (v3) -- ($(v3)!1pt!(v0)$); \draw[line width=.3pt] (v1) -- ($(v1)!1pt!(v10)$); \draw[line width=.3pt] (v10) -- ($(v10)!1pt!(v1)$); \draw[line width=.3pt] (v1) -- ($(v1)!1pt!(v11)$); \draw[line width=.3pt] (v11) -- ($(v11)!1pt!(v1)$); \draw[line width=.3pt] (v2) -- ($(v2)!1pt!(v9)$); \draw[line width=.3pt] (v9) -- ($(v9)!1pt!(v2)$); \draw[line width=.3pt] (v2) -- ($(v2)!1pt!(v11)$); \draw[line width=.3pt] (v11) -- ($(v11)!1pt!(v2)$); \draw[line width=.3pt] (v3) -- ($(v3)!1pt!(v7)$); \draw[line width=.3pt] (v7) -- ($(v7)!1pt!(v3)$); \draw[line width=.3pt] (v3) -- ($(v3)!1pt!(v8)$); \draw[line width=.3pt] (v8) -- ($(v8)!1pt!(v3)$); \draw[line width=.3pt] (v4) -- ($(v4)!1pt!(v6)$); \draw[line width=.3pt] (v6) -- ($(v6)!1pt!(v4)$); \draw[line width=.3pt] (v4) -- ($(v4)!1pt!(v7)$); \draw[line width=.3pt] (v7) -- ($(v7)!1pt!(v4)$); \draw[line width=.3pt] (v4) -- ($(v4)!1pt!(v10)$); \draw[line width=.3pt] (v10) -- ($(v10)!1pt!(v4)$); \draw[line width=.3pt] (v5) -- ($(v5)!1pt!(v6)$); \draw[line width=.3pt] (v6) -- ($(v6)!1pt!(v5)$); \draw[line width=.3pt] (v5) -- ($(v5)!1pt!(v8)$); \draw[line width=.3pt] (v8) -- ($(v8)!1pt!(v5)$); \draw[line width=.3pt] (v5) -- ($(v5)!1pt!(v10)$); \draw[line width=.3pt] (v10) -- ($(v10)!1pt!(v5)$); \draw[line width=.3pt] (v6) -- ($(v6)!1pt!(v9)$); \draw[line width=.3pt] (v9) -- ($(v9)!1pt!(v6)$); \draw[line width=.3pt] (v7) -- ($(v7)!1pt!(v11)$); \draw[line width=.3pt] (v11) -- ($(v11)!1pt!(v7)$); \draw[line width=.3pt] (v8) -- ($(v8)!1pt!(v9)$); \draw[line width=.3pt] (v9) -- ($(v9)!1pt!(v8)$); \filldraw (v0) circle (.3pt); \filldraw (v1) circle (.3pt); \filldraw (v2) circle (.3pt); \filldraw (v3) circle (.3pt); \filldraw (v4) circle (.3pt); \filldraw (v5) circle (.3pt); \filldraw (v6) circle (.3pt); \filldraw (v7) circle (.3pt); \filldraw (v8) circle (.3pt); \filldraw (v9) circle (.3pt); \filldraw (v10) circle (.3pt); \filldraw (v11) circle (.3pt); \end{tikzpicture} & G_{5,13} ^\star &= \begin{tikzpicture}[x=1.5em,y=1.5em,baseline={([yshift=-.7ex]current bounding box.center)}] \coordinate (v0) at (-1.000000,0.073634); \coordinate (v1) at (0.162942,0.191717); \coordinate (v2) at (-0.742315,-0.616628); \coordinate (v3) at (-0.771877,0.684628); \coordinate (v4) at (0.734709,0.530147); \coordinate (v5) at (1.000000,-0.037729); \coordinate (v6) at (0.067880,-0.692482); \coordinate (v7) at (0.020683,1.000000); \coordinate (v8) at (-0.095617,-0.218483); \coordinate (v9) at (0.698577,-0.803969); \coordinate (v10) at (0.002434,-1.000000); \coordinate (v11) at (0.013241,0.654745); \draw[preaction={draw, white, line width=.9pt, -},line width=.3pt] (v0) -- (v1); \draw[preaction={draw, white, line width=.9pt, -},line width=.3pt] (v0) -- (v2); \draw[preaction={draw, white, line width=.9pt, -},line width=.3pt] (v0) -- (v3); \draw[preaction={draw, white, line width=.9pt, -},line width=.3pt] (v1) -- (v9); \draw[preaction={draw, white, line width=.9pt, -},line width=.3pt] (v1) -- (v11); \draw[preaction={draw, white, line width=.9pt, -},line width=.3pt] (v2) -- (v10); \draw[preaction={draw, white, line width=.9pt, -},line width=.3pt] (v2) -- (v11); \draw[preaction={draw, white, line width=.9pt, -},line width=.3pt] (v3) -- (v7); \draw[preaction={draw, white, line width=.9pt, -},line width=.3pt] (v3) -- (v8); \draw[preaction={draw, white, line width=.9pt, -},line width=.3pt] (v4) -- (v5); \draw[preaction={draw, white, line width=.9pt, -},line width=.3pt] (v4) -- (v6); \draw[preaction={draw, white, line width=.9pt, -},line width=.3pt] (v4) -- (v7); \draw[preaction={draw, white, line width=.9pt, -},line width=.3pt] (v5) -- (v8); \draw[preaction={draw, white, line width=.9pt, -},line width=.3pt] (v5) -- (v9); \draw[preaction={draw, white, line width=.9pt, -},line width=.3pt] (v6) -- (v8); \draw[preaction={draw, white, line width=.9pt, -},line width=.3pt] (v6) -- (v10); \draw[preaction={draw, white, line width=.9pt, -},line width=.3pt] (v7) -- (v11); \draw[preaction={draw, white, line width=.9pt, -},line width=.3pt] (v9) -- (v10); \draw[line width=.3pt] (v0) -- ($(v0)!1pt!(v1)$); \draw[line width=.3pt] (v1) -- ($(v1)!1pt!(v0)$); \draw[line width=.3pt] (v0) -- ($(v0)!1pt!(v2)$); \draw[line width=.3pt] (v2) -- ($(v2)!1pt!(v0)$); \draw[line width=.3pt] (v0) -- ($(v0)!1pt!(v3)$); \draw[line width=.3pt] (v3) -- ($(v3)!1pt!(v0)$); \draw[line width=.3pt] (v1) -- ($(v1)!1pt!(v9)$); \draw[line width=.3pt] (v9) -- ($(v9)!1pt!(v1)$); \draw[line width=.3pt] (v1) -- ($(v1)!1pt!(v11)$); \draw[line width=.3pt] (v11) -- ($(v11)!1pt!(v1)$); \draw[line width=.3pt] (v2) -- ($(v2)!1pt!(v10)$); \draw[line width=.3pt] (v10) -- ($(v10)!1pt!(v2)$); \draw[line width=.3pt] (v2) -- ($(v2)!1pt!(v11)$); \draw[line width=.3pt] (v11) -- ($(v11)!1pt!(v2)$); \draw[line width=.3pt] (v3) -- ($(v3)!1pt!(v7)$); \draw[line width=.3pt] (v7) -- ($(v7)!1pt!(v3)$); \draw[line width=.3pt] (v3) -- ($(v3)!1pt!(v8)$); \draw[line width=.3pt] (v8) -- ($(v8)!1pt!(v3)$); \draw[line width=.3pt] (v4) -- ($(v4)!1pt!(v5)$); \draw[line width=.3pt] (v5) -- ($(v5)!1pt!(v4)$); \draw[line width=.3pt] (v4) -- ($(v4)!1pt!(v6)$); \draw[line width=.3pt] (v6) -- ($(v6)!1pt!(v4)$); \draw[line width=.3pt] (v4) -- ($(v4)!1pt!(v7)$); \draw[line width=.3pt] (v7) -- ($(v7)!1pt!(v4)$); \draw[line width=.3pt] (v5) -- ($(v5)!1pt!(v8)$); \draw[line width=.3pt] (v8) -- ($(v8)!1pt!(v5)$); \draw[line width=.3pt] (v5) -- ($(v5)!1pt!(v9)$); \draw[line width=.3pt] (v9) -- ($(v9)!1pt!(v5)$); \draw[line width=.3pt] (v6) -- ($(v6)!1pt!(v8)$); \draw[line width=.3pt] (v8) -- ($(v8)!1pt!(v6)$); \draw[line width=.3pt] (v6) -- ($(v6)!1pt!(v10)$); \draw[line width=.3pt] (v10) -- ($(v10)!1pt!(v6)$); \draw[line width=.3pt] (v7) -- ($(v7)!1pt!(v11)$); \draw[line width=.3pt] (v11) -- ($(v11)!1pt!(v7)$); \draw[line width=.3pt] (v9) -- ($(v9)!1pt!(v10)$); \draw[line width=.3pt] (v10) -- ($(v10)!1pt!(v9)$); \filldraw (v0) circle (.3pt); \filldraw (v1) circle (.3pt); \filldraw (v2) circle (.3pt); \filldraw (v3) circle (.3pt); \filldraw (v4) circle (.3pt); \filldraw (v5) circle (.3pt); \filldraw (v6) circle (.3pt); \filldraw (v7) circle (.3pt); \filldraw (v8) circle (.3pt); \filldraw (v9) circle (.3pt); \filldraw (v10) circle (.3pt); \filldraw (v11) circle (.3pt); \end{tikzpicture} & \end{align*}
have different periods,
\begin{align*} P_{5,5} = P_{5,13} = -3\,\zeta(3)^3 + 4\, \zeta(3) &= 0.473405\cdots   \\
P_{5,9} = \frac{147}{16}\, \zeta(7) -5\, \zeta(5) -3 \,\zeta(3)&= 0.473400\cdots   \end{align*}
and the same Hepp bound\footnote{We thank Erik Panzer for sharing his Hepp bound computations in six-dimensional $\varphi^3$ theory.},
\begin{align*} H_{5,5} = H_{5,9} = H_{5,13} = \frac{59607}{8},  \end{align*}
where $H_{\ell,n}$ was computed as defined in \cite[eq.~(1.5)]{Panzer:2019yxl}.
The difference between the numerical values of the two periods is remarkably small. %

The table of $\phi^3$ periods in the ancillary material also includes the Hepp bound for each Feynman period (see Appendix~\ref{app}).

\section{Feynman periods and graphical functions}
\label{feynpergf}
\subsection{Momentum-space Feynman periods}
\label{sec:momperiods}
A massless Euclidean  momentum space two-point Feynman integral in $D$-dimensional spacetime associated to a graph $\td{G}$ with edges $\sE_{\td{G}}$ and a set of edge weights $\td{\bb \nu} = (\td{\nu}_1,\ldots,\td{\nu}_{|\sE_{\td{G}}|})$ and total in- and outgoing momentum $Q \in \RR^D$ is given by
\begin{align} \label{eq:feynmom} \td{I}_{\td{G}}(Q) = \int \left( \prod_{\ell} \frac{\dd^D k_\ell}{\pi^{\frac{D}{2}}} \right) \prod_{e \in \sE_{\td{G}}} \frac{1}{D_e^{\td{\nu}_e}}, \end{align}
where we integrate over $\RR^D$ for each loop momentum $k_\ell$. The factors $D_e^{-\td{\nu}_e}=q_e( \bb k, Q)^{-2\td{\nu}_e}$ in the integrand are the (weighted) massless Euclidean momentum-space Feynman propagators with $q_e( \bb k, Q)$ being the total loop or external momentum flowing through the edge $e$. Because the graph ${\td{G}}$ only has two external legs and no internal masses, the (convergent) integral on the right hand side can only depend on a power of $Q^2$ by standard symmetry arguments. 

The associated \emph{momentum-space Feynman period} is the following quotient which is independent of $Q$, 
as can easily be checked by dimensional analysis of~\eqref{eq:feynmom}:
\begin{align} \label{eq:pspaceperiod} \td{P}_{\td{G}} = \frac{\td{I}_{\td{G}}(Q)}{ (Q^2)^{-\td{\omega}_{\td{G}}} }, \end{align}
where 
$\td{\omega}_{\td{G}} = \sum_{e\in \sE_{\td{G}}} \td{\nu}_e - \frac{D}{2}h_1({\td{G}})$ 
is the \emph{momentum-space superficial degree of divergence} of ${\td{G}}$ and $h_1({\td{G}})$ is the loop number of ${\td{G}}$. 

\subsection{Parametric representation of Feynman periods}
\label{sec:parametric}

To formulate the \emph{parametric representation} (see, e.g.~\cite{Brown:2008um,gfe}) of Feynman periods we define the graph
${\td{G}'}$ as the graph obtained from ${\td{G}}$ by connecting its two external legs to form a new edge $e=|\sE_{\td{G}}|+1$.
We get

\begin{align} \label{eq:parametric} \td{P}_{\td{G}} = \frac{\Gamma\left( \frac{D}{2}\right)}{ \Gamma\left( \frac{D}{2} - \td{\omega}_{\td{G}}\right) \prod_{e\in \sE_{\td{G}}} \Gamma(\td{\nu}_e) } \int_{x_e \geq 0} \frac{\prod_{e=1}^{|\sE_{\td{G}}|} x_e^{\td{\nu}_e}}{(\Psi_{\td{G}'})^{\frac{D}{2}}} \Big|_{x_{|\sE_{\td{G}}|+1}=1} \frac{ \dd x_1 \cdots \dd x_{|\sE_{\td{G}}|} } { x_1 \cdots x_{|\sE_{\td{G}}|} }, \end{align}
where $\Psi_{{\td{G}'}}$ is the  \emph{Kirchhof-polynomial} in the variables $x_1,\ldots,x_{|\sE_{\td{G}}|+1}$. It is given by
\begin{align} \Psi_{{\td{G}'}} = \sum_{T \subset {\td{G}'}} \prod_{e \not \in T} x_e, \end{align}
where the sum is over all \emph{spanning trees} of ${\td{G}'}$.

The fact that many different propagator graphs can be \emph{closed} to form the same graph $G'$ is useful in many contexts. It is also known as the \emph{cut-and-glue} identity~\cite{Chetyrkin:1980pr,Chetyrkin:1981qh,Baikov:2010hf}.

The parametric representation connects the theory of Feynman periods to the theory of canonical invariant forms for graph complex computations~\cite{Brown:2021umn}.

As is often the case when studying a family of numbers, it is helpful to generalize to functions which have the numbers under inspection as certain special values. This way, tools from analysis can be applied. A particularly natural such generalization in the case of Feynman periods are \emph{graphical functions}.

\subsection{Graphical functions}
\label{sec:graphicalfunctions}
We will briefly review the basics of the graphical function method. For details and proofs we refer to \cite{gfe}.
Graphical functions are specially parameterized Euclidean massless position-space Feynman integrals associated to graphs with three external vertices. For a graph $G$ with external vertices $x_a,x_b,x_c$ and a set of position-space edge weights $\nu_{1}, \ldots, \nu_{|\sE_G|}$ %
such a Feynman integral reads
\begin{align} \label{feynmanint} I_G(x_a,x_b,x_c) = \int \left(\prod_{v\in \sVGint} \frac{\dd^D y_v}{\pi^{D/2}} \right) \prod_{e=\{v,w\} \in \sE_G} \frac{1}{\| y_v -y_w\|^{(D-2)\nu_e}}, \end{align}
where we integrate over $\RR^D$ for each \emph{internal vertex} in $\sVGint$. The integrand is the product of the massless position space propagators. Each propagator is a power of the Euclidean distance between the two incident vertices of the associated edge. We identify the variable $y_v$ with the respective external variable $x_a,x_b$ or $x_c$ if the incident vertex $v$ is external. 
In contrast to the momentum-space case, the position-space edge weights $\nu_1,\ldots,\nu_{|\sE_G|}$ are \emph{additative}: If the graph $G$ has two parallel edges $u,v$ with weights $\nu_{u}$
and $\nu_v$ both incident to the same pair of vertices, then we can replace these edges by a single edge with weight $\nu_u + \nu_v$ (and vice versa).

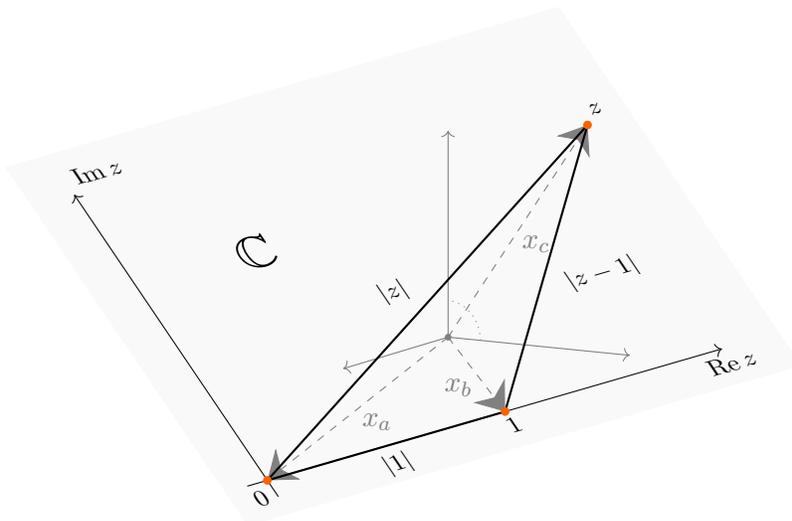
\begin{figure}[ht]
\tdplotsetmaincoords{80}{120}
\begin{center}
\begin{tikzpicture}[tdplot_main_coords,scale=.62] \draw[thin, ->,black!50] (0,0,0) -- (4.5,0,0); \draw[thin, ->,black!50] (0,0,0) -- (0,4.5,0); \draw[thin, ->,black!50] (0,0,0) -- (0,0,4.5); \tdplotsetrotatedcoords{0}{90}{0} \draw[dotted,black!50,tdplot_rotated_coords] (0,.8,0) arc (90:180:.8); \pgfmathsetmacro{\Ox}{32} \pgfmathsetmacro{\Oy}{14} \pgfmathsetmacro{\Oz}{3} \pgfmathsetmacro{\Onex}{-5} \pgfmathsetmacro{\Oney}{3} \pgfmathsetmacro{\Onez}{1} \pgfmathsetmacro{\Zx}{-12} \pgfmathsetmacro{\Zy}{1} \pgfmathsetmacro{\Zz}{6} \tdplotcrossprod(\Onex,\Oney,\Onez)(\Zx,\Zy,\Zz) \pgfmathsetmacro{\rx}{\tdplotresx} \pgfmathsetmacro{\ry}{\tdplotresy} \pgfmathsetmacro{\rz}{\tdplotresz} \tdplotcrossprod(\rx,\ry,\rz)(\Onex,\Oney,\Onez) \pgfmathsetmacro{\tx}{\tdplotresx/100} \pgfmathsetmacro{\ty}{\tdplotresy/100} \pgfmathsetmacro{\tz}{\tdplotresz/100} \pgfmathsetmacro{\dplane}{\rx*\Ox + \ry*\Oy + \rz*\Oz} \pgfmathsetmacro{\OneLen}{{sqrt(\Onex*\Onex+\Oney*\Oney+\Onez*\Onez)}} \pgfmathsetmacro{\iLen}{{sqrt(\tx*\tx+\ty*\ty+\tz*\tz)}} \tikzset{perspective/.style= {canvas is plane={O(0,0,0)x(\Onex/\OneLen,\Oney/\OneLen,\Onez/\OneLen)y(\tx/\iLen,\ty/\iLen,\tz/\iLen)}} } \pgfmathsetmacro{\axisscale}{2} \tikzset{perspective2/.style= {canvas is plane={O(0,0,0)x(\axisscale*\Onex/\OneLen,\axisscale*\Oney/\OneLen,\axisscale*\Onez/\OneLen)y(\axisscale*\tx/\iLen,\axisscale*\ty/\iLen,\axisscale*\tz/\iLen)}} } \pgfmathsetmacro{\axisovershoot}{2} \pgfmathsetmacro{\axisundershoot}{.5} \coordinate (v0) at (\Ox,\Oy,\Oz); \coordinate (v1) at ([shift={(\Onex,\Oney,\Onez)}]v0); \coordinate (vz) at ([shift={(\Zx,\Zy,\Zz)}]v0); \coordinate (vi) at ([shift={(\tx,\ty,\tz)}]v0); \filldraw[black!50] (0,0,0) circle (1.7pt); \draw[dashed,-{Stealth[length=10pt, width=15pt]},black!50] (0,0,0) -- node[inner sep=1.5pt,below right] {$x_a$} (v0); \draw[dashed,-{Stealth[length=10pt, width=15pt]},black!50] (0,0,0) -- node[inner sep=1.5pt,below left] {$x_b$} (v1); \draw[dashed,-{Stealth[length=10pt, width=15pt]},black!50] (0,0,0) -- node[inner sep=1.5pt,below right] {$x_c$} (vz); \draw[thin,->] ($(v0)!-\axisundershoot/\OneLen!(v1)$) -- ($(v0)!{1+2.7*\axisovershoot/\OneLen}!(v1)$) node[perspective,anchor=north,opacity=1]{$\Re z$}; \draw[thin,->] ($(v0)!-\axisundershoot/\iLen!(vi)$) -- ($(v0)!{\OneLen/\iLen+1.4*\axisovershoot/\iLen}!(vi)$) node[perspective,anchor=south west,opacity=1]{$\Im z$}; \coordinate (C1) at ($(v0)!{2*\OneLen/\iLen}!(vi)$); \node[perspective2] (C) at ($(C1)!{.5}!(v1)$) {$\CC$}; \pgfmathsetmacro{\xscale}{1.7} \coordinate (R0) at (${1 + 2.1*\axisundershoot/\OneLen + 2*\axisundershoot/\iLen}*(v0) - 2.1*\axisundershoot/\OneLen*(v1)- 2*\axisundershoot/\iLen*(vi)$); \coordinate (R1) at (${ - \xscale*2*\axisovershoot/\OneLen + 2*\axisundershoot/\iLen}*(v0) + {1 + 2*\xscale*\axisovershoot/\OneLen}*(v1)- 2*\axisundershoot/\iLen*(vi)$); \coordinate (R2) at (${ - \OneLen/\iLen - 2*\xscale*\axisovershoot/\OneLen - 2*\axisovershoot/\iLen}*(v0) + {1 + 2*\xscale*\axisovershoot/\OneLen}*(v1)+ {\OneLen/\iLen + 2*\axisovershoot/\iLen}*(vi)$); \coordinate (R3) at (${1 - \OneLen/\iLen + 2.1*\axisundershoot/\OneLen - 2*\axisovershoot/\iLen}*(v0) - {2.1*\axisundershoot/\OneLen}*(v1)+ {\OneLen/\iLen + 2*\axisovershoot/\iLen}*(vi)$); \draw[opacity = .05,fill,black!50] (R0) -- (R1) -- (R2) -- (R3); \draw[thick] (v0) -- node[perspective,below]{$|1|$} (v1); \draw[thick] (v0) -- node[perspective,above left]{$|z|$} (vz); \draw[thick] (vz) -- node[perspective,below right]{$|z-1|$} (v1); \filldraw (v0) circle(1.7pt) node[below left,perspective] {$0$}; \filldraw (v1) circle(1.7pt) node[below,perspective] {$1$}; \filldraw (vz) circle(1.7pt) node[above right,perspective] {$z$}; \filldraw[col1] (v0) circle(2.3pt); \filldraw[col1] (v1) circle(2.3pt); \filldraw[col1] (vz) circle(2.3pt); \end{tikzpicture}
\end{center}
\caption{The vectors $x_a,x_b,x_c\in\RR^D$ span a plane that is identified with the complex plane $\CC$ by requiring $x_a,x_b$ to coincide with $0,1\in \CC$.
The position of the vector $x_c$ inside the plane determines the value $z\in\CC$ (up to conjugation). Comparing ratios of squared side lengths of the triangles $x_a,x_b,x_c \in \RR^D$
and $0,1,z \in \CC$ gives the relations in \eqref{eqinvs}.}
\label{fig:Ctriangle}
\end{figure}

The value of $I_G(x_a,x_b,x_c)$ is invariant under Poincare transformations and scalings of the $D$-dimensional vectors $x_a,x_b,x_c$. 
For an efficient parameterization that resolves these symmetries we identify the affine plane that is spanned by the 
three vectors $x_a,x_b,x_c \in \RR^D$ with the complex plane $\CC$, associating $x_a$ to $0 \in \CC$, $x_b$ to $1\in \CC$ and $x_c$ to a free parameter $z \in \CC$.
By identification of the standard metric on $\CC$ with the Euclidean metric in $\RR^D$ we get the identities,
\begin{align} \label{eqinvs} z\zz &= \frac{\|x_c - x_a\|^2}{\|x_a - x_b\|^2}, &&& (1-z)(1-\zz) &= \frac{\|x_c - x_b\|^2}{\|x_a - x_b\|^2}, \end{align}
which are evident from Figure~\ref{fig:Ctriangle}.
Up to a trivial factor, the value of $I_G(x_a,x_b,x_c)$ can only depend on $z$.
Therefore, we can define a function $f_G:\CC \rightarrow \RR$ that captures the nontrivial dependence of $I_G$ on the invariants~\eqref{eqinvs},
\begin{align} \label{eq:gfdef} f_G(z) = \frac{ I_G(x_a,x_b,x_c) }{ (\|x_a - x_b\|^2)^{-\omega_G} } \end{align}
where $\omega_G = \frac{(D-2)}{2} \sum_{e\in \sE_G} \nu_e - \frac{D}{2} \VGint$ is the \emph{position-space superficial degree of divergence} of $G$.
The function $f_G(z)$ is the \emph{graphical function} associated to the graph $G$. 
Note that, this definition is completely analogous to the one from the momentum-space Feynman period in~\eqref{eq:pspaceperiod}.

Provided that UV and IR singularities are properly regulated (for instance by dimensional regularization) the graphical function is uniquely defined for all graphs $G$ with three external vertices and arbitrary edge weights. 

Figure~\ref{fig:Ctriangle} suggests to change the labels $x_a,x_b,x_c$ for the external vertices of the Feynman graph $G$ to $0,1,z$. This identification will also be useful to capture the symmetries of graphical functions under special Möbius transformations (see Section~\ref{sec:perm}).

A graphical function $f_G(z)$ is a \emph{single-valued} real analytic function on $\CC \setminus \{0,1\}$ \cite{Golz:2015rea}.
(Note that graphical functions only have singularities at 0, 1, and $\infty$.)
The graphical functions that are going to be discussed in this article can all be expressed in the function space of \emph{generalized single-valued hyperlogarithms} (GSVHs).
GSVHs are single-valued iterated integrals of differential forms
$ \frac{\dd z}{a z \zz + b z + c\zz + d} $
with $a,b,c,d \in \CC$. The theory of this function space has been developed by the second author in~\cite{Schnetz:2021ebf}.
The function space of GSVHs with singularities at $0,1,\infty$ is closed under addition, multiplication, holomorphic and  anti-holomorphic differentiation ($\partial_z$ and $\partial_{\zz}$), holomorphic and anti-holomorphic integration ($\int \dd z$ and $\int \dd \zz$), and special Möbius transformations generated by $z\mapsto \frac{1}{z}$ and $z\mapsto 1-z$.

The key feature of the graphical function method is the direct correspondence between the graph $G$ and the complex function $f_G$,
\begin{align*} G=\,& \begin{tikzpicture}[baseline={([yshift=-.7ex]0,0)}] \coordinate (v) at (0,0); \useasboundingbox ({-2*\rada},{-2*\rada}) rectangle ({2*\rada},{2*\rada}); \coordinate[label=above left:$1$] (v1) at ([shift=(120:\rada)]v); \coordinate[label=below left:$0$] (v0) at ([shift=(240:\rada)]v); \coordinate[label=right:$z$] (vz) at ([shift=(0:\rada)]v); \coordinate (vm) at ([shift=(-90:{2*\rada})]v); \draw[pattern=north west lines, pattern color=black!20] (v1) arc (150:210:{sqrt(3)*\rada}) arc (-90:-30:{sqrt(3)*\rada}) arc (30:90:{sqrt(3)*\rada}); \filldraw[col1] (v0) circle (1.7pt); \filldraw[col1] (v1) circle (1.7pt); \filldraw[col1] (vz) circle (1.7pt); \end{tikzpicture} & & \begin{tikzpicture}[baseline={([yshift=-.7ex]0,0)}] \useasboundingbox ({-2*\rada},{-2*\rada}) rectangle ({2*\rada},{2*\rada}); \draw[dashed,<->] ({-.7*\rada},0) -- ({.7*\rada},0); \end{tikzpicture} & & \begin{tikzpicture}[baseline={([yshift=-.9ex]0,0)}] \coordinate (v) at (0,0); \useasboundingbox ({-2*\rada},{-2*\rada}) rectangle ({2*\rada},{2*\rada}); \node (0,0) {$ f_{G}(z) $}; \end{tikzpicture} \end{align*}
As indicated, we will use \textbf{\color{col1} orange} for the external vertices $0,1,z$ to distinguish them from the \textbf{black} internal vertices over which we integrate.

\subsection{Position-space Feynman periods}
\label{sec:positionperiod}

In analogy to momentum-space, a position-space Feynman period is the value of a massless Euclidean position-space two-point Feynman integral up to a scale dependence. 
Specializing from three-point to two-point functions works transparently in position space: The massless Feynman integral corresponding to a graph $G$ with two external vertices is the special case of \eqref{feynmanint} where the right hand side does not dependent on the value of $x_c$. Hence, $I_G$ is only a function of $x_a$ and $x_b$. It follows that $f_G(z)$ defined in \eqref{eq:gfdef} is a \emph{constant} function. This constant is the position-space period: $P_G = f_G(z)$. 
So, we can depict the relationship between graphs, graphical functions and position-space Feynman periods as follows:
\begin{align} \label{eq:period_def} G=\,& \begin{tikzpicture}[baseline={([yshift=-.7ex]0,0)}] \coordinate (v) at (0,0); \useasboundingbox ({-2*\rada},{-2*\rada}) rectangle ({2*\rada},{2*\rada}); \coordinate[label=above:$1$] (v1) at ([shift=(120:\rada)]v); \coordinate[label=below:$0$] (v0) at ([shift=(240:\rada)]v); \coordinate[label=right:$z$] (vz) at ([shift=(0:\rada)]v); \coordinate (vm) at ([shift=(-90:{2*\rada})]v); \draw[pattern=north west lines, pattern color=black!20] (v1) arc (135:225:{\rada*sqrt(3)/sqrt(2)}) arc (-45:45:{\rada*sqrt(3)/sqrt(2)}); \filldraw[col1] (v0) circle (1.7pt); \filldraw[col1] (v1) circle (1.7pt); \filldraw[col1] (vz) circle (1.7pt); \end{tikzpicture} & & \begin{tikzpicture}[baseline={([yshift=-.7ex]0,0)}] \useasboundingbox ({-2*\rada},{-2*\rada}) rectangle ({2*\rada},{2*\rada}); \draw[dashed,<->] ({-.7*\rada},0) -- ({.7*\rada},0); \end{tikzpicture} & & \begin{tikzpicture}[baseline={([yshift=-.9ex]0,0)}] \coordinate (v) at (0,0); \useasboundingbox ({-2*\rada},{-2*\rada}) rectangle ({2*\rada},{2*\rada}); \node (0,0) {$ P_G = f_{G}(z) $}; \end{tikzpicture} \end{align}
where the left hand side represents the Feynman period as a graph with two external vertices $0,1$ and an isolated external vertex $z$.

Momentum- and position-space Feynman periods are equivalent up to certain pre\-factors that depend on the dimension and the edge weights. 
If $G$ is a graph with two external vertices labeled $0,1$ and $\td{G}$ is the same graph
with legs attached to those distinguished vertices, then the 
momentum and position-space Feynman periods $\td{P}_{\td{G}}$ and $P_G$ are related by
\begin{align} \label{eq:momposperiod} \Gamma\left(\frac{D}{2} - \omega_G\right) \left( \prod_{e} \Gamma( \lambda \nu_e ) \right) P_G = \Gamma\left(\frac{D}{2} - \td{\omega}_{\td{G}}\right) \left( \prod_{e} \Gamma( \td{\nu}_e ) \right) \td{P}_{\td{G}}, \end{align}
where $\lambda = \frac{D-2}{2}$. The momentum-space edge weights $\td{\nu}_e$ and the position-space edge weights $\nu_e$ are related by
$ \lambda \nu_e = \frac{D}{2} - \td{\nu}_e. $
As $P_G$ and $\td{P}_{\td G}$ are meromorphic functions in $\nu_1,\ldots,\nu_{|\sE_G|}$ and $D$ (see~\cite{Speer:1975dc}), the above relation holds for these functions and not only for specific (integer) values of $\nu_1,\ldots, \nu_{|\sE_G|}$ and $D$. This way, the identity is also useful in singular cases such as $\lambda\nu_e = -1$.%

\section{Transformation rules}
\label{sec:trafo}

In this section we describe the main transformation rules for graphical functions that lie at the heart of the method. Each transformation rule consists of a combinatorial operation on the graph $G$ and an analytic operation on the associated function $f_G(z)$ such that the correspondence between graph and function is left intact. In fact, a large class of graphs can be obtained by starting with the empty graph corresponding to the trivial graphical function, 
\begin{align} \label{eq:trivialgf} G_1=\,& \begin{tikzpicture}[baseline={([yshift=-.7ex]0,0)}] \coordinate (v) at (0,0); \useasboundingbox ({-2*\rada},{-2*\rada}) rectangle ({2*\rada},{2*\rada}); \coordinate[label=above left:$1$] (v1) at ([shift=(120:\rada)]v); \coordinate[label=below left:$0$] (v0) at ([shift=(240:\rada)]v); \coordinate[label=right:$z$] (vz) at ([shift=(0:\rada)]v); \coordinate (vm) at ([shift=(-90:{2*\rada})]v); \filldraw[col1] (v0) circle (1.7pt); \filldraw[col1] (v1) circle (1.7pt); \filldraw[col1] (vz) circle (1.7pt); \end{tikzpicture} & & \begin{tikzpicture}[baseline={([yshift=-.7ex]0,0)}] \useasboundingbox ({-2*\rada},{-2*\rada}) rectangle ({2*\rada},{2*\rada}); \draw[dashed,<->] ({-.7*\rada},0) -- ({.7*\rada},0); \end{tikzpicture} & & \begin{tikzpicture}[baseline={([yshift=-.9ex]0,0)}] \coordinate (v) at (0,0); \useasboundingbox ({-2*\rada},{-2*\rada}) rectangle ({2*\rada},{2*\rada}); \node (0,0) {$ f_{G_1}(z) =1 $}; \end{tikzpicture} \end{align}
and applying successive transformations. Graphs that can be obtained this way are called \emph{constructible}. For instance, all ladder or zig-zag graphs belong to this class.

\subsection{Adding edges between external vertices}
\label{sec:addedges}
A simple transformation rule that is still very handy for manipulating graphical functions 
is \emph{adding an edge between external vertices}: 
\begin{align*} G=\,& \begin{tikzpicture}[baseline={([yshift=-.7ex]0,0)}] \coordinate (v) at (0,0); \useasboundingbox ({-2.5*\rada},{-2*\rada}) rectangle ({2.5*\rada},{2*\rada}); \coordinate[label=above left:$1$] (v1) at ([shift=(120:\rada)]v); \coordinate[label=below left:$0$] (v0) at ([shift=(240:\rada)]v); \coordinate[label=right:$z$] (vz) at ([shift=(0:\rada)]v); \coordinate (vm) at ([shift=(-90:{2*\rada})]v); \draw[pattern=north west lines, pattern color=black!20] (v1) arc (150:210:{sqrt(3)*\rada}) arc (-90:-30:{sqrt(3)*\rada}) arc (30:90:{sqrt(3)*\rada}); \filldraw[col1] (v0) circle (1.7pt); \filldraw[col1] (v1) circle (1.7pt); \filldraw[col1] (vz) circle (1.7pt); \end{tikzpicture} & & \begin{tikzpicture}[baseline={([yshift=-.9ex]0,0)}] \coordinate (v) at (0,0); \useasboundingbox ({-7*\rada},{-2*\rada}) rectangle ({7*\rada},{2*\rada}); \node (0,0) {$ f_{G}(z)=\, \displaystyle (z\zz)^{\lambda \nu_{0z}} ((1-z)(1-\zz))^{\lambda \nu_{1z}} f_{G'}(z) $}; \end{tikzpicture} \\
& \begin{tikzpicture}[baseline={([yshift=-.7ex]0,0)}] \useasboundingbox ({-2.5*\rada},{-.7*\rada}) rectangle ({2.5*\rada},{.7*\rada}); \draw[<->] (0,{-.7*\rada}) -- (0,{.7*\rada}); \end{tikzpicture} & & \begin{tikzpicture}[baseline={([yshift=-.7ex]0,0)}] \useasboundingbox ({-7*\rada},{-.7*\rada}) rectangle ({7*\rada},{.7*\rada}); \draw[<->] (0,{-.7*\rada}) -- (0,{.7*\rada}); \end{tikzpicture} \notag \\
G'=\,& \begin{tikzpicture}[baseline={([yshift=-.7ex]0,0)}] \coordinate (v) at (0,0); \useasboundingbox ({-2.5*\rada},{-2*\rada}) rectangle ({2.5*\rada},{2*\rada}); \coordinate[label=above left:$1$] (v1) at ([shift=(120:\rada)]v); \coordinate[label=below left:$0$] (v0) at ([shift=(240:\rada)]v); \coordinate[label=right:$z$] (vz) at ([shift=(0:\rada)]v); \coordinate (vm) at ([shift=(90:{2*\rada})]v); \draw[pattern=north west lines, pattern color=black!20] (v1) arc (150:210:{sqrt(3)*\rada}) arc (-90:-30:{sqrt(3)*\rada}) arc (30:90:{sqrt(3)*\rada}); \draw (v1) arc (90:270:{sqrt(3)/2*\rada}) node[pos=.5,left] {$\nu_{01}$}; \draw (vz) arc (-30:150:{sqrt(3)/2*\rada}) node[pos=.5,right] {$\nu_{1z}$}; \draw (v0) arc (-150:30:{sqrt(3)/2*\rada}) node[pos=.5,right] {$\nu_{0z}$}; \filldraw[col1] (v0) circle (1.7pt); \filldraw[col1] (v1) circle (1.7pt); \filldraw[col1] (vz) circle (1.7pt); \end{tikzpicture} & & \begin{tikzpicture}[baseline={([yshift=-.9ex]0,0)}] \coordinate (v) at (0,0); \useasboundingbox ({-7*\rada},{-2*\rada}) rectangle ({7*\rada},{2*\rada}); \node (0,0) {$ \displaystyle f_{G'}(z)=\, (z\zz)^{-\lambda \nu_{0z}} ((1-z)(1-\zz))^{-\lambda \nu_{1z}} f_G(z) $}; \end{tikzpicture}  \end{align*}
where $\lambda = \frac{D}{2}-1$.
Adding or removing an edge of arbitrary weight between a pair of external vertices of the graph (depicted on the left hand side)
only changes the graphical function (depicted on the right hand side) by a factor. If the exponents $\lambda \nu_{e}$ are integers, 
then this factor is a rational function in $z$ and $\zz$. 
Note that adding an edge between the external edges $0$ and $1$ leaves the function unchanged. This operation only modifies the total weight of the function 
as ${\omega_{G'} = \lambda ( \nu_{01} + \nu_{0z} + \nu_{1z} ) + \omega_G}$. 

The validity of this transformation rule can easily be proved using the 
Feynman integral representation~\eqref{feynmanint} and \eqref{eq:gfdef} with the parameterization~\eqref{eqinvs}.
\subsection{Appending an external edge}
\label{sec:appendedge}
The most important transformation rule is \emph{appending an edge of weight $1$} to the external vertex $z$ while creating a new internal vertex. The utility of this operation stems from the increase in complexity of the associated Feynman integral while passing from the graph $G$ to $G'$. The graph $G'$ has one more internal vertex and the associated integral therefore features one more integration over $D$-dimensional spacetime in the position-space Feynman integral representation~\eqref{feynmanint}. 
Here are both sides of this transformation rule:
\begin{align} G=\,& \begin{tikzpicture}[baseline={([yshift=-.7ex]0,0)}] \coordinate (v) at (0,0); \useasboundingbox ({-1.5*\rada},{-2*\rada}) rectangle ({3.5*\rada},{2*\rada}); \coordinate[label=above left:$1$] (v1) at ([shift=(120:\rada)]v); \coordinate[label=below left:$0$] (v0) at ([shift=(240:\rada)]v); \coordinate[label=right:$z$] (vz) at ([shift=(0:\rada)]v); \coordinate (vm) at ([shift=(-90:{2*\rada})]v); \draw[pattern=north west lines, pattern color=black!20] (v1) arc (150:210:{sqrt(3)*\rada}) arc (-90:-30:{sqrt(3)*\rada}) arc (30:90:{sqrt(3)*\rada}); \filldraw[col1] (v0) circle (1.7pt); \filldraw[col1] (v1) circle (1.7pt); \filldraw[col1] (vz) circle (1.7pt); \end{tikzpicture} & & \begin{tikzpicture}[baseline={([yshift=-.3ex]0,0)}] \coordinate (v) at (0,0); \useasboundingbox ({-7*\rada},{-2*\rada}) rectangle ({7*\rada},{2*\rada}); \node (0,0) { $ \displaystyle f_{G}(z)=\, -\Gamma(\lambda) \, \frac{1}{(z-\zz)^{\lambda}} \Delta_{\lambda-1} (z-\zz)^{\lambda} f_{G'}(z) $ }; \end{tikzpicture} \label{eqappend_diff} \\
& \begin{tikzpicture}[baseline={([yshift=-.7ex]0,0)}] \useasboundingbox ({-2.5*\rada},{-.7*\rada}) rectangle ({2.5*\rada},{.7*\rada}); \draw[<->] (0,{-.7*\rada}) -- (0,{.7*\rada}); \end{tikzpicture} & & \begin{tikzpicture}[baseline={([yshift=-.7ex]0,0)}] \useasboundingbox ({-7*\rada},{-.7*\rada}) rectangle ({7*\rada},{.7*\rada}); \draw[<->] (0,{-.7*\rada}) -- (0,{.7*\rada}); \end{tikzpicture} \notag \\
G'=\,& \begin{tikzpicture}[baseline={([yshift=-.7ex]0,0)}] \coordinate (v) at (0,0); \useasboundingbox ({-1.5*\rada},{-2*\rada}) rectangle ({3.5*\rada},{2*\rada}); \coordinate[label=above left:$1$] (v1) at ([shift=(120:\rada)]v); \coordinate[label=below left:$0$] (v0) at ([shift=(240:\rada)]v); \coordinate (vx) at ([shift=(0:\rada)]v); \coordinate[label=right:$z$] (vz) at ([shift=(0:{(1+sqrt(3))*\rada})]v); \coordinate (vm) at ([shift=(90:{2*\rada})]v); \draw (vx) -- (vz); \draw[pattern=north west lines, pattern color=black!20] (v1) arc (150:210:{sqrt(3)*\rada}) arc (-90:-30:{sqrt(3)*\rada}) arc (30:90:{sqrt(3)*\rada}); \filldraw[col1] (v0) circle (1.7pt); \filldraw[col1] (v1) circle (1.7pt); \filldraw (vx) circle (1.7pt); \filldraw[col1] (vz) circle (1.7pt); \end{tikzpicture} & & \begin{tikzpicture}[baseline={([yshift=-.3ex]0,0)}] \coordinate (v) at (0,0); \useasboundingbox ({-7*\rada},{-2*\rada}) rectangle ({7*\rada},{2*\rada}); \node (0,0) {$ \displaystyle f_{G'}(z)=\, -\frac{1}{\Gamma(\lambda)} \, \frac{1}{(z-\zz)^{\lambda}} \mathcal{I}_{\lambda-1} (z-\zz)^{\lambda} f_{G}(z) $}; \end{tikzpicture} \label{eqappend_int} \end{align}
where $\lambda = \frac{D}{2}-1$. The effective Laplacian on graphical functions is given by 
$\Delta_{\lambda-1} = \partial_z \partial_{\zz} + \frac{\lambda(\lambda-1)}{(z-\zz)^2}$.
The integration operator $\mathcal I_{\lambda-1}$  which inverts the Laplacian reads
\begin{align*} \mathcal I_{\lambda-1} &= \sum^{\lambda-1}_{k,\ell=0} c_{\lambda-1,k,\ell} \frac{1}{(z-\zz)^k} \int_{\textrm{sv}} \dd z (z-\zz)^{k+\ell} \int_{\textrm{sv}} \dd \zz \frac{1}{(z-\zz)^\ell} \end{align*}
with coefficients $c_{n,k,\ell} = (-1)^{n+k+\ell} \frac{(n+k)!(n+\ell)!}{(n-k)!(n-\ell)!(k+\ell)!k!\ell!}$. The single valued integration operators $\int_{\text{sv}} \dd z$ in the definition of $\mathcal I_{\lambda-1}$ have to be interpreted properly on a suitable function space. A substantial difficulty is to get control over the kernel of the differential operator $\Delta_{\lambda-1}$.
This kernel is reflected in the ambiguity of the integral operators $\int_{\textrm{sv}} \dd z$ and $\int_{\textrm{sv}} \dd\zz$ by rational (anti-)holomorphic functions.
A major result in the theory of graphical functions is that the kernel of $\Delta_{\lambda-1}$ is trivial in the space of graphical functions and that it can be controlled by an efficient algorithm,
see~\cite[Sec.~6]{gfe} for details.

It is a straightforward exercise in variable transformations to prove \eqref{eqappend_diff} %
(see~\cite[Sec.~1]{gfe}). The explicit form for $f_{G'}(z)$ in \eqref{eqappend_int} is substantially harder to prove. This formula is one of the main results of~\cite{gfe} (see Theorem~34).

Appending an edge with weight $\nu_e \neq 1$ is also possible in many cases. 
For instance, there exist simple transformation rules the can be used to append edges of weight ${\nu_e = \frac{k}{\lambda}}$ where $k$ is any integer $k < \lambda$~\cite[Sec.~6]{gfe}.

The transformation rule also holds in modified form in near integer dimensions such as $D=4-2\varepsilon$, where we can interpret $f_G$ and $f_{G'}$ as Laurent expansions in the dimensional regularization parameter $\varepsilon$.
There even exists a variant of this transformation rule that admits $G$ and $G'$ to be divergent for $\varepsilon \rightarrow 0$. The divergences can be regulated and the integrations can still be performed. Details of this procedure will be described in~\cite{7loops}.

In essence, this transformation rule is a sophisticated version of the \emph{differential equation} \emph{method} \cite{Kotikov:1990kg,Remiddi:1997ny} for the computation of a Feynman integral.
It does not only provide a differential equation (an inhomogeneous second-order partial differential equation), but comes with a full, unique, explicit, proved, and calculable solution.
The details of this solution including an algorithm for its construction are in \cite{gfe}. A well-tested implementation is \cite{Shlog}.

\subsection{Example: Computation of the claw graph}
\label{exclaw}
We can start with the trivial graphical function and the associated graph $G_1$ as in~\eqref{eq:trivialgf} and add external edges between external vertices by the transformation rule in Section~\ref{sec:addedges}. If we restrict ourselves to $D=4$ and edge weights $\nu_{01}=0$, $\nu_{0z}=1$ and $\nu_{1z}=1$, we get
\begin{align*} G_2=\,& \begin{tikzpicture}[baseline={([yshift=-.7ex]0,0)}] \coordinate (v) at (0,0); \useasboundingbox ({-2*\rada},{-2*\rada}) rectangle ({2*\rada},{2*\rada}); \coordinate[label=above left:$1$] (v1) at ([shift=(120:\rada)]v); \coordinate[label=below left:$0$] (v0) at ([shift=(240:\rada)]v); \coordinate[label=right:$z$] (vz) at ([shift=(0:\rada)]v); \draw (v0) -- (vz); \draw (v1) -- (vz); \filldraw[col1] (v0) circle (1.7pt); \filldraw[col1] (v1) circle (1.7pt); \filldraw[col1] (vz) circle (1.7pt); \end{tikzpicture} & & \begin{tikzpicture}[baseline={([yshift=-.7ex]0,0)}] \useasboundingbox ({-2*\rada},{-2*\rada}) rectangle ({2*\rada},{2*\rada}); \draw[dashed,<->] ({-.7*\rada},0) -- ({.7*\rada},0); \end{tikzpicture} & & \begin{tikzpicture}[baseline={([yshift=-.9ex]0,0)}] \coordinate (v) at (0,0); \useasboundingbox ({-6.5*\rada},{-2*\rada}) rectangle ({6.5*\rada},{2*\rada}); \node (0,0) {$ \displaystyle f_{G_2}(z) =\frac{f_{G_1}(z)}{z\zz(1-z)(1-\zz)} =\frac{1}{z\zz(1-z)(1-\zz)} $}; \end{tikzpicture}. \end{align*}
We continue by appending an edge of weight $1$ to the vertex $z$ as described in Section~\ref{sec:appendedge},
\begin{align*} G_3=\,& \begin{tikzpicture}[baseline={([yshift=-.7ex]0,0)}] \coordinate (v) at (0,0); \useasboundingbox ({-2*\rada},{-2*\rada}) rectangle ({2*\rada},{2*\rada}); \coordinate[label=above left:$1$] (v1) at ([shift=(120:\rada)]v); \coordinate[label=below left:$0$] (v0) at ([shift=(240:\rada)]v); \coordinate[label=right:$z$] (vz) at ([shift=(0:\rada)]v); \draw (v0) -- (v); \draw (v1) -- (v); \draw (v) -- (vz); \filldraw (v) circle (1.7pt); \filldraw[col1] (v0) circle (1.7pt); \filldraw[col1] (v1) circle (1.7pt); \filldraw[col1] (vz) circle (1.7pt); \end{tikzpicture} & & \begin{tikzpicture}[baseline={([yshift=-.7ex]0,0)}] \useasboundingbox ({-2*\rada},{-2*\rada}) rectangle ({2*\rada},{2*\rada}); \draw[dashed,<->] ({-.7*\rada},0) -- ({.7*\rada},0); \end{tikzpicture} & & \begin{tikzpicture}[baseline={([yshift=-.9ex]0,0)}] \coordinate (v) at (0,0); \useasboundingbox ({-6.5*\rada},{-2.5*\rada}) rectangle ({6.5*\rada},{2.5*\rada}); \node (0,0) {$ \displaystyle \begin{aligned} f_{G_3}(z) &= - \frac{1}{z-\zz} \int_{\textrm{sv}} \dd z \int_{\textrm{sv}} \dd \zz (z-\zz) f_{G_2}(z) \\
&=- \frac{1}{z-\zz} \int_{\textrm{sv}} \dd z \int_{\textrm{sv}} \dd \zz \frac{z-\zz}{ z \zz (1-z)(1-\zz) } \\
&= \frac{4i D(z)}{z-\zz} \end{aligned} $}; \end{tikzpicture}, \end{align*}
where the integration operator $\mathcal I_{\lambda-1}$ is 
$ \frac{1}{z-\zz} \int_{\textrm{sv}} \dd z \int_{\textrm{sv}} \dd \zz (z-\zz) $ in the case $\lambda = 1= (D-2)/2$ and 
$D(z)$ is the \emph{Bloch--Wigner} dilog $D(z) = \Im(\Li_2(z)+\log(1-z) \log |z|)$ with $\Li_2(z) = \sum_{k =1}^\infty \frac{z^k}{k^2}$.
The integration operations can be performed algorithmically in the function space of GSVHs. Explicitly, the required computer algebra can be performed conveniently using the \texttt{Maple} package \texttt{HyperlogProcedures}~\cite{Shlog}.

\subsection{Example: The claw graph in all even dimensions \texorpdfstring{$\geq6$}{>=6}}
Similarly, we can evaluate the claw graph, $G_3$, in any even dimension larger than $4$ using the same combination of transformation rules. For the $D$ dimensional evaluation of $G_3$,
\begin{align} \begin{aligned} \label{eq:clawD} f_{G_3}^{(D)}(z) &= \frac{1}{(\lambda-1)\Gamma(\lambda)}\int_0^1\frac{(z\zz)^{1-\lambda}-t^{2\lambda-2}}{(1-tz)^\lambda(1-t\zz)^\lambda}\,\dd t, \end{aligned} \end{align}
for all integers $\lambda = (D-2)/2 \geq 2$ (i.e.~$D \geq 6$ and even) which follows from (3.16) in~\cite{Schnetz:2013hqa}.

As the residues 
 of the integrand at $t=1/z$ and $t=1/\zz$ vanish, it follows that 
$ f_{G_3}^{(D)}(z) $ is a rational function in $z,\zz$ for even $D \geq 6$ (see~\cite[Example 4]{gfe}). 

\subsection{Example: Generalized ladder graphs in all dimensions \texorpdfstring{$> 2$}{>2}}
\label{sec:ladder}
In interesting family of graphs that we can evaluate by iterating the 
procedure from the previous examples
are the \emph{generalized ladders},
\begin{align} L_{n}^{(\lambda)} = \quad \begin{tikzpicture}[baseline={([yshift=-.7ex]0,0)}] \useasboundingbox ({-3*\rada},{-2*\rada}) rectangle ({3*\rada},{2*\rada}); \coordinate[label=above:$0$] (v0) at (0,\rada); \coordinate[label=left:$1$] (v1) at ({-2.5*\rada},0); \coordinate[label=right:$z$] (vz) at ({2.5*\rada},0); \coordinate (v2) at ({-1.5*\rada},0); \coordinate (v3) at ({-1.0*\rada},0); \coordinate (v4) at ({-0.5*\rada},0); \coordinate (v5) at ({ 0.0*\rada},0); \coordinate (v6) at ({ 0.5*\rada},0); \coordinate (v7) at ({ 1.0*\rada},0); \coordinate (v8) at ({ 1.5*\rada},0); \draw (v1) -- (vz); \draw[dashed] (v2) -- (v0); \draw[dashed] (v4) -- (v0); \draw[dashed,black!50] (v0) -- ($(v0)!.7!(v6)$); \draw[dashed,black!50] (v0) -- ($(v0)!.7!(v8)$); \filldraw (v2) circle (1.7pt); \filldraw (v4) circle (1.7pt); \filldraw[black!50] (v6) circle (1.7pt); \filldraw[black!50] (v8) circle (1.7pt); \filldraw[col1] (v0) circle (1.7pt); \filldraw[col1] (v1) circle (1.7pt); \filldraw[col1] (vz) circle (1.7pt); \draw [decorate,decoration={brace,amplitude=10pt}] ({1.5*\rada},-{.5*\rada}) -- ({-1.5*\rada},-{.5*\rada}) node [black,midway,yshift=-0.7cm] {$n$ \text{steps}}; \end{tikzpicture} \end{align}
where the solid edges $\solidedge$ have weight $1$  and the dashed edges $\dashededge$ have weight $\frac{1}{\lambda} = \frac{2}{D-2}$.
The example from Section~\ref{exclaw} corresponds to the case $n =1$ and $D=4 \Rightarrow \lambda = (D-2)/2 = 1$. We can iterate the procedure described there and keep appending edges of weight $1$ together with edges $0z$ of weight $1/\lambda$ in general even dimensions $D= 2(\lambda +1)$. What we will find is in line with the following closed form expression for the generalized ladders,
\begin{align} f_{L_n}^{(\lambda)}(z)=&\,\frac{1}{\Gamma(\lambda)^{n+2}}\sum_{k=0}^n\genfrac(){0pt}{}{2n-k}{n}\frac{(-\log(z\zz))^k}{k!}\partial_z^{\lambda-1}\partial_\zz^{\lambda-1} \frac{z^{\lambda-1}{\mathrm{Li}}_{2n-k}(z)-\zz^{\lambda-1}{\mathrm{Li}}_{2n-k}(\zz)}{z-\zz}. \end{align}
The $D=4\Rightarrow \lambda=1$ case of this formula has been obtained in~\cite{Usyukina:1993ch}. The more general form presented here can also be derived using integrability methods~\cite{Isaev:2003tk}.
We proved this formula using Gegenbauer methods~(see~\cite[Examples 71, 80 and 86]{gfe}). The Gegenbauer method for the evaluation of Feynman integrals was originally put forward in~\cite{Chetyrkin:1980pr}.
Rewriting the expression for $f_{L_n}^{(\lambda)}(z)$ yields a formula that holds for arbitrary dimensions $D > 2$ by analytic continuation,
\begin{align} f_{L_n}^{(\lambda)}(z)=& -\frac{1}{\Gamma(\lambda)^nn!(n-1)!}\int_0^1\frac{t^{\lambda-1}\log^{n-1}t\log^{n-1}(tz\zz)\log(t^2z\zz)}{(1-tz)^\lambda(1-t\zz)^\lambda}\dd t. \end{align}

\subsection{Products of graphical functions}
Another simple transformation rule is the \emph{product rule}. If the underlying Feynman graph splits into two parts upon removal of the three external vertices, then the graphical function factorizes into a product of the graphical functions as illustrated below.
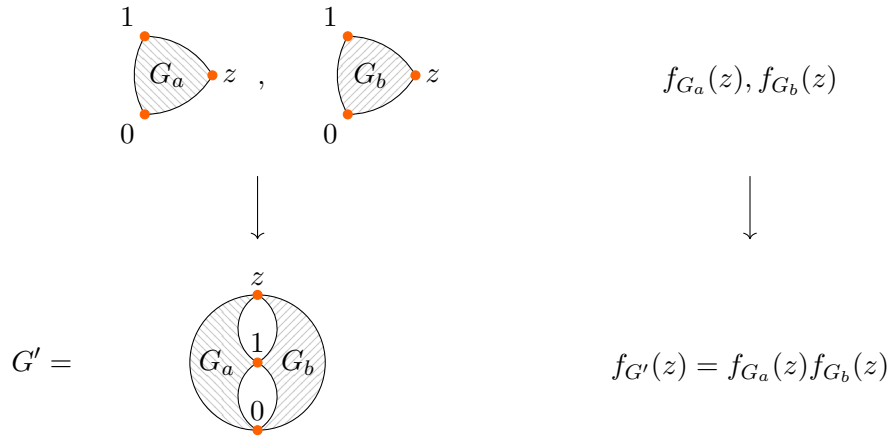
\begin{align*} & \begin{tikzpicture}[baseline={([yshift=-.7ex]0,0)}] \coordinate (v) at (0,0); \useasboundingbox ({-2*\rada},{-2*\rada}) rectangle ({2*\rada},{2*\rada}); \coordinate[label=above left:$1$] (v1) at ([shift=(120:\rada)]v); \coordinate[label=below left:$0$] (v0) at ([shift=(240:\rada)]v); \coordinate[label=right:$z$] (vz) at ([shift=(0:\rada)]v); \coordinate (vm) at ([shift=(-90:{2*\rada})]v); \draw[pattern=north west lines, pattern color=black!20] (v1) arc (150:210:{sqrt(3)*\rada}) arc (-90:-30:{sqrt(3)*\rada}) arc (30:90:{sqrt(3)*\rada}); \node (G) at (v) {$G_a$}; \filldraw[col1] (v0) circle (1.7pt); \filldraw[col1] (v1) circle (1.7pt); \filldraw[col1] (vz) circle (1.7pt); \end{tikzpicture} , ~ \begin{tikzpicture}[baseline={([yshift=-.7ex]0,0)}] \coordinate (v) at (0,0); \useasboundingbox ({-2*\rada},{-2*\rada}) rectangle ({2*\rada},{2*\rada}); \coordinate[label=above left:$1$] (v1) at ([shift=(120:\rada)]v); \coordinate[label=below left:$0$] (v0) at ([shift=(240:\rada)]v); \coordinate[label=right:$z$] (vz) at ([shift=(0:\rada)]v); \coordinate (vm) at ([shift=(-90:{2*\rada})]v); \draw[pattern=north east lines, pattern color=black!20] (v1) arc (150:210:{sqrt(3)*\rada}) arc (-90:-30:{sqrt(3)*\rada}) arc (30:90:{sqrt(3)*\rada}); \node (G) at (v) {$G_b$}; \filldraw[col1] (v0) circle (1.7pt); \filldraw[col1] (v1) circle (1.7pt); \filldraw[col1] (vz) circle (1.7pt); \end{tikzpicture} & & \begin{tikzpicture}[baseline={([yshift=-.3ex]0,0)}] \coordinate (v) at (0,0); \useasboundingbox ({-4*\rada},{-.7*\rada}) rectangle ({4*\rada},{.7*\rada}); \node (0,0) {$ \displaystyle f_{G_a}(z) , f_{G_b}(z) $}; \end{tikzpicture} \\
& \begin{tikzpicture}[baseline={([yshift=-.7ex]0,0)}] \useasboundingbox ({-4*\rada},{-.7*\rada}) rectangle ({4*\rada},{.7*\rada}); \draw[<-] (0,{-.7*\rada}) -- (0,{.7*\rada}); \end{tikzpicture} & & \begin{tikzpicture}[baseline={([yshift=-.7ex]0,0)}] \useasboundingbox ({-4*\rada},{-.7*\rada}) rectangle ({4*\rada},{.7*\rada}); \draw[<-] (0,{-.7*\rada}) -- (0,{.7*\rada}); \end{tikzpicture} \notag \\
G'=\,& \begin{tikzpicture}[baseline={([yshift=-.7ex]0,0)}] \coordinate (v) at (0,0); \useasboundingbox ({-4*\rada},{-2.5*\rada}) rectangle ({4*\rada},{2.5*\rada}); \coordinate[label=above:$z$] (vz) at (0,{1.5*\rada}); \coordinate[label=above:$1$] (v1) at (0,0); \coordinate[label=above:$0$] (v0) at (0,-{1.5*\rada}); \draw[pattern=north west lines, pattern color=black!20] (v0) arc (-120:-240:{1.5*\rada/sqrt(3)}) arc (-120:-240:{1.5*\rada/sqrt(3)}) arc (90:270:{1.5*\rada}); \draw[pattern=north east lines, pattern color=black!20] (vz) arc (60:-60:{1.5*\rada/sqrt(3)}) arc (60:-60:{1.5*\rada/sqrt(3)}) arc (-90:90:{1.5*\rada}); \node (G1) at (-{.9*\rada},0) {$G_a$}; \node (G2) at ({.9*\rada},0) {$G_b$}; \filldraw[col1] (v0) circle (1.7pt); \filldraw[col1] (v1) circle (1.7pt); \filldraw[col1] (vz) circle (1.7pt); \end{tikzpicture} & & \begin{tikzpicture}[baseline={([yshift=-.3ex]0,0)}] \coordinate (v) at (0,0); \useasboundingbox ({-4*\rada},{-.7*\rada}) rectangle ({4*\rada},{.7*\rada}); \node (0,0) {$ \displaystyle f_{G'}(z)= f_{G_a}(z) f_{G_b}(z) $}; \end{tikzpicture} \end{align*}
This property is evident on the level of the position-space Feynman integral~\eqref{feynmanint} which factorizes if the graph splits along external vertices. 

\subsection{Internalizing a vertex}
\label{sec:internalizevertex}
The next rule transforms a graphical function into a Feynman period. It does so by integrating over the external vertex $z$ to make it internal. %
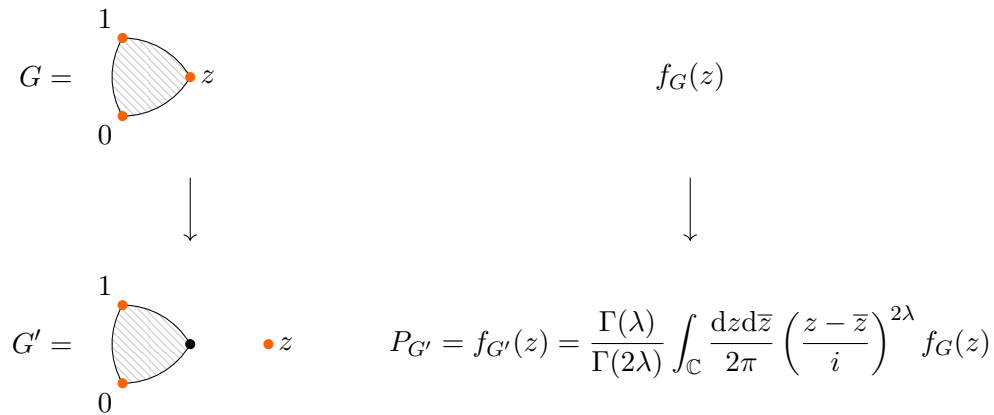
\begin{align*} G=\,& \begin{tikzpicture}[baseline={([yshift=-.7ex]0,0)}] \coordinate (v) at (0,0); \useasboundingbox ({-1.5*\rada},{-2*\rada}) rectangle ({3.5*\rada},{2*\rada}); \coordinate[label=above left:$1$] (v1) at ([shift=(120:\rada)]v); \coordinate[label=below left:$0$] (v0) at ([shift=(240:\rada)]v); \coordinate[label=right:$z$] (vz) at ([shift=(0:\rada)]v); \coordinate (vm) at ([shift=(-90:{2*\rada})]v); \draw[pattern=north west lines, pattern color=black!20] (v1) arc (150:210:{sqrt(3)*\rada}) arc (-90:-30:{sqrt(3)*\rada}) arc (30:90:{sqrt(3)*\rada}); \filldraw[col1] (v0) circle (1.7pt); \filldraw[col1] (v1) circle (1.7pt); \filldraw[col1] (vz) circle (1.7pt); \end{tikzpicture} & & \begin{tikzpicture}[baseline={([yshift=-.7ex]0,0)}] \coordinate (v) at (0,0); \useasboundingbox ({-7*\rada},{-2*\rada}) rectangle ({7*\rada},{2*\rada}); \node (0,0) {$ f_{G}(z)$}; \end{tikzpicture} \\
& \begin{tikzpicture}[baseline={([yshift=-.7ex]0,0)}] \useasboundingbox ({-2.5*\rada},{-.7*\rada}) rectangle ({2.5*\rada},{.7*\rada}); \draw[<-] (0,{-.7*\rada}) -- (0,{.7*\rada}); \end{tikzpicture} & & \begin{tikzpicture}[baseline={([yshift=-.7ex]0,0)}] \useasboundingbox ({-7*\rada},{-.7*\rada}) rectangle ({7*\rada},{.7*\rada}); \draw[<-] (0,{-.7*\rada}) -- (0,{.7*\rada}); \end{tikzpicture} \notag \\
G'=\,& \begin{tikzpicture}[baseline={([yshift=-.7ex]0,0)}] \coordinate (v) at (0,0); \useasboundingbox ({-1.5*\rada},{-2*\rada}) rectangle ({3.5*\rada},{2*\rada}); \coordinate[label=above left:$1$] (v1) at ([shift=(120:\rada)]v); \coordinate[label=below left:$0$] (v0) at ([shift=(240:\rada)]v); \coordinate (vx) at ([shift=(0:\rada)]v); \coordinate[label=right:$z$] (vz) at ([shift=(0:{(1+sqrt(3))*\rada})]v); \coordinate (vm) at ([shift=(90:{2*\rada})]v); \draw[pattern=north west lines, pattern color=black!20] (v1) arc (150:210:{sqrt(3)*\rada}) arc (-90:-30:{sqrt(3)*\rada}) arc (30:90:{sqrt(3)*\rada}); \filldraw[col1] (v0) circle (1.7pt); \filldraw[col1] (v1) circle (1.7pt); \filldraw (vx) circle (1.7pt); \filldraw[col1] (vz) circle (1.7pt); \end{tikzpicture} & & \begin{tikzpicture}[baseline={([yshift=-.9ex]0,0)}] \coordinate (v) at (0,0); \useasboundingbox ({-7*\rada},{-2*\rada}) rectangle ({7*\rada},{2*\rada}); \node (0,0) {$ \displaystyle \begin{gathered} P_{G'} = f_{G'}(z) = \frac{\Gamma(\lambda)}{\Gamma(2 \lambda)} \int_\CC \frac{\dd z \dd \zz}{2 \pi} \left(\frac{z-\zz}{i}\right)^{2\lambda} f_G(z) \end{gathered} $}; \end{tikzpicture}  \end{align*}
The integration can be performed (using Stokes' theorem) by calculating residues in the functions space of GSVHs, see \cite[Sec.~2.8]{Schnetz:2013hqa}.
This transformation rule can be proved by performing the integral over the external $z$ vertex in position space and translating the resulting expression into the $z,\zz$ variables (see~\cite[Proposition~20]{gfe}).

\subsection{Example: A simple Feynman period computation}
\label{sec:periodexample}
Consider the following momentum-space two-point function Feynman graph with two loops in $D=4$ with unit edge weights,
\begin{align*} \td{G}= \begin{tikzpicture}[baseline={([yshift=-.7ex]0,0)}] \coordinate (v) at (0,0); \useasboundingbox ({-2.5*\rada},{-2*\rada}) rectangle ({3*\rada},{2*\rada}); \coordinate (vi) at ([shift=(0:\rada)]v); \coordinate (vo) at ([shift=(180:\rada)]v); \coordinate (v1) at ([shift=(90:\rada)]v); \coordinate (v2) at ([shift=(-90:\rada)]v); \coordinate[label=right:$-Q$] (li) at ([shift=(0:{1.5*\rada})]v); \coordinate[label=left:$Q$] (lo) at ([shift=(180:{1.5*\rada})]v); \draw (v) circle (\rada); \draw (v1) -- (v2); \draw (vi) -- (li); \draw (vo) -- (lo); \filldraw (vi) circle (1.7pt); \filldraw (vo) circle (1.7pt); \filldraw (v1) circle (1.7pt); \filldraw (v2) circle (1.7pt); \end{tikzpicture} & & \begin{tikzpicture}[baseline={([yshift=-.7ex]0,0)}] \coordinate (v) at (0,0); \useasboundingbox ({-8*\rada},{-2*\rada}) rectangle ({8*\rada},{2*\rada}); \node (0,0) {$ \displaystyle \begin{aligned} \td{I}_{\td{G}}(Q) &= \int \frac{\dd^4 k_1 \dd^4 k_2}{\pi^{D}} \frac{1} {k_1^2 k_2^2 (k_1+k_2)^2 (k_1+Q)^2 (k_2-Q)^2} \\
&= Q^{-2} \td{P}_{\td{G}} \end{aligned} $}; \end{tikzpicture} \end{align*}
We can translate this into a position-space Feynman period computation: 
By $\lambda = (D-2)/2 =1$, $\nu_e = \frac{D}{2} - \td{\nu}_e = 1$ and~\eqref{eq:momposperiod}, we have $\td{P}_{\td{G}} = P_G$. Recall that $G$ is just the graph $\td{G}$ where the vertices with attached legs are replaced with the external vertices $0$ and $1$: 
\begin{align*} G= \begin{tikzpicture}[baseline={([yshift=-.7ex]0,0)}] \coordinate (v) at (0,0); \useasboundingbox ({-2*\rada},{-1.5*\rada}) rectangle ({2*\rada},{1.5*\rada}); \coordinate[label=right:$1$] (vi) at ([shift=(0:\rada)]v); \coordinate[label=left:$0$] (vo) at ([shift=(180:\rada)]v); \coordinate (v1) at ([shift=(90:\rada)]v); \coordinate (v2) at ([shift=(-90:\rada)]v); \coordinate[label=right:$z$] (vz) at ([shift=(45:{1.5*\rada})]v); \draw (v) circle (\rada); \draw (v1) -- (v2); \filldraw[col1] (vi) circle (1.7pt); \filldraw[col1] (vo) circle (1.7pt); \filldraw (v1) circle (1.7pt); \filldraw (v2) circle (1.7pt); \filldraw[col1] (vz) circle (1.7pt); \end{tikzpicture}, \end{align*}
where we added the isolated vertex $z$ to highlight the translation between position-space Feynman periods and graphical functions. 
We can compute $P_G$ recursively by starting with the result of the example in Section~\ref{exclaw}:
\begin{align*} G_3=\,& \begin{tikzpicture}[baseline={([yshift=-.7ex]0,0)}] \coordinate (v) at (0,0); \useasboundingbox ({-2*\rada},{-1.6*\rada}) rectangle ({2*\rada},{1.6*\rada}); \coordinate[label=above left:$1$] (v1) at ([shift=(120:\rada)]v); \coordinate[label=below left:$0$] (v0) at ([shift=(240:\rada)]v); \coordinate[label=right:$z$] (vz) at ([shift=(0:\rada)]v); \draw (v0) -- (v); \draw (v1) -- (v); \draw (v) -- (vz); \filldraw (v) circle (1.7pt); \filldraw[col1] (v0) circle (1.7pt); \filldraw[col1] (v1) circle (1.7pt); \filldraw[col1] (vz) circle (1.7pt); \end{tikzpicture} & & \begin{tikzpicture}[baseline={([yshift=-.7ex]0,0)}] \useasboundingbox ({-2*\rada},{-1.5*\rada}) rectangle ({2*\rada},{1.5*\rada}); \draw[dashed,<->] ({-.7*\rada},0) -- ({.7*\rada},0); \end{tikzpicture} & & \begin{tikzpicture}[baseline={([yshift=-.9ex]0,0)}] \coordinate (v) at (0,0); \useasboundingbox ({-6.5*\rada},{-1.5*\rada}) rectangle ({6.5*\rada},{1.5*\rada}); \node (0,0) {$ \displaystyle \begin{aligned} f_{G_3}(z) &= \frac{4i D(z)}{z-\zz} \end{aligned} $}; \end{tikzpicture} \\
& \begin{tikzpicture}[baseline={([yshift=-.7ex]0,0)}] \coordinate (v) at (0,0); \useasboundingbox ({-2*\rada},{-1.5*\rada}) rectangle ({2*\rada},{1.5*\rada}); \draw[<-] (0,{-.7*\rada}) -- (0,{.7*\rada}); \end{tikzpicture} & & \begin{tikzpicture}[baseline={([yshift=-.7ex]0,0)}] \useasboundingbox ({-2*\rada},{-1.5*\rada}) rectangle ({2*\rada},{1.5*\rada}); \node (v) at (0,0) {add external edges (Sec.~\ref{sec:addedges})}; \end{tikzpicture} & & \begin{tikzpicture}[baseline={([yshift=-.9ex]0,0)}] \useasboundingbox ({-6.5*\rada},{-1.5*\rada}) rectangle ({6.5*\rada},{1.5*\rada}); \draw[<-] (0,{-.7*\rada}) -- (0,{.7*\rada}); \end{tikzpicture} \\
G_4=\,& \begin{tikzpicture}[baseline={([yshift=-.7ex]0,0)}] \coordinate (v) at (0,0); \useasboundingbox ({-2*\rada},{-1.5*\rada}) rectangle ({2*\rada},{1.5*\rada}); \coordinate[label=above left:$1$] (v1) at ([shift=(120:\rada)]v); \coordinate[label=below left:$0$] (v0) at ([shift=(240:\rada)]v); \coordinate[label=right:$z$] (vz) at ([shift=(0:\rada)]v); \draw (v0) -- (v); \draw (v1) -- (v); \draw (v) -- (vz); \draw (vz) arc (0:120:\rada); \draw (vz) arc (0:-120:\rada); \filldraw (v) circle (1.7pt); \filldraw[col1] (v0) circle (1.7pt); \filldraw[col1] (v1) circle (1.7pt); \filldraw[col1] (vz) circle (1.7pt); \end{tikzpicture} & & \begin{tikzpicture}[baseline={([yshift=-.7ex]0,0)}] \useasboundingbox ({-2*\rada},{-1.5*\rada}) rectangle ({2*\rada},{1.5*\rada}); \draw[dashed,<->] ({-.7*\rada},0) -- ({.7*\rada},0); \end{tikzpicture} & & \begin{tikzpicture}[baseline={([yshift=-.9ex]0,0)}] \coordinate (v) at (0,0); \useasboundingbox ({-6.5*\rada},{-1.5*\rada}) rectangle ({6.5*\rada},{1.5*\rada}); \node (0,0) {$ \displaystyle \begin{aligned} f_{G_4}(z) &= \frac{f_{G_3}(z)}{z \zz (1-z) (1-\zz)} \\
&= \frac{4i D(z)}{z \zz (z-\zz) (1-z) (1-\zz)} \end{aligned} $}; \end{tikzpicture} \\
& \begin{tikzpicture}[baseline={([yshift=-.7ex]0,0)}] \coordinate (v) at (0,0); \useasboundingbox ({-2*\rada},{-1.5*\rada}) rectangle ({2*\rada},{1.5*\rada}); \draw[<-] (0,{-.7*\rada}) -- (0,{.7*\rada}); \end{tikzpicture} & & \begin{tikzpicture}[baseline={([yshift=-.7ex]0,0)}] \useasboundingbox ({-2*\rada},{-1.5*\rada}) rectangle ({2*\rada},{1.5*\rada}); \node (v) at (0,0) {internalize $z$ (Sec.~\ref{sec:internalizevertex})}; \end{tikzpicture} & & \begin{tikzpicture}[baseline={([yshift=-.9ex]0,0)}] \useasboundingbox ({-6.5*\rada},{-1.5*\rada}) rectangle ({6.5*\rada},{1.5*\rada}); \draw[<-] (0,{-.7*\rada}) -- (0,{.7*\rada}); \end{tikzpicture} \\
G_5=\,& \begin{tikzpicture}[baseline={([yshift=-.7ex]0,0)}] \coordinate (v) at (0,0); \useasboundingbox ({-2*\rada},{-1.6*\rada}) rectangle ({2*\rada},{1.6*\rada}); \coordinate[label=above left:$1$] (v1) at ([shift=(120:\rada)]v); \coordinate[label=below left:$0$] (v0) at ([shift=(240:\rada)]v); \coordinate (vm) at ([shift=(0:\rada)]v); \coordinate[label=right:$z$] (vz) at ([shift=(45:{1.5*\rada})]v); \draw (v0) -- (v); \draw (v1) -- (v); \draw (v) -- (vm); \draw (vm) arc (0:120:\rada); \draw (vm) arc (0:-120:\rada); \filldraw (v) circle (1.7pt); \filldraw[col1] (v0) circle (1.7pt); \filldraw[col1] (v1) circle (1.7pt); \filldraw (vm) circle (1.7pt); \filldraw[col1] (vz) circle (1.7pt); \end{tikzpicture} & & \begin{tikzpicture}[baseline={([yshift=-.7ex]0,0)}] \useasboundingbox ({-2*\rada},{-1.5*\rada}) rectangle ({2*\rada},{1.5*\rada}); \draw[dashed,<->] ({-.7*\rada},0) -- ({.7*\rada},0); \end{tikzpicture} & & \begin{tikzpicture}[baseline={([yshift=-.9ex]0,0)}] \coordinate (v) at (0,0); \useasboundingbox ({-6.5*\rada},{-1.5*\rada}) rectangle ({6.5*\rada},{1.5*\rada}); \node (0,0) {$ \displaystyle \begin{aligned} P_{G_5} &= \int_\CC \frac{\dd z \dd \zz}{2 \pi} \left(\frac{z-\zz}{i}\right)^2 f_{G_4}(z) \\
&= \int_\CC \frac{\dd z \dd \zz}{2 \pi} \frac{4 (z-\zz) D(z)}{i z \zz (1-z) (1-\zz)} = 6 \zeta(3) \end{aligned} $}; \end{tikzpicture} \end{align*}
where we applied all transformation rules with $\lambda= (D-2)/2 =1$. Obviously, $G = G_5$ and we recovered $P_G = \td{P}_{\td{G}} = 6 \zeta(3)$ by performing the integration in the last step. Again all computations can be performed conveniently using  \texttt{HyperlogProcedures}~\cite{Shlog}.

\subsection{Example: The wheel Feynman periods in all dimensions \texorpdfstring{$\geq 3$}{>=3}}
\label{sec:wheels}
A similar procedure may be applied to convert the ladder graphs $L_{n}^{(\lambda)}$ from the example in Section~\ref{sec:ladder} into Feynman periods. Consider the following family of \emph{generalized wheel} graphs in $D$ dimensions, 
\begin{align} W_{n}^{(\lambda)} = \quad \begin{tikzpicture}[baseline={([yshift=-.7ex]0,0)}] \useasboundingbox ({-3*\rada},{-2*\rada}) rectangle ({3*\rada},{2*\rada}); \coordinate[label=above:$0$] (v0) at (0,\rada); \coordinate[label=left:$1$] (v1) at ({-2.5*\rada},0); \coordinate (vz) at ({2.5*\rada},0); \coordinate[label=right:$z$] (vzz) at ({2.5*\rada},\rada); \coordinate (v2) at ({-1.5*\rada},0); \coordinate (v3) at ({-1.0*\rada},0); \coordinate (v4) at ({-0.5*\rada},0); \coordinate (v5) at ({ 0.0*\rada},0); \coordinate (v6) at ({ 0.5*\rada},0); \coordinate (v7) at ({ 1.0*\rada},0); \coordinate (v8) at ({ 1.5*\rada},0); \draw (v1) -- (vz); \draw[dashed] (v1) -- (v0); \draw[dashed] (v2) -- (v0); \draw[dashed] (v4) -- (v0); \draw[dashed] (vz) -- (v0); \draw[dashed,black!50] (v0) -- ($(v0)!.7!(v6)$); \draw[dashed,black!50] (v0) -- ($(v0)!.7!(v8)$); \draw (vz) to[out=-45,in=-135,looseness=.5] (v1); \filldraw (v2) circle (1.7pt); \filldraw (v4) circle (1.7pt); \filldraw[black!50] (v6) circle (1.7pt); \filldraw[black!50] (v8) circle (1.7pt); \filldraw[col1] (v0) circle (1.7pt); \filldraw[col1] (v1) circle (1.7pt); \filldraw[col1] (vzz) circle (1.7pt); \filldraw (vz) circle (1.7pt); \draw [decorate,decoration={brace,amplitude=10pt}] ({2.5*\rada},-{.5*\rada}) -- ({-2.5*\rada},-{.5*\rada}) node [black,midway,yshift=-0.7cm] {$n$ \text{spokes}}; \end{tikzpicture}, \end{align}
where the solid edges $\solidedge$ have weight $1$  and the dashed edges $\dashededge$ have weight $\frac{1}{\lambda} = \frac{2}{D-2}$ analogous to Section~\ref{sec:ladder}. Clearly, these graphs can be obtained from the ladder graphs $L_{n}^{(\lambda)}$ by adding edges (Section~\ref{sec:addedges}) and internalizing the $z$ vertex (Section~\ref{sec:internalizevertex}).
The resulting doubly infinite family of Feynman periods follows the remarkably simple formula,
\begin{align} \label{eq:wheels} P_{W_{n}}^{(\lambda)}=\frac{\genfrac(){0pt}{}{2n-2}{n-1}}{\Gamma(\lambda)^{n-1}}\sum_{k=\lambda}^\infty\frac{\binom{k+\lambda-1}{2\lambda -1}}{k^{2n-2}},\qquad n-2\geq\lambda, \end{align}
as can be checked by computing $P_{W_{n}}^{(\lambda)}$ for a couple of values of $n$ and $\lambda = (D-2)/2$ using~\texttt{HyperlogProcedures}~\cite{Shlog}. This formula can be proved for all $n$ and integer $D \geq 3$ (with $D \leq 2n-2$ for convergence) using the Gegenbauer method~\cite[Example~87]{gfe}. The sum in~\eqref{eq:wheels} evaluates to a linear combination of odd Riemann-$\zeta$-values for even $D$ and even $\zeta$-values for odd $D$ (where it ranges over half integers $k$). 
The even $D$ case is especially interesting, as a linear combination of the numbers $P_{W_n}^{(\lambda)}$ allows the certification of the nontrivial pairing of a cycle and a cocycle in the commutative graph homology~\cite[Section~10.4]{Brown:2021umn}. An injection of this graph homology into the cohomology of $\mathcal M_g$, the moduli space of curves of genus $g$, using the \emph{tropical Torelli map}~\cite{chan2021tropical} maps the odd $g$-loop wheel class to a nontrivial class in $H^{4g-6}(\mathcal M_g,\QQ)$. 
This hints for a rather intricate connection between Feynman periods and the cohomology of $\mathcal M_g$. The family of classes which is certified by the wheel graphs is part of the only known infinite family of unstable cohomology classes of $\mathcal M_g$. 
This is an especially dire situation as the dimension of the cohomology of $\mathcal M_g$ is known to grow quite rapidly~\cite{harer1986euler}.
 It seems promising to look for families of Feynman graphs that evaluate to interesting Feynman periods as candidates for new generators of cohomology classes of the moduli space of curves.

We note that for this and similar applications of Feynman periods \emph{analytic} evaluation is crucial. Mere numerical evaluation of the periods, e.g.\ with Monte Carlo methods as in~\cite{borinsky2020tropical}, is not sufficient to deduce rigorous statements.

\subsection{Completion of graphical functions: Conformal four-point integrals}
\label{sec:completegf}
To formulate the remaining transformation rules, it is convenient to introduce another slightly modified graphical representation of the functions. We refer to this modified graphical representation as the \emph{completion} of the Feynman graph \cite{Schnetz:2013hqa}. Completed graphical functions are \emph{conformal four-point integrals} and completion allows the full utilization of the hidden conformal invariance of the underlying massless Feynman integral. By introducing a new auxiliary external vertex $\infty$, completion makes this conformal invariance manifest. It must be emphasized that completion is not a transformation of the graph, but merely a change in the graphical representation of the corresponding function. We can always go back and forth between the completed and the uncompleted representation without changing the associated analytic expression:
\begin{align*} G=\,& \begin{tikzpicture}[baseline={([yshift=-.7ex]0,0)}] \coordinate (v) at (0,0); \useasboundingbox ({-2*\rada},{-2*\rada}) rectangle ({2.5*\rada},{2*\rada}); \coordinate[label=above left:$1$] (v1) at ([shift=(120:\rada)]v); \coordinate[label=below left:$0$] (v0) at ([shift=(240:\rada)]v); \coordinate[label=right:$z$] (vz) at ([shift=(0:\rada)]v); \coordinate (vm) at ([shift=(-90:{2*\rada})]v); \draw[pattern=north west lines, pattern color=black!20] (v1) arc (150:210:{sqrt(3)*\rada}) arc (-90:-30:{sqrt(3)*\rada}) arc (30:90:{sqrt(3)*\rada}); \filldraw[col1] (v0) circle (1.7pt); \filldraw[col1] (v1) circle (1.7pt); \filldraw[col1] (vz) circle (1.7pt); \end{tikzpicture} & & \begin{tikzpicture}[baseline={([yshift=-.7ex]0,0)}] \coordinate (v) at (0,0); \useasboundingbox ({-2*\rada},{-2*\rada}) rectangle ({2*\rada},{2*\rada}); \draw[<->] (-\rada,0) -- (\rada,0); \end{tikzpicture} & \overline{G}=\,& \begin{tikzpicture}[baseline={([yshift=-.7ex]0,0)}] \coordinate (v) at (0,0); \useasboundingbox ({-2*\rada},{-2*\rada}) rectangle ({4*\rada},{2*\rada}); \coordinate[label=above left:$1$] (v1) at ([shift=(120:\rada)]v); \coordinate[label=below left:$0$] (v0) at ([shift=(240:\rada)]v); \coordinate[label=above right:$z$] (vz) at ([shift=(0:\rada)]v); \coordinate[label=right:$\infty$] (voo) at ([shift=(0:{(1+sqrt(3))*\rada})]v); \draw[dashed] (voo) arc (45:105:{sqrt(6+3*sqrt(3))*\rada}); \draw[opacity=.3,dashed] (voo) arc (55:115:{sqrt(6+3*sqrt(3))*\rada}); \draw[opacity=.3,dashed] (voo) arc (65:125:{sqrt(6+3*sqrt(3))*\rada}); \draw[opacity=.3,dashed] (voo) arc (-65:-125:{sqrt(6+3*sqrt(3))*\rada}); \draw[opacity=.3,dashed] (voo) arc (-55:-115:{sqrt(6+3*sqrt(3))*\rada}); \draw[dashed] (voo) arc (-45:-105:{sqrt(6+3*sqrt(3))*\rada}); \draw[dashed] (voo) -- (vz); \draw[dashed] (v1) arc (90:270:{sqrt(3)/2*\rada}); \draw[preaction={fill, white},pattern=north west lines, pattern color=black!20] (v1) arc (150:210:{sqrt(3)*\rada}) arc (-90:-30:{sqrt(3)*\rada}) arc (30:90:{sqrt(3)*\rada}); \coordinate[label=above left:$1$] (v1) at ([shift=(120:\rada)]v); \coordinate[label=below left:$0$] (v0) at ([shift=(240:\rada)]v); \coordinate[label={[fill=white]above right:$z$}] (vz) at ([shift=(0:\rada)]v); \filldraw[col2] (v0) circle (1.7pt); \filldraw[col2] (v1) circle (1.7pt); \filldraw[col2] (vz) circle (1.7pt); \filldraw[col2] (voo) circle (1.7pt); \end{tikzpicture} = \begin{tikzpicture}[baseline={([yshift=-.7ex]0,0)}] \coordinate (v) at (0,0); \useasboundingbox ({-2*\rada},{-2*\rada}) rectangle ({2*\rada},{2*\rada}); \coordinate[label=above right:$1$] (v1) at ([shift=(45:\rada)]v); \coordinate[label=above left:$0$] (v0) at ([shift=(135:\rada)]v); \coordinate[label=below left:$\infty$] (vz) at ([shift=(225:\rada)]v); \coordinate[label=below right:$z$] (voo) at ([shift=(315:\rada)]v); \draw[pattern=crosshatch, pattern color=black!20] (v1) arc (67.5:112.5:{sqrt(2+sqrt(2))*\rada}) arc ({90+67.5}:{90+112.5}:{sqrt(2+sqrt(2))*\rada}) arc ({180+67.5}:{180+112.5}:{sqrt(2+sqrt(2))*\rada}) arc ({270+67.5}:{270+112.5}:{sqrt(2+sqrt(2))*\rada}); \filldraw[col2] (v0) circle (1.7pt); \filldraw[col2] (v1) circle (1.7pt); \filldraw[col2] (vz) circle (1.7pt); \filldraw[col2] (voo) circle (1.7pt); \end{tikzpicture} \\
& \begin{tikzpicture}[baseline={([yshift=-.9ex]0,0)}] \coordinate (v) at (0,0); \useasboundingbox ({-2.25*\rada},{-2*\rada}) rectangle ({2.25*\rada},{2*\rada}); \node (0,0) {$ \displaystyle f_G(z) $}; \end{tikzpicture} & & \begin{tikzpicture}[baseline={([yshift=-.9ex]0,0)}] \coordinate (v) at (0,0); \useasboundingbox ({-2*\rada},{-2*\rada}) rectangle ({2*\rada},{2*\rada}); \node (0,0) {$ \displaystyle = $}; \end{tikzpicture} & & \begin{tikzpicture}[baseline={([yshift=-.9ex]0,0)}] \coordinate (v) at (0,0); \useasboundingbox ({-3*\rada},{-2*\rada}) rectangle ({3*\rada},{2*\rada}); \node (0,0) {$ \displaystyle f_{\overline{G}}(z) $}; \end{tikzpicture} \end{align*}
We distinguish between uncompleted and completed graphs by using \textbf{\color{col2} blue} for the four external vertices $0,1,z$ and $\infty$ of completed graphs. 
To complete a graph $G$ we
\begin{enumerate}
\item
 add a new external vertex $\infty$ to the graph,
\item
add  edges from $\infty$ to each internal vertex with a weight such that the total sum of the incident edge weights at the internal vertex equals $2 D/(D-2)$, 
\item
add an edge between $\infty$ and the external vertex $z$ with a weight such that the incident edge weight sum at $z$ is equal to $0$, 
\item
add edges between the three external vertices $0,1$ and $\infty$ such that the incident edge weight sum at each external vertex is $0$. 
\end{enumerate}
Following this procedure we obtain a unique completed graph $\overline{G}$ with weighted internal vertex degree of $2D/(D-2)$ and weighted external vertex degree of $0$~\cite[Prop.~9]{gfe}. 
Translating the completed graph $\overline{G}$ back into an uncompleted graph without the $\infty$ vertex is not entirely unique as the uncompleted graph $G$ can have an edge of arbitrary weight between the external vertices $0$ and $1$. 
However, such edges are insignificant as they do not change the expression for the associated function.
Hence, we can define the function $f_{\overline{G}}(z)$  to be equal to the function $f_G(z)$ associated to the uncompleted graph.

\subsection{Permutations of external vertices}
\label{sec:perm}
With completion we can fully employ the conformal invariance of the underlying Feynman integral. Such a conformal four-point integral only depends on the cross-ratio of the four external positions. Double-transpositions of these four external vertices leaves the cross-ratios invariant. Hence, also the graphical function does not change and we get the transformation rule,
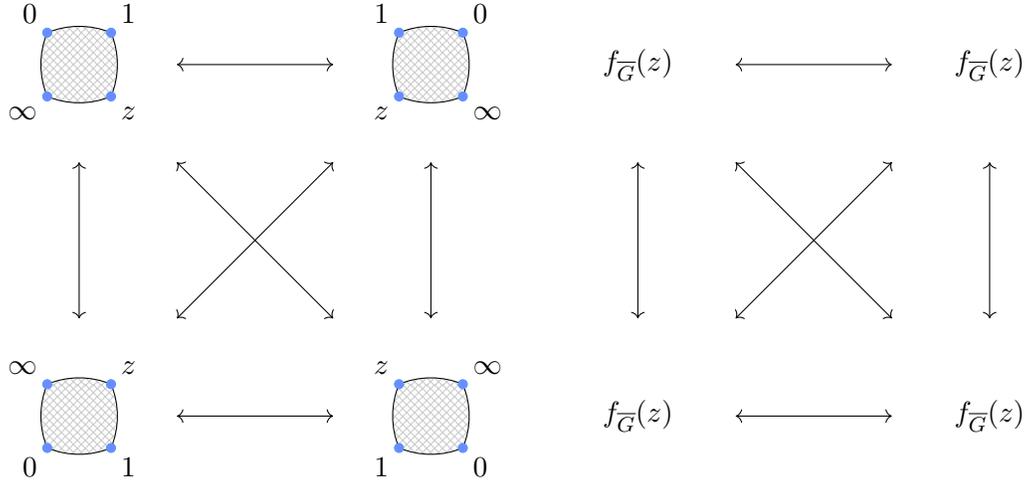
\begin{align*} \,& \end{align*}
Permutations of the external vertices that are not double transpositions do not leave the function invariant. The latter is transformed by a Möbius transformation of the argument. 
The full group of permutations of the external vertices is generated by double transpositions and permutations of $0,1,\infty$,
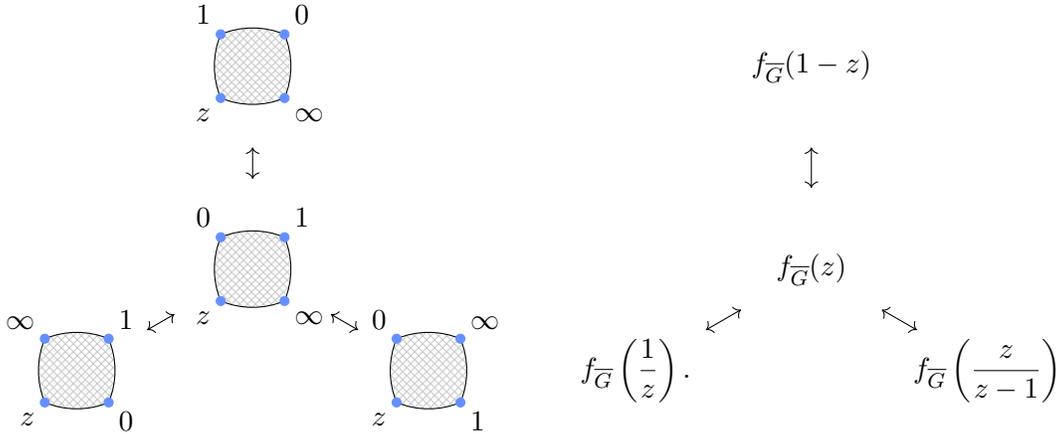
\begin{align*} %
 \end{align*}
The transformation rules can be proved by using the conformal symmetry of the underlying Feynman integral (see~\cite[Section~2.4]{gfe}). The Möbius transformations generated by $z \mapsto 1-z$, $z\mapsto \frac{1}{z}$ (and $z\mapsto \frac{z}{z-1}$) close in the function space of GSVHs.

\subsection{Example: Bootstrapping graphical functions using completion}

\begin{figure}
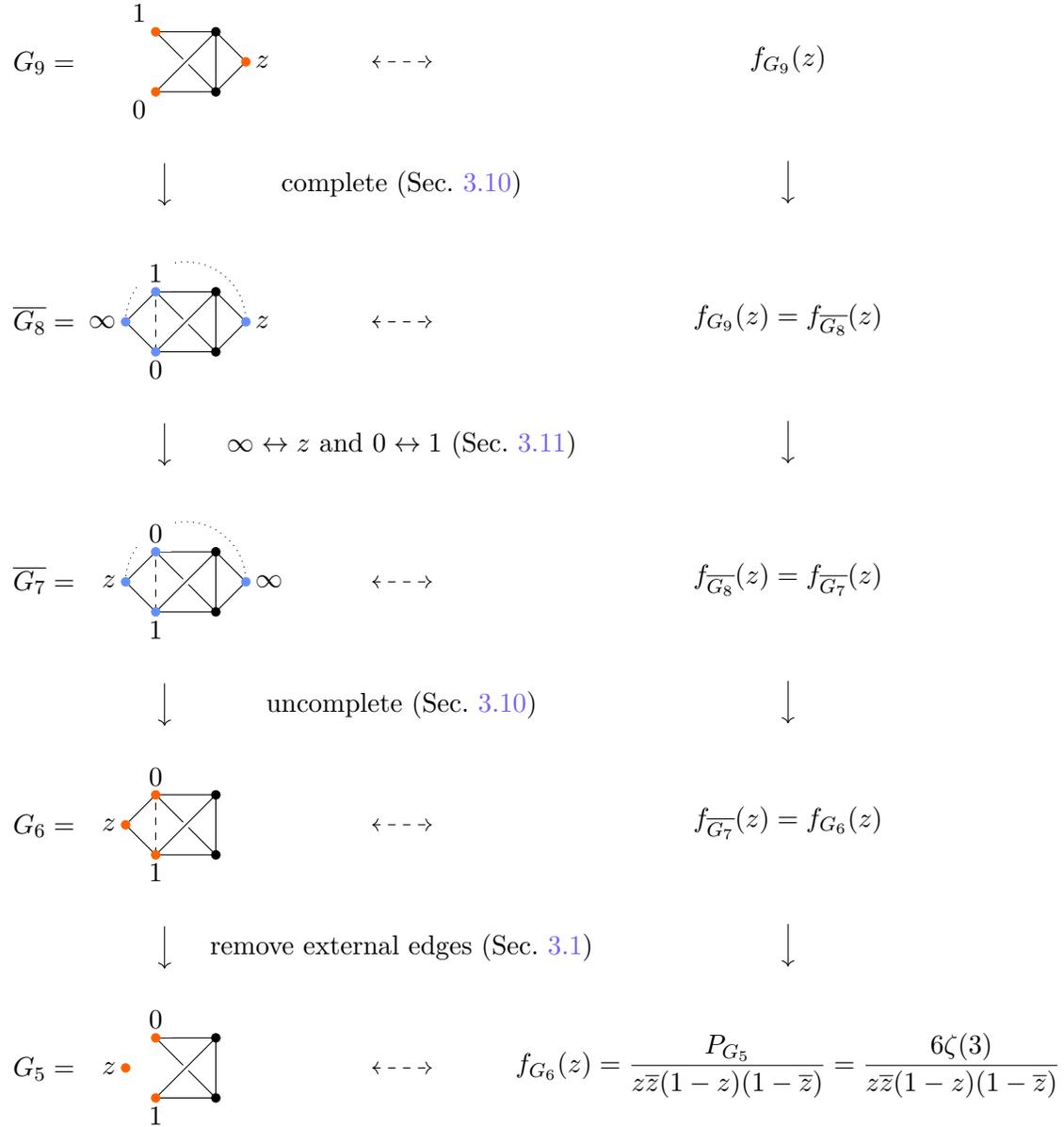

\begin{align*} G_9=\,& %
 \end{align*}
\caption{
Bootstrapping the graphical function $f_{G_9}(z)$ using its conformal symmetry.
The combinatorics of this bootstrap is dictated by the reduction rules. 
The solid edges $\solidedge$ have weight $1$, the dotted edges $\dottededge$ have weight $-2$ and the dashed edges $\dashededge$ have weight $-3$.
}
\label{fig:bootstrap}
\end{figure}

Completion is a powerful tool as the conformal symmetry allows one to bootstrap graphical functions of quite complicated graphs. 
To compute a certain Feynman period, it might for instance be necessary to know the graphical function associated to the graph
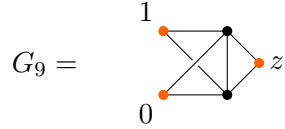
\begin{align*} G_9=\,& \begin{tikzpicture}[baseline={([yshift=-.7ex]0,0)}] \coordinate (v) at (0,0); \useasboundingbox ({-2.5*\rada},{-1.6*\rada}) rectangle ({2.5*\rada},{1.6*\rada}); \coordinate[label=above left:$1$] (v1) at ([shift=(135:\rada)]v); \coordinate[label=below left:$0$] (v0) at ([shift=(225:\rada)]v); \coordinate (va) at ([shift=(45:\rada)]v); \coordinate (vb) at ([shift=(-45:\rada)]v); \coordinate[label=right:$z$] (vz) at ([shift=(0:{sqrt(2)*\rada})]v); \draw (v1) -- (vb); \draw[preaction={draw, white, line width=3pt, -}] (v0) -- (va); \draw (va) -- (vb); \draw (v0) -- (vb); \draw (v1) -- (va); \draw (vz) -- (va); \draw (vz) -- (vb); \filldraw (va) circle (1.7pt); \filldraw (vb) circle (1.7pt); \filldraw[col1] (v0) circle (1.7pt); \filldraw[col1] (v1) circle (1.7pt); \filldraw[col1] (vz) circle (1.7pt); \end{tikzpicture} \end{align*}
in four-dimensional spacetime.
The expression for this graphical function is completely fixed by the conformal symmetry and the Feynman period computation of the example in Section~\ref{sec:periodexample}. The combinatorics of the reduction of the graphical function $G_9$ to the Feynman period computation $P_{G_5}$ in Section~\ref{sec:periodexample} is illustrated in Figure~\ref{fig:bootstrap}.

\subsection{Feynman period completion}
\label{sec:period_completion}
Completion can immediately be extended to Feynman periods by their definition as constant graphical functions in~\eqref{eq:period_def}, but it is convenient to use another slightly modified version of completion for periods. 
To do so, we start with a Feynman period graph $G$ with external vertices $0,1$ and an isolated vertex $z$ and complete this graph as described in Section~\ref{sec:completegf}:
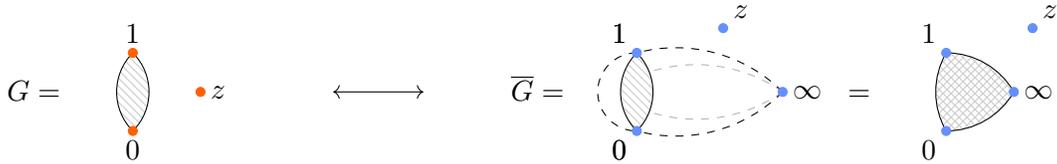
\begin{align*} G=\,& \begin{tikzpicture}[baseline={([yshift=-.7ex]0,0)}] \coordinate (v) at (0,0); \useasboundingbox ({-2*\rada},{-2*\rada}) rectangle ({2*\rada},{2*\rada}); \coordinate[label=above:$1$] (v1) at ([shift=(120:\rada)]v); \coordinate[label=below:$0$] (v0) at ([shift=(240:\rada)]v); \coordinate[label=right:$z$] (vz) at ([shift=(0:\rada)]v); \coordinate (vm) at ([shift=(-90:{2*\rada})]v); \draw[pattern=north west lines, pattern color=black!20] (v1) arc (135:225:{\rada*sqrt(3)/sqrt(2)}) arc (-45:45:{\rada*sqrt(3)/sqrt(2)}); \filldraw[col1] (v0) circle (1.7pt); \filldraw[col1] (v1) circle (1.7pt); \filldraw[col1] (vz) circle (1.7pt); \end{tikzpicture} & & \begin{tikzpicture}[baseline={([yshift=-.7ex]0,0)}] \coordinate (v) at (0,0); \useasboundingbox ({-2*\rada},{-2*\rada}) rectangle ({2*\rada},{2*\rada}); \draw[<->] (-\rada,0) -- (\rada,0); \end{tikzpicture} & \overline{G}=\,& \begin{tikzpicture}[baseline={([yshift=-.7ex]0,0)}] \coordinate (v) at (0,0); \useasboundingbox ({-2*\rada},{-2*\rada}) rectangle ({4*\rada},{2*\rada}); \coordinate[label=above left:$1$] (v1) at ([shift=(120:\rada)]v); \coordinate[label=below left:$0$] (v0) at ([shift=(240:\rada)]v); \coordinate[label=above right:$z$] (vz) at ([shift=(45:{2*\rada})]v); \coordinate[label=right:$\infty$] (voo) at ([shift=(0:{(1+sqrt(3))*\rada})]v); \draw[dashed] (voo) arc (45:105:{sqrt(6+3*sqrt(3))*\rada}); \draw[opacity=.3,dashed] (voo) arc (55:115:{sqrt(6+3*sqrt(3))*\rada}); \draw[opacity=.3,dashed] (voo) arc (-55:-115:{sqrt(6+3*sqrt(3))*\rada}); \draw[dashed] (voo) arc (-45:-105:{sqrt(6+3*sqrt(3))*\rada}); \draw[dashed] (v1) arc (90:270:{sqrt(3)/2*\rada}); \draw[preaction={fill, white},pattern=north west lines, pattern color=black!20] (v1) arc (135:225:{\rada*sqrt(3)/sqrt(2)}) arc (-45:45:{\rada*sqrt(3)/sqrt(2)}); \coordinate[label=above left:$1$] (v1) at ([shift=(120:\rada)]v); \coordinate[label=below left:$0$] (v0) at ([shift=(240:\rada)]v); \filldraw[col2] (v0) circle (1.7pt); \filldraw[col2] (v1) circle (1.7pt); \filldraw[col2] (vz) circle (1.7pt); \filldraw[col2] (voo) circle (1.7pt); \end{tikzpicture} = \begin{tikzpicture}[baseline={([yshift=-.7ex]0,0)}] \coordinate (v) at (0,0); \useasboundingbox ({-2*\rada},{-2*\rada}) rectangle ({2*\rada},{2*\rada}); \coordinate[label=above left:$1$] (v1) at ([shift=(120:\rada)]v); \coordinate[label=below left:$0$] (v0) at ([shift=(240:\rada)]v); \coordinate[label=right:$\infty$] (voo) at ([shift=(0:\rada)]v); \coordinate[label=above right:$z$] (vz) at ([shift=(45:{2*\rada})]v); \draw[pattern=crosshatch, pattern color=black!20] (v1) arc (150:210:{sqrt(3)*\rada}) arc (-90:-30:{sqrt(3)*\rada}) arc (30:90:{sqrt(3)*\rada}); \filldraw[col2] (v0) circle (1.7pt); \filldraw[col2] (v1) circle (1.7pt); \filldraw[col2] (vz) circle (1.7pt); \filldraw[col2] (voo) circle (1.7pt); \end{tikzpicture}                                  \end{align*}
where the isolated external vertex $z$ is left untouched by the completion procedure as it already had weight $0$ from the start. On the resulting completed graph $\overline{G}$ we apply the following \emph{irreversible} procedure. We add a triangle of edges, each of weight $D/(D-2)$, between the vertices $0,1,\infty$, remove the isolated vertex $z$ and \emph{forget} the labels $0,1,\infty$ in the resulting graph $G^\star$ as illustrated below
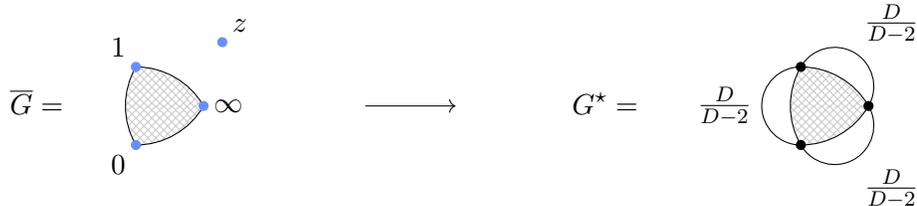
\begin{align*} \overline{G}=\,& \begin{tikzpicture}[baseline={([yshift=-.7ex]0,0)}] \coordinate (v) at (0,0); \useasboundingbox ({-2*\rada},{-2*\rada}) rectangle ({2*\rada},{2*\rada}); \coordinate[label=above left:$1$] (v1) at ([shift=(120:\rada)]v); \coordinate[label=below left:$0$] (v0) at ([shift=(240:\rada)]v); \coordinate[label=right:$\infty$] (voo) at ([shift=(0:\rada)]v); \coordinate[label=above right:$z$] (vz) at ([shift=(45:{2*\rada})]v); \draw[pattern=crosshatch, pattern color=black!20] (v1) arc (150:210:{sqrt(3)*\rada}) arc (-90:-30:{sqrt(3)*\rada}) arc (30:90:{sqrt(3)*\rada}); \filldraw[col2] (v0) circle (1.7pt); \filldraw[col2] (v1) circle (1.7pt); \filldraw[col2] (vz) circle (1.7pt); \filldraw[col2] (voo) circle (1.7pt); \end{tikzpicture} & & \begin{tikzpicture}[baseline={([yshift=-.7ex]0,0)}] \coordinate (v) at (0,0); \useasboundingbox ({-2*\rada},{-2*\rada}) rectangle ({2*\rada},{2*\rada}); \draw[->] (-\rada,0) -- (\rada,0); \end{tikzpicture} & G^\star=\,& \begin{tikzpicture}[baseline={([yshift=-.7ex]0,0)}] \coordinate (v) at (0,0); \useasboundingbox ({-4*\rada},{-2*\rada}) rectangle ({4*\rada},{2*\rada}); \coordinate (v1) at ([shift=(120:\rada)]v); \coordinate (v0) at ([shift=(240:\rada)]v); \coordinate (vz) at ([shift=(0:\rada)]v); \draw[pattern=crosshatch, pattern color=black!20] (v1) arc (150:210:{sqrt(3)*\rada}) arc (-90:-30:{sqrt(3)*\rada}) arc (30:90:{sqrt(3)*\rada}); \draw (v1) arc (90:270:{sqrt(3)/2*\rada}) node[pos=.5,left] {$\frac{D}{D-2}$}; \draw (vz) arc (-30:150:{sqrt(3)/2*\rada}) node[pos=.5,above right] {$\frac{D}{D-2}$}; \draw (v0) arc (-150:30:{sqrt(3)/2*\rada}) node[pos=.5,below right] {$\frac{D}{D-2}$}; \filldraw (v0) circle (1.7pt); \filldraw (v1) circle (1.7pt); \filldraw (vz) circle (1.7pt); \end{tikzpicture} \end{align*}
The graph $G^\star$ is a \emph{weight-regular} graph, i.e.\ each vertex has total weight $\frac{2D}{D-2}$ without any distinction between internal or external vertices.  A surprising and extremely useful property of the period completion is that the Feynman period of a graph $G$ only depends on the period completed graph $G^\star$:
\begin{align} \label{eq:period_completion} P_G = P_{\overline{G}} = P_{G^\star} \end{align}
This property follows from the full conformal symmetry of the underlying integrals (see~\cite[Theorem~2.7]{Schnetz:2008mp}).
In contrast to the completion of graphical functions there are many 
Feynman periods that period complete to the same graph $G^\star$.
The period completion property~\eqref{eq:period_completion} therefore, severely constraints the number of Feynman periods and allows one to bootstrap their value in many cases. 

For both period and graphical function completion we remark that the procedure has the best utility  while working in a fixed integer dimension. While working with dimensional regularization in a non-integer dimension, the second step in the completion procedure in Section~\ref{sec:completegf} usually creates many new edges. Typically, the vertex $\infty$ becomes connected to each other vertex. These additional edges impede the utility of the transformation rules which are presented in this article. However, there exists a collection of new rules which make completion a powerful tool even in non-integer dimensions \cite{7loops}.

\subsection{Factorization}
Another transformation rule that operates on completed graphs is the 
\emph{factorization rule}. It generalizes the \emph{uniqueness} identity on Feynman graphs~\cite{Kazakov:1983dyk}.
The signature of this transformation rule is simple: Whenever the completed graph 
has a three-vertex cut with all the external vertices being isolated on one side of the cut, as indicated below, then the associated expression for the graphical function
factorizes into the product of the completed graphical function on the external side and the completed graphical function on the internal side of the cut. The latter is only a constant graphical function and hence a period.
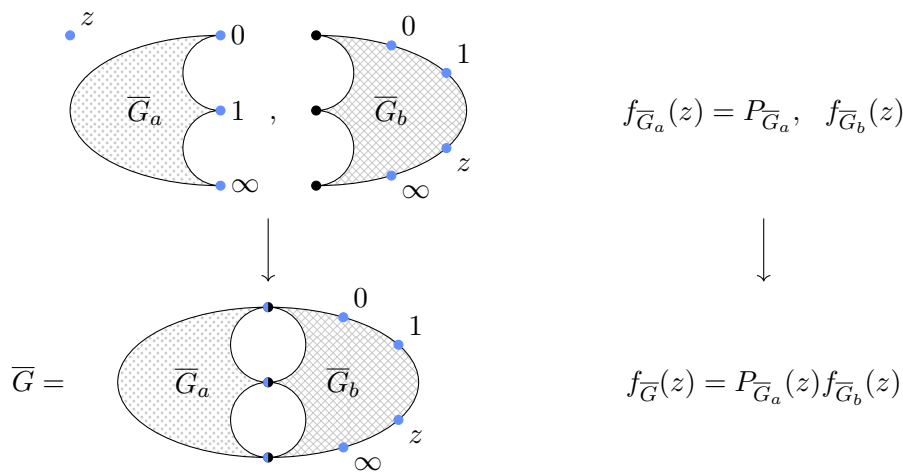
\begin{align*} & \begin{tikzpicture}[baseline={([yshift=-.7ex]0,0)}] \coordinate[label=right:$0$] (va) at (0,1); \coordinate[label=right:$1$] (vb) at (0,0); \coordinate[label=right:$\infty$] (vc) at (0,-1); \coordinate[label=above right:$z$] (voo) at (-2,1); \draw[pattern=crosshatch dots, pattern color=black!20] (va) arc (90:270:2 and 1) arc (270:90:.5) arc (270:90:.5); \node (G1) at (-1,0) {$\overline{G}_a$}; \filldraw[col2] (va) circle (1.7pt); \filldraw[col2] (vb) circle (1.7pt); \filldraw[col2] (vc) circle (1.7pt); \filldraw[col2] (voo) circle (1.7pt); \end{tikzpicture} , \quad \begin{tikzpicture}[baseline={([yshift=-.7ex]0,0)}] \coordinate (wa) at (0,1); \coordinate (wb) at (0,0); \coordinate (wc) at (0,-1); \filldraw (wa) circle (1.7pt); \filldraw (wb) circle (1.7pt); \filldraw (wc) circle (1.7pt); \draw[pattern=crosshatch, pattern color=black!20] (wa) arc (90:-90:2 and 1) arc (-90:90:.5) arc (-90:90:.5); \node (G2) at (1,0) {$\overline{G}_b$}; \coordinate[label=above right:$0$] (v0) at (60:2 and 1); \coordinate[label=above right:$1$] (v1) at (30:2 and 1); \coordinate[label=below right:$z$] (vz) at (-30:2 and 1); \coordinate[label=below right:$\infty$] (voo) at (-60:2 and 1); \filldraw[col2] (v0) circle (1.7pt); \filldraw[col2] (v1) circle (1.7pt); \filldraw[col2] (vz) circle (1.7pt); \filldraw[col2] (voo) circle (1.7pt); \end{tikzpicture} & & \begin{tikzpicture}[baseline={([yshift=-.3ex]0,0)}] \coordinate (v) at (0,0); \useasboundingbox ({-4*\rada},{-.7*\rada}) rectangle ({4*\rada},{.7*\rada}); \node (0,0) {$ \displaystyle f_{\overline{G}_a}(z) = P_{\overline{G}_a} , ~~ f_{\overline{G}_b}(z) $}; \end{tikzpicture} \\
& \begin{tikzpicture}[baseline={([yshift=-.7ex]0,0)}] \useasboundingbox ({-4.5*\rada},{-.7*\rada}) rectangle ({4.5*\rada},{.7*\rada}); \draw[<-] (0,{-.7*\rada}) -- (0,{.7*\rada}); \end{tikzpicture} & & \begin{tikzpicture}[baseline={([yshift=-.7ex]0,0)}] \useasboundingbox ({-4*\rada},{-.7*\rada}) rectangle ({4*\rada},{.7*\rada}); \draw[<-] (0,{-.7*\rada}) -- (0,{.7*\rada}); \end{tikzpicture} \notag \\
 \overline{G}= & \begin{tikzpicture}[baseline={([yshift=-.7ex]0,0)}] \useasboundingbox ({-4.5*\rada},{-2*\rada}) rectangle ({4.5*\rada},{2*\rada}); \coordinate (va) at (0,1); \coordinate (vb) at (0,0); \coordinate (vc) at (0,-1); \draw[pattern=crosshatch dots, pattern color=black!20] (va) arc (90:270:2 and 1) arc (270:90:.5) arc (270:90:.5); \draw[pattern=crosshatch, pattern color=black!20] (va) arc (90:-90:2 and 1) arc (-90:90:.5) arc (-90:90:.5); \node (G1) at (-1,0) {$\overline{G}_a$}; \node (G2) at (1,0) {$\overline{G}_b$}; \coordinate[label=above right:$0$] (v0) at (60:2 and 1); \coordinate[label=above right:$1$] (v1) at (30:2 and 1); \coordinate[label=below right:$z$] (vz) at (-30:2 and 1); \coordinate[label=below right:$\infty$] (voo) at (-60:2 and 1); \filldraw[col2] (v0) circle (1.7pt); \filldraw[col2] (v1) circle (1.7pt); \filldraw[col2] (vz) circle (1.7pt); \filldraw[col2] (voo) circle (1.7pt); \filldraw (va) circle (1.7pt); \filldraw (vb) circle (1.7pt); \filldraw (vc) circle (1.7pt); \filldraw[col2] ([yshift=1.7pt]va) arc (90:270:1.7pt); \filldraw[col2] ([yshift=1.7pt]vb) arc (90:270:1.7pt); \filldraw[col2] ([yshift=1.7pt]vc) arc (90:270:1.7pt); \end{tikzpicture} & & \begin{tikzpicture}[baseline={([yshift=-.3ex]0,0)}] \coordinate (v) at (0,0); \useasboundingbox ({-4*\rada},{-.7*\rada}) rectangle ({4*\rada},{.7*\rada}); \node (0,0) {$ \displaystyle f_{\overline{G}}(z) = P_{\overline{G}_a}(z) f_{\overline{G}_b}(z) $}; \end{tikzpicture} \end{align*}
This identity follows from the conformal invariance of the respective integrals in coordinate space. See \cite[Section~4.3]{gfe} for an elaborate proof. 
Especially interesting are cases where the period $P_{G_a}$
evaluates to $1$. In this case, we have an additional identity that directly relates the functions of different graphs.
This phenomenon is important for graphical functions in six-dimensional spacetime where we get the
degree preserving $\Delta$-Y identity. This identity will be discussed in detail in Section~\ref{sec:deltaY}.

\subsection{Twist}
\label{sec:twist}
If a completed graph has a four-vertex cut with one side of the cut containing all external vertices, then we can perform another transformation: the \emph{twist}. 
To perform a twist on such a graph we have to ensure that both sides of the cut are individually completed graphs. This can be achieved for every four-vertex cut: We can always add auxiliary pairs of positive and negative weight edges (which sum to zero) in between the cut-vertices, which we assign to either of the cut graphs. 
If the graphs on each side of the cut are completed, we can detach one of the sides of the cut, perform a double transposition on the external vertices of the detached side and attach the `twisted' graph again (see illustration below). This operation leaves the associated function invariant.
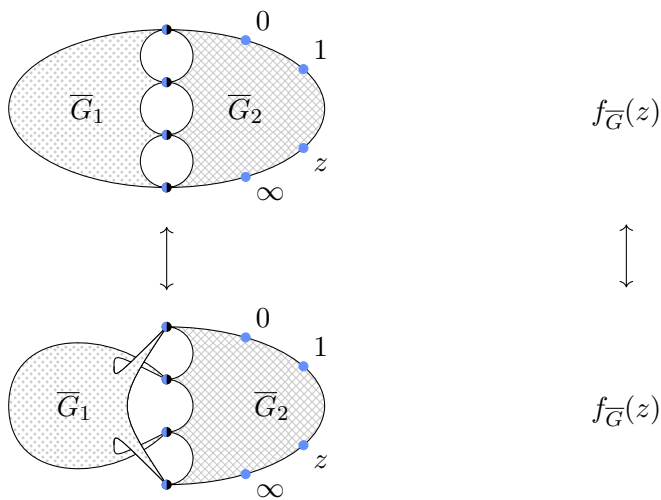
\begin{align*}  & \begin{tikzpicture}[baseline={([yshift=-1.3ex]current bounding box.center)},scale=.7] \useasboundingbox (-4,{-2}) rectangle (4,2); \coordinate (va) at (0,1.5); \coordinate (vb) at (0,.5); \coordinate (vm) at (0,0); \coordinate (vc) at (0,-.5); \coordinate (vd) at (0,-1.5); \draw[pattern=crosshatch dots, pattern color=black!20] (va) arc (90:270:3 and 1.5) arc (270:90:.5) arc (270:90:.5) arc (270:90:.5); \draw[pattern=crosshatch, pattern color=black!20] (va) arc (90:-90:3 and 1.5) arc (-90:90:.5) arc (-90:90:.5) arc (-90:90:.5); \node (G1) at (-1.5,0) {$\overline{G}_1$}; \node (G2) at (+1.5,0) {$\overline{G}_2$}; \coordinate[label=above right:$0$] (v0) at ([shift={(vm)}]60:3 and 1.5); \coordinate[label=above right:$1$] (v1) at ([shift={(vm)}]30:3 and 1.5); \coordinate[label=below right:$z$] (vz) at ([shift={(vm)}]-30:3 and 1.5); \coordinate[label=below right:$\infty$] (voo) at ([shift={(vm)}]-60:3 and 1.5); \filldraw[col2] (v0) circle ({1.7pt/.7}); \filldraw[col2] (v1) circle ({1.7pt/.7}); \filldraw[col2] (vz) circle ({1.7pt/.7}); \filldraw[col2] (voo) circle ({1.7pt/.7}); \filldraw (va) circle ({1.7pt/.7}); \filldraw (vb) circle ({1.7pt/.7}); \filldraw (vc) circle ({1.7pt/.7}); \filldraw (vd) circle ({1.7pt/.7}); \filldraw[col2] ([yshift={1.7pt/.7}]va) arc (90:270:{1.7pt/.7}); \filldraw[col2] ([yshift={1.7pt/.7}]vb) arc (90:270:{1.7pt/.7}); \filldraw[col2] ([yshift={1.7pt/.7}]vc) arc (90:270:{1.7pt/.7}); \filldraw[col2] ([yshift={1.7pt/.7}]vd) arc (90:270:{1.7pt/.7}); \end{tikzpicture} & & \begin{tikzpicture}[baseline={([yshift=-.9ex]0,0)},scale=.7] \coordinate (v) at (0,0); \useasboundingbox ({-1*\rada},{-1*\rada}) rectangle ({1*\rada},{1*\rada}); \node (0,0) {$ f_{\overline{G}}(z) $}; \end{tikzpicture} \\
& \begin{tikzpicture}[baseline={([yshift=-.7ex]0,0)},scale=.7] \coordinate (v) at (0,0); \useasboundingbox (-4,{-\rada}) rectangle (4,\rada); \draw[<->] (0,-\rada) -- (0,\rada); \end{tikzpicture} & & \begin{tikzpicture}[baseline={([yshift=-.9ex]0,0)},scale=.7] \coordinate (v) at (0,0); \useasboundingbox ({-1*\rada},{-1*\rada}) rectangle ({1*\rada},{1*\rada}); \draw[<->] (0,-\rada) -- (0,\rada); \end{tikzpicture} \\
 & \begin{tikzpicture}[baseline={([yshift=-1.3ex]current bounding box.center)},scale=.7] \useasboundingbox (-4,{-2}) rectangle (4,2); \coordinate (va) at (0,1.5); \coordinate (vb) at (0,.5); \coordinate (vm) at (0,0); \coordinate (vc) at (0,-.5); \coordinate (vd) at (0,-1.5); \draw[pattern=crosshatch dots, pattern color=black!20] (vb) .. controls (-1,1.5) and (-3,1.5) .. (-3,0) .. controls (-3,-1.5) and (-1,-1.5) .. (vc) .. controls (-1,-1) .. (-1,-.75) .. controls (-1,-.5) .. (vd) .. controls (-1,0) .. (va) .. controls (-1,.5) .. (-1,.75) .. controls (-1,1) .. (vb); \draw[preaction={fill, white},pattern=crosshatch dots, pattern color=black!20] (-1,-.75) .. controls (-1,-.5) .. (vd) .. controls (-1,0) .. (va) .. controls (-1,.5) .. (-1,.75); \draw[pattern=crosshatch, pattern color=black!20] (va) arc (90:-90:3 and 1.5) arc (-90:90:.5) arc (-90:90:.5) arc (-90:90:.5); \node (G1) at (-1.75,0) {$\overline{G}_1$}; \node (G2) at (+2,0) {$\overline{G}_2$}; \coordinate[label=above right:$0$] (v0) at ([shift={(vm)}]60:3 and 1.5); \coordinate[label=above right:$1$] (v1) at ([shift={(vm)}]30:3 and 1.5); \coordinate[label=below right:$z$] (vz) at ([shift={(vm)}]-30:3 and 1.5); \coordinate[label=below right:$\infty$] (voo) at ([shift={(vm)}]-60:3 and 1.5); \filldraw[col2] (v0) circle ({1.7pt/.7}); \filldraw[col2] (v1) circle ({1.7pt/.7}); \filldraw[col2] (vz) circle ({1.7pt/.7}); \filldraw[col2] (voo) circle ({1.7pt/.7}); \filldraw (va) circle ({1.7pt/.7}); \filldraw (vb) circle ({1.7pt/.7}); \filldraw (vc) circle ({1.7pt/.7}); \filldraw (vd) circle ({1.7pt/.7}); \filldraw[col2] ([yshift={1.7pt/.7}]va) arc (90:270:{1.7pt/.7}); \filldraw[col2] ([yshift={1.7pt/.7}]vb) arc (90:270:{1.7pt/.7}); \filldraw[col2] ([yshift={1.7pt/.7}]vc) arc (90:270:{1.7pt/.7}); \filldraw[col2] ([yshift={1.7pt/.7}]vd) arc (90:270:{1.7pt/.7}); \end{tikzpicture} & & \begin{tikzpicture}[baseline={([yshift=-.9ex]0,0)},scale=.7] \coordinate (v) at (0,0); \useasboundingbox ({-1*\rada},{-1*\rada}) rectangle ({1*\rada},{1*\rada}); \node (0,0) {$ f_{\overline{G}}(z) $}; \end{tikzpicture} \end{align*}
The twist identity was already introduced in~\cite{Drummond:2006rz} for conformal four-point integrals.
The proof of this identity follows from the conformal invariance and the invariance of conformal four-point integrals under double transposition of the external vertices (Section~\ref{sec:perm}). For a full proof in the context of periods see~\cite[Theorem~2.11]{Schnetz:2008mp} or \cite[Section~8.1]{gfe}.

\subsection{Other general transformation rules}
\label{sec:othertrafo}
We discussed the most significant transformation rules for graphical functions and conformal four-point integrals.
There are more such rules which are useful in some cases: 
\begin{enumerate}
\item
\emph{Planar duality}, where a subgraph from a four-vertex-cut (as for the twist) is replaced with its planar dual (see~\cite[Section~8.2]{gfe}). 
\item
\emph{Gegenbauer factorization}, which transforms a graph with a three-vertex cut, where not all external vertices have to lie on one side of the cut (see \cite[Section~7]{gfe}).
\item
\emph{External differentiation} which can be used to simplify graphs with multiple edges attached to the external vertices (see~\cite[Section~11]{gfe}).
\item
\emph{Integration by parts} (IBP) which works as the usual integration by parts identities in momentum space, but has a combinatorial interpretation in the context of graphical functions (see \cite[Section~9]{gfe}). The shape of the IBP identity depends heavily on the dimension of spacetime.  
We will discuss the six-dimensional version of this identity explicitly in Section~\ref{sec:IBP}. 
\end{enumerate}

\subsection{Other methods to obtain analytic expressions for graphical functions}
\label{sec:external}

Starting with the trivial graph and applying different transformation rules repeatedly allows one to construct analytic expressions for a huge family of graphs. However, not all graphs are accessible via these transformation rules.

A simple conformal four-point integral that cannot be computed from the trivial graph $G_1$~\eqref{eq:trivialgf} in $D=4$ using the transformation rules from Section~\ref{sec:addedges}-\ref{sec:twist} is 
\begin{align*} \def\scale{1} \begin{tikzpicture}[baseline={([yshift=-.7ex]0,0)}] \coordinate (vm) at (0,0); \coordinate[label=above left:$z$] (vz) at (-\scale, \scale); \coordinate[label=above right:$1$] (v1) at ( \scale, \scale); \coordinate[label=below left:$\infty$] (voo) at (-\scale,-\scale); \coordinate[label=below right:$0$] (v0) at (\scale, -\scale); \coordinate (v3) at ([shift=(90:{\scale/2})]vm); \coordinate (v2) at ([shift=(210:{\scale/2})]vm); \coordinate (v4) at ([shift=(330:{\scale/2})]vm); \draw (vz) -- (v2); \draw (voo) -- (v4); \draw (voo) -- (v2); \draw (v0) -- (v4); \draw (v1) -- (v4); \draw (vz) -- (v3); \draw[preaction={draw, white, line width=3pt, -}] (v0) -- (v2); \draw (v1) -- (v3); \draw (v0) -- ($(v0)!.2!(v4)$); \draw (v2) -- (v3) -- (v4); \filldraw[col2] (v0) circle (1.7pt); \filldraw[col2] (v1) circle (1.7pt); \filldraw[col2] (vz) circle (1.7pt); \filldraw[col2] (voo) circle (1.7pt); \filldraw (v2) circle (1.7pt); \filldraw (v3) circle (1.7pt); \filldraw (v4) circle (1.7pt); \end{tikzpicture}, \end{align*}
where edges between external vertices have been omitted as they do not influence the constructability of the graph: We can add and remove edges between external edges arbitrarily (see Section~\ref{sec:addedges}). The associated function can still be computed using Gegenbauer factorization~(\cite[Section~7]{gfe}). This way the graph above is a new nontrivial starting point
(a kernel) for the transformation rules and another infinite family of graphical functions and conformal four-point integrals can be obtained recursively. 

In general, conformal four-point integrals from complementary methods are very valuable. A rich source of external graphical functions is Erik Panzer's \texttt{HyperInt}~\cite{Panzer:2014caa}, which operates on the graphical  functions' parametric representation~\cite{Golz:2015rea}, relying on linear reducibility~\cite{Brown:2008um} and therefore works complementary to the methods described here.

An interesting example of a conformal four-point integral, for which no closed form expression is known, is
\begin{align} \label{eq:elliptic} \def\scale{1} \begin{tikzpicture}[baseline={([yshift=-.7ex]0,0)}] \coordinate (vm) at (0,0); \coordinate[label=above left:$z$] (vz) at (-\scale, \scale); \coordinate[label=above right:$1$] (v1) at ( \scale, \scale); \coordinate[label=below left:$\infty$] (voo) at (-\scale,-\scale); \coordinate[label=below right:$0$] (v0) at (\scale, -\scale); \coordinate (v2) at ({ \scale/3}, { \scale/3}); \coordinate (v3) at ({-\scale/3}, { \scale/3}); \coordinate (v4) at ({-\scale/3}, {-\scale/3}); \coordinate (v5) at ({ \scale/3}, {-\scale/3}); \draw (vz) -- (v2); \draw (voo) -- (v3); \draw (v0) -- (v4); \draw (v1) -- (v5); \draw[preaction={draw, white, line width=3pt, -}] (vz) -- (v4); \draw[preaction={draw, white, line width=3pt, -}] (voo) -- (v5); \draw[preaction={draw, white, line width=3pt, -}] (v0) -- (v2); \draw[preaction={draw, white, line width=3pt, -}] (v1) -- (v3); \draw (vz) -- ($(vz)!.2!(v2)$); \draw (voo) -- ($(voo)!.2!(v3)$); \draw (v0) -- ($(v0)!.2!(v4)$); \draw (v1) -- ($(v1)!.2!(v5)$); \draw (v2) -- (v3) -- (v4) -- (v5) -- (v2); \filldraw[col2] (v0) circle (1.7pt); \filldraw[col2] (v1) circle (1.7pt); \filldraw[col2] (vz) circle (1.7pt); \filldraw[col2] (voo) circle (1.7pt); \filldraw (v2) circle (1.7pt); \filldraw (v3) circle (1.7pt); \filldraw (v4) circle (1.7pt); \filldraw (v5) circle (1.7pt); \end{tikzpicture}, \end{align}
where again edges between external vertices have been omitted. 
There is no (known) way to use the (known) transformation rules to simply this graph. 
The associated parametric Feynman integral is not linear reducible in the sense of~\cite{Brown:2008um} and \texttt{HyperInt} fails.
In fact, the function associated to this graph is conjectured to be \emph{elliptic}. A closed form expression for this function (together with a good handle on elliptic functions)
would be especially interesting, as a Feynman period that can be evaluated from it provides the most interesting check on the coaction conjecture in $\phi^4$
theory~\cite{PanzerSchnetz:2017coact} to date.

Many conformal four-point integrals that have been computed using integrability methods fall into the class of nontrivial starting points for the transformation rules. 

An especially rich resource are the fishnet graphs in $D=4$,
\begin{align*} \def\scale{.5} \begin{tikzpicture}[baseline={([yshift=-.7ex]0,0)}] \coordinate[label=above:$0$] (v0) at (0, {4*\scale}); \coordinate[label=left:$1$] (v1) at ({-4*\scale}, 0); \coordinate[label=right:$z$] (vz) at ({4*\scale},0); \coordinate[label=below:$\infty$] (voo) at (0, {-4*\scale}); \coordinate (vlt0) at ({-2*\scale}, {2*\scale}); \coordinate (vlt1) at ({-1*\scale}, {2*\scale}); \coordinate (vlt2) at ({-2*\scale}, {1*\scale}); \coordinate (vlt3) at ({-1*\scale}, {1*\scale}); \coordinate (vrt0) at ({2*\scale}, {2*\scale}); \coordinate (vrt1) at ({1*\scale}, {2*\scale}); \coordinate (vrt2) at ({2*\scale}, {1*\scale}); \coordinate (vrt3) at ({1*\scale}, {1*\scale}); \coordinate (vlb0) at ({-2*\scale}, {-2*\scale}); \coordinate (vlb1) at ({-1*\scale}, {-2*\scale}); \coordinate (vlb2) at ({-2*\scale}, {-1*\scale}); \coordinate (vlb3) at ({-1*\scale}, {-1*\scale}); \coordinate (vrb0) at ({2*\scale}, {-2*\scale}); \coordinate (vrb1) at ({1*\scale}, {-2*\scale}); \coordinate (vrb2) at ({2*\scale}, {-1*\scale}); \coordinate (vrb3) at ({1*\scale}, {-1*\scale}); \draw (v0) -- (vlt0); \draw (v0) -- (vlt1); \draw (v0) -- (vrt0); \draw (v0) -- (vrt1); \draw (v1) -- (vlt0); \draw (v1) -- (vlt2); \draw (v1) -- (vlb0); \draw (v1) -- (vlb2); \draw (voo) -- (vlb0); \draw (voo) -- (vlb1); \draw (voo) -- (vrb0); \draw (voo) -- (vrb1); \draw (vz) -- (vrt0); \draw (vz) -- (vrt2); \draw (vz) -- (vrb0); \draw (vz) -- (vrb2); \draw (vlt0) -- (vlt1) -- (vlt3) -- (vlt2) -- (vlt0); \draw (vrt0) -- (vrt1) -- (vrt3) -- (vrt2) -- (vrt0); \draw (vlb0) -- (vlb1) -- (vlb3) -- (vlb2) -- (vlb0); \draw (vrb0) -- (vrb1) -- (vrb3) -- (vrb2) -- (vrb0); \draw[black!50] (vlt2) -- ($(vlt2)!.3!(vlb2)$); \draw[black!50] (vlb2) -- ($(vlb2)!.3!(vlt2)$); \draw[black!50] (vlt3) -- ($(vlt3)!.3!(vlb3)$); \draw[black!50] (vlb3) -- ($(vlb3)!.3!(vlt3)$); \draw[black!50] (vrt2) -- ($(vrt2)!.3!(vrb2)$); \draw[black!50] (vrb2) -- ($(vrb2)!.3!(vrt2)$); \draw[black!50] (vrt3) -- ($(vrt3)!.3!(vrb3)$); \draw[black!50] (vrb3) -- ($(vrb3)!.3!(vrt3)$); \draw[black!50] (vlt1) -- ($(vlt1)!.3!(vrt1)$); \draw[black!50] (vrt1) -- ($(vrt1)!.3!(vlt1)$); \draw[black!50] (vlt3) -- ($(vlt3)!.3!(vrt3)$); \draw[black!50] (vrt3) -- ($(vrt3)!.3!(vlt3)$); \draw[black!50] (vlb1) -- ($(vlb1)!.3!(vrb1)$); \draw[black!50] (vrb1) -- ($(vrb1)!.3!(vlb1)$); \draw[black!50] (vlb3) -- ($(vlb3)!.3!(vrb3)$); \draw[black!50] (vrb3) -- ($(vrb3)!.3!(vlb3)$); \draw[black!50] (v0) -- ($(v0)!.15!(vlb3)$); \draw[black!50] (v0) -- ($(v0)!.15!(vrb3)$); \draw[black!50] (v1) -- ($(v1)!.15!(vrt3)$); \draw[black!50] (v1) -- ($(v1)!.15!(vrb3)$); \draw[black!50] (vz) -- ($(vz)!.15!(vlt3)$); \draw[black!50] (vz) -- ($(vz)!.15!(vlb3)$); \draw[black!50] (voo) -- ($(voo)!.15!(vlt3)$); \draw[black!50] (voo) -- ($(voo)!.15!(vrt3)$); \node[black!50] (BT) at (0,{.5*1.5}) {$\bullet \bullet \bullet$}; \node[black!50] (BL) at ({-.5*1.5},0) {\rotatebox{90}{$\bullet \bullet \bullet$}}; \node[black!50] (BB) at (0,{-.5*1.5}) {$\bullet \bullet \bullet$}; \node[black!50] (BR) at ({.5*1.5},0) {\rotatebox{90}{$\bullet \bullet \bullet$}}; \filldraw[col2] (voo) circle (1.7pt); \filldraw[col2] (v1) circle (1.7pt); \filldraw[col2] (v0) circle (1.7pt); \filldraw[col2] (vz) circle (1.7pt); \filldraw (vlt0) circle (1.7pt); \filldraw (vlt1) circle (1.7pt); \filldraw (vlt2) circle (1.7pt); \filldraw (vlt3) circle (1.7pt); \filldraw (vrt0) circle (1.7pt); \filldraw (vrt1) circle (1.7pt); \filldraw (vrt2) circle (1.7pt); \filldraw (vrt3) circle (1.7pt); \filldraw (vlb0) circle (1.7pt); \filldraw (vlb1) circle (1.7pt); \filldraw (vlb2) circle (1.7pt); \filldraw (vlb3) circle (1.7pt); \filldraw (vrb0) circle (1.7pt); \filldraw (vrb1) circle (1.7pt); \filldraw (vrb2) circle (1.7pt); \filldraw (vrb3) circle (1.7pt); \end{tikzpicture}~~, \end{align*}
which generalize the ladder graphs from~\cite{Usyukina:1993ch}. 
Even though the transformation rules~(Section~\ref{sec:addedges}--\ref{sec:appendedge}) are sufficient to construct the ladder graphs from the empty graph (see Section~\ref{sec:ladder}), the more general fishnet graphs cannot be evaluated using any (known) combination of transformation rules.

A general formula for the fishnet graphs, which was conjectured in~\cite{Basso:2017jwq}, has recently been proved \cite{Basso:2021omx}. This formula is also implemented in \texttt{HyperlogProcedures}~\cite{Shlog}. This way, each fishnet graph provides a new  nontrivial starting point for the transformation rules from Section~\ref{sec:addedges}--\ref{sec:othertrafo} to be applied. 

\section{Application to \texorpdfstring{$\phi^3$}{phi3} theory in six-dimensional spacetime}
\label{phi3}

In this section, we will go through the workflow of using conformal four-point integrals and graphical functions to tackle a concrete QFT computation: We will apply our methods to $\phi^3$ theory in $D=6$ dimensional spacetime. We will focus on the computation of the \emph{primitive} part of $\phi^3$ theory. That means we will only compute Feynman periods of $3$-regular graphs in $D=6$ and leave Feynman graphs in $\phi^3$ theory aside that have subdivergences. 

The results of this computation up to $5$ loops have already been used in~\cite{Borinsky:2021jdb} to make predictions for critical exponents in percolation theory and for the Lee--Yang edge singularity.

A finite graphical function or conformal four-point integral in $D=6$ with a rational integrand can have edge weights: $1, \tfrac12, 0,-\tfrac12, -1, -\tfrac32, \ldots$
Otherwise, the associated Feynman integral~\eqref{feynmanint} would feature roots in the integrand or be divergent. 
In a completed graph corresponding to a conformal four-point function in $D=6$, every internal vertex has total weight $3$. The Feynman period completed graphs, as described in Section~\ref{sec:period_completion}, have total weight $3$ at each vertex in $D=6$.%

With these restrictions two identities that have been discussed in the previous section take a particularly simple form.

\subsection{\texorpdfstring{$\Delta$-Y}{Delta-Y} transform in \texorpdfstring{$D=6$}{D=6}}
\label{sec:deltaY}
Graphical functions, conformal four-point functions and Feynman periods in $D=6$ are left invariant if the associated graph is modified using a variation of the classical \emph{$\Delta$-Y transform}~\cite{kennelly1899equivalence}. The $\Delta$-Y transform has applications in electrical networks, statistical physics and combinatorics in general (for a discussion of the literature see \cite{Jeffries:2021aog}). In the context of Feynman integrals this identity is known as the \emph{uniqueness relation} \cite{Kazakov:1983dyk}. 

Our $\Delta$-Y transform on the level of graphical functions, conformal four-point functions and Feynman periods in $D=6$ modifies 
the underlying graph by replacing an internal vertex with exactly three incident weight $1$ edges (the `Y') with a triangle of weight $\tfrac12$ edges (the `$\Delta$') such that the total degrees of the adjacent vertices are preserved. The functions or numbers associated to the two graphs are equal.
This transformation rule can be illustrated as follows:
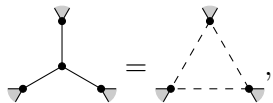
\begin{align*} \begin{tikzpicture}[baseline={([yshift=-0.7ex]0,0)}] \coordinate (vm) at (0, 0); \coordinate (v1) at ([shift=(-30:.6)]vm); \coordinate (v2) at ([shift=(90:.6)]vm); \coordinate (v3) at ([shift=(210:.6)]vm); \filldraw[black!20] (v2) -- ([shift=(60:.2)]v2) arc (60:120:.2) -- (v2); \draw (v2) -- ([shift=(60:.2)]v2); \draw (v2) -- ([shift=(120:.2)]v2); \filldraw[black!20] (v1) -- ([shift=(0:.2)]v1) arc (0:-60:.2) -- (v1); \draw (v1) -- ([shift=(0:.2)]v1); \draw (v1) -- ([shift=(-60:.2)]v1); \filldraw[black!20] (v3) -- ([shift=(180:.2)]v3) arc (180:240:.2) -- (v3); \draw (v3) -- ([shift=(180:.2)]v3); \draw (v3) -- ([shift=(240:.2)]v3); \filldraw (v1) circle (1.3pt); \filldraw (v2) circle (1.3pt); \filldraw (v3) circle (1.3pt); \filldraw (vm) circle (1.3pt); \draw (vm) -- (v1); \draw (vm) -- (v2); \draw (vm) -- (v3); \end{tikzpicture} = \begin{tikzpicture}[baseline={([yshift=-0.7ex]0,0)}] \coordinate (vm) at (0, 0); \coordinate (v1) at ([shift=(-30:.6)]vm); \coordinate (v2) at ([shift=(90:.6)]vm); \coordinate (v3) at ([shift=(210:.6)]vm); \filldraw[black!20] (v2) -- ([shift=(60:.2)]v2) arc (60:120:.2) -- (v2); \draw (v2) -- ([shift=(60:.2)]v2); \draw (v2) -- ([shift=(120:.2)]v2); \filldraw[black!20] (v1) -- ([shift=(0:.2)]v1) arc (0:-60:.2) -- (v1); \draw (v1) -- ([shift=(0:.2)]v1); \draw (v1) -- ([shift=(-60:.2)]v1); \filldraw[black!20] (v3) -- ([shift=(180:.2)]v3) arc (180:240:.2) -- (v3); \draw (v3) -- ([shift=(180:.2)]v3); \draw (v3) -- ([shift=(240:.2)]v3); \filldraw (v1) circle (1.3pt); \filldraw (v2) circle (1.3pt); \filldraw (v3) circle (1.3pt); \draw[dashed] (v1) -- (v2); \draw[dashed] (v2) -- (v3); \draw[dashed] (v1) -- (v3); \end{tikzpicture}, \end{align*}
where solid edges $\solidedge$ have weight $1$ and dashed edges $\dashededge$
have weight $\frac12$. This is a \emph{local} graph identity. We have to think of the graphs to be inserted into the same spot of a larger ambient graph at the gray indicated attachment points.
The ambient graphs on both sides of the equation have equal graphical functions, conformal four-point integrals or Feynman periods.

In~\cite{Jeffries:2021aog} the equivalence classes of graphs under this $\Delta$-Y relation have been studied extensively. The $\Delta$-Y transform yields very interesting equivalence classes of analytic objects. The size of these equivalence classes is often infinite~\cite{Jeffries:2021aog}. Many graphs reduce to the trivial graph by application of the $\Delta$-Y identity alone.

\subsection{IBP identity in \texorpdfstring{$D=6$}{D=6}}
\label{sec:IBP}
The most useful IBP identity for graphical functions, conformal four-point integrals and Feynman periods in $D=6$ (the general form is discussed in \cite[Section~9]{gfe}) takes the following simple form.
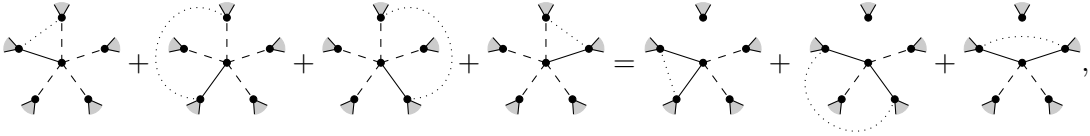
\begin{align*} \begin{tikzpicture}[baseline={([yshift=-0.7ex]0,0)}] \coordinate (vm) at (0, 0); \coordinate (v1) at ([shift=(90:.6)]vm); \coordinate (v2) at ([shift=(162:.6)]vm); \coordinate (v3) at ([shift=(234:.6)]vm); \coordinate (v4) at ([shift=(306:.6)]vm); \coordinate (v5) at ([shift=(18:.6)]vm); \filldraw[black!20] (v1) -- ([shift=(60:.2)]v1) arc (60:120:.2) -- (v1); \draw (v1) -- ([shift=(60:.2)]v1); \draw (v1) -- ([shift=(120:.2)]v1); \filldraw[black!20] (v2) -- ([shift=(132:.2)]v2) arc (132:192:.2) -- (v2); \draw (v2) -- ([shift=(132:.2)]v2); \draw (v2) -- ([shift=(192:.2)]v2); \filldraw[black!20] (v3) -- ([shift=(204:.2)]v3) arc (204:264:.2) -- (v3); \draw (v3) -- ([shift=(204:.2)]v3); \draw (v3) -- ([shift=(264:.2)]v3); \filldraw[black!20] (v4) -- ([shift=(276:.2)]v4) arc (276:336:.2) -- (v4); \draw (v4) -- ([shift=(274:.2)]v4); \draw (v4) -- ([shift=(336:.2)]v4); \filldraw[black!20] (v5) -- ([shift=(348:.2)]v5) arc (-12:48:.2) -- (v5); \draw (v5) -- ([shift=(348:.2)]v5); \draw (v5) -- ([shift=(48:.2)]v5); \filldraw (v1) circle (1.3pt); \filldraw (v2) circle (1.3pt); \filldraw (v3) circle (1.3pt); \filldraw (v4) circle (1.3pt); \filldraw (v5) circle (1.3pt); \filldraw (vm) circle (1.3pt); \draw[dashed] (vm) -- (v1); \draw (vm) -- (v2); \draw[dashed] (vm) -- (v3); \draw[dashed] (vm) -- (v4); \draw[dashed] (vm) -- (v5); \draw[dotted] (v1) -- (v2); \end{tikzpicture} + \hspace{-3ex} \begin{tikzpicture}[baseline={([yshift=-0.7ex]0,0)}] \coordinate (vm) at (0, 0); \coordinate (v1) at ([shift=(90:.6)]vm); \coordinate (v2) at ([shift=(162:.6)]vm); \coordinate (v3) at ([shift=(234:.6)]vm); \coordinate (v4) at ([shift=(306:.6)]vm); \coordinate (v5) at ([shift=(18:.6)]vm); \filldraw[black!20] (v1) -- ([shift=(60:.2)]v1) arc (60:120:.2) -- (v1); \draw (v1) -- ([shift=(60:.2)]v1); \draw (v1) -- ([shift=(120:.2)]v1); \filldraw[black!20] (v2) -- ([shift=(132:.2)]v2) arc (132:192:.2) -- (v2); \draw (v2) -- ([shift=(132:.2)]v2); \draw (v2) -- ([shift=(192:.2)]v2); \filldraw[black!20] (v3) -- ([shift=(204:.2)]v3) arc (204:264:.2) -- (v3); \draw (v3) -- ([shift=(204:.2)]v3); \draw (v3) -- ([shift=(264:.2)]v3); \filldraw[black!20] (v4) -- ([shift=(276:.2)]v4) arc (276:336:.2) -- (v4); \draw (v4) -- ([shift=(274:.2)]v4); \draw (v4) -- ([shift=(336:.2)]v4); \filldraw[black!20] (v5) -- ([shift=(348:.2)]v5) arc (-12:48:.2) -- (v5); \draw (v5) -- ([shift=(348:.2)]v5); \draw (v5) -- ([shift=(48:.2)]v5); \filldraw (v1) circle (1.3pt); \filldraw (v2) circle (1.3pt); \filldraw (v3) circle (1.3pt); \filldraw (v4) circle (1.3pt); \filldraw (v5) circle (1.3pt); \filldraw (vm) circle (1.3pt); \draw[dashed] (vm) -- (v1); \draw[dashed] (vm) -- (v2); \draw (vm) -- (v3); \draw[dashed] (vm) -- (v4); \draw[dashed] (vm) -- (v5); \draw[dotted] (v1) .. controls ([shift=(126:1.5)]vm) and ([shift=(198:1.5)]vm) .. (v3); \end{tikzpicture} + \begin{tikzpicture}[baseline={([yshift=-0.7ex]0,0)}] \coordinate (vm) at (0, 0); \coordinate (v1) at ([shift=(90:.6)]vm); \coordinate (v2) at ([shift=(162:.6)]vm); \coordinate (v3) at ([shift=(234:.6)]vm); \coordinate (v4) at ([shift=(306:.6)]vm); \coordinate (v5) at ([shift=(18:.6)]vm); \filldraw[black!20] (v1) -- ([shift=(60:.2)]v1) arc (60:120:.2) -- (v1); \draw (v1) -- ([shift=(60:.2)]v1); \draw (v1) -- ([shift=(120:.2)]v1); \filldraw[black!20] (v2) -- ([shift=(132:.2)]v2) arc (132:192:.2) -- (v2); \draw (v2) -- ([shift=(132:.2)]v2); \draw (v2) -- ([shift=(192:.2)]v2); \filldraw[black!20] (v3) -- ([shift=(204:.2)]v3) arc (204:264:.2) -- (v3); \draw (v3) -- ([shift=(204:.2)]v3); \draw (v3) -- ([shift=(264:.2)]v3); \filldraw[black!20] (v4) -- ([shift=(276:.2)]v4) arc (276:336:.2) -- (v4); \draw (v4) -- ([shift=(274:.2)]v4); \draw (v4) -- ([shift=(336:.2)]v4); \filldraw[black!20] (v5) -- ([shift=(348:.2)]v5) arc (-12:48:.2) -- (v5); \draw (v5) -- ([shift=(348:.2)]v5); \draw (v5) -- ([shift=(48:.2)]v5); \filldraw (v1) circle (1.3pt); \filldraw (v2) circle (1.3pt); \filldraw (v3) circle (1.3pt); \filldraw (v4) circle (1.3pt); \filldraw (v5) circle (1.3pt); \filldraw (vm) circle (1.3pt); \draw[dashed] (vm) -- (v1); \draw[dashed] (vm) -- (v2); \draw[dashed] (vm) -- (v3); \draw (vm) -- (v4); \draw[dashed] (vm) -- (v5); \draw[dotted] (v1) .. controls ([shift=(54:1.5)]vm) and ([shift=(-18:1.5)]vm) .. (v4); \end{tikzpicture} \hspace{-3ex} + \begin{tikzpicture}[baseline={([yshift=-0.7ex]0,0)}] \coordinate (vm) at (0, 0); \coordinate (v1) at ([shift=(90:.6)]vm); \coordinate (v2) at ([shift=(162:.6)]vm); \coordinate (v3) at ([shift=(234:.6)]vm); \coordinate (v4) at ([shift=(306:.6)]vm); \coordinate (v5) at ([shift=(18:.6)]vm); \filldraw[black!20] (v1) -- ([shift=(60:.2)]v1) arc (60:120:.2) -- (v1); \draw (v1) -- ([shift=(60:.2)]v1); \draw (v1) -- ([shift=(120:.2)]v1); \filldraw[black!20] (v2) -- ([shift=(132:.2)]v2) arc (132:192:.2) -- (v2); \draw (v2) -- ([shift=(132:.2)]v2); \draw (v2) -- ([shift=(192:.2)]v2); \filldraw[black!20] (v3) -- ([shift=(204:.2)]v3) arc (204:264:.2) -- (v3); \draw (v3) -- ([shift=(204:.2)]v3); \draw (v3) -- ([shift=(264:.2)]v3); \filldraw[black!20] (v4) -- ([shift=(276:.2)]v4) arc (276:336:.2) -- (v4); \draw (v4) -- ([shift=(274:.2)]v4); \draw (v4) -- ([shift=(336:.2)]v4); \filldraw[black!20] (v5) -- ([shift=(348:.2)]v5) arc (-12:48:.2) -- (v5); \draw (v5) -- ([shift=(348:.2)]v5); \draw (v5) -- ([shift=(48:.2)]v5); \filldraw (v1) circle (1.3pt); \filldraw (v2) circle (1.3pt); \filldraw (v3) circle (1.3pt); \filldraw (v4) circle (1.3pt); \filldraw (v5) circle (1.3pt); \filldraw (vm) circle (1.3pt); \draw[dashed] (vm) -- (v1); \draw[dashed] (vm) -- (v2); \draw[dashed] (vm) -- (v3); \draw[dashed] (vm) -- (v4); \draw (vm) -- (v5); \draw[dotted] (v1) -- (v5); \end{tikzpicture} = \begin{tikzpicture}[baseline={([yshift=-0.7ex]0,0)}] \coordinate (vm) at (0, 0); \coordinate (v1) at ([shift=(90:.6)]vm); \coordinate (v2) at ([shift=(162:.6)]vm); \coordinate (v3) at ([shift=(234:.6)]vm); \coordinate (v4) at ([shift=(306:.6)]vm); \coordinate (v5) at ([shift=(18:.6)]vm); \filldraw[black!20] (v1) -- ([shift=(60:.2)]v1) arc (60:120:.2) -- (v1); \draw (v1) -- ([shift=(60:.2)]v1); \draw (v1) -- ([shift=(120:.2)]v1); \filldraw[black!20] (v2) -- ([shift=(132:.2)]v2) arc (132:192:.2) -- (v2); \draw (v2) -- ([shift=(132:.2)]v2); \draw (v2) -- ([shift=(192:.2)]v2); \filldraw[black!20] (v3) -- ([shift=(204:.2)]v3) arc (204:264:.2) -- (v3); \draw (v3) -- ([shift=(204:.2)]v3); \draw (v3) -- ([shift=(264:.2)]v3); \filldraw[black!20] (v4) -- ([shift=(276:.2)]v4) arc (276:336:.2) -- (v4); \draw (v4) -- ([shift=(274:.2)]v4); \draw (v4) -- ([shift=(336:.2)]v4); \filldraw[black!20] (v5) -- ([shift=(348:.2)]v5) arc (-12:48:.2) -- (v5); \draw (v5) -- ([shift=(348:.2)]v5); \draw (v5) -- ([shift=(48:.2)]v5); \filldraw (v1) circle (1.3pt); \filldraw (v2) circle (1.3pt); \filldraw (v3) circle (1.3pt); \filldraw (v4) circle (1.3pt); \filldraw (v5) circle (1.3pt); \filldraw (vm) circle (1.3pt); \draw (vm) -- (v2); \draw (vm) -- (v3); \draw[dashed] (vm) -- (v4); \draw[dashed] (vm) -- (v5); \draw[dotted] (v2) -- (v3); \end{tikzpicture} + \hspace{-3ex} \begin{tikzpicture}[baseline={([yshift=-0.7ex]0,0)}] \coordinate (vm) at (0, 0); \coordinate (v1) at ([shift=(90:.6)]vm); \coordinate (v2) at ([shift=(162:.6)]vm); \coordinate (v3) at ([shift=(234:.6)]vm); \coordinate (v4) at ([shift=(306:.6)]vm); \coordinate (v5) at ([shift=(18:.6)]vm); \filldraw[black!20] (v1) -- ([shift=(60:.2)]v1) arc (60:120:.2) -- (v1); \draw (v1) -- ([shift=(60:.2)]v1); \draw (v1) -- ([shift=(120:.2)]v1); \filldraw[black!20] (v2) -- ([shift=(132:.2)]v2) arc (132:192:.2) -- (v2); \draw (v2) -- ([shift=(132:.2)]v2); \draw (v2) -- ([shift=(192:.2)]v2); \filldraw[black!20] (v3) -- ([shift=(204:.2)]v3) arc (204:264:.2) -- (v3); \draw (v3) -- ([shift=(204:.2)]v3); \draw (v3) -- ([shift=(264:.2)]v3); \filldraw[black!20] (v4) -- ([shift=(276:.2)]v4) arc (276:336:.2) -- (v4); \draw (v4) -- ([shift=(274:.2)]v4); \draw (v4) -- ([shift=(336:.2)]v4); \filldraw[black!20] (v5) -- ([shift=(348:.2)]v5) arc (-12:48:.2) -- (v5); \draw (v5) -- ([shift=(348:.2)]v5); \draw (v5) -- ([shift=(48:.2)]v5); \filldraw (v1) circle (1.3pt); \filldraw (v2) circle (1.3pt); \filldraw (v3) circle (1.3pt); \filldraw (v4) circle (1.3pt); \filldraw (v5) circle (1.3pt); \filldraw (vm) circle (1.3pt); \draw (vm) -- (v2); \draw[dashed] (vm) -- (v3); \draw (vm) -- (v4); \draw[dashed] (vm) -- (v5); \draw[dotted] (v2) .. controls ([shift=(198:1.5)]vm) and ([shift=(270:1.5)]vm) .. (v4); \end{tikzpicture} + \begin{tikzpicture}[baseline={([yshift=-0.7ex]0,0)}] \coordinate (vm) at (0, 0); \coordinate (v1) at ([shift=(90:.6)]vm); \coordinate (v2) at ([shift=(162:.6)]vm); \coordinate (v3) at ([shift=(234:.6)]vm); \coordinate (v4) at ([shift=(306:.6)]vm); \coordinate (v5) at ([shift=(18:.6)]vm); \filldraw[black!20] (v1) -- ([shift=(60:.2)]v1) arc (60:120:.2) -- (v1); \draw (v1) -- ([shift=(60:.2)]v1); \draw (v1) -- ([shift=(120:.2)]v1); \filldraw[black!20] (v2) -- ([shift=(132:.2)]v2) arc (132:192:.2) -- (v2); \draw (v2) -- ([shift=(132:.2)]v2); \draw (v2) -- ([shift=(192:.2)]v2); \filldraw[black!20] (v3) -- ([shift=(204:.2)]v3) arc (204:264:.2) -- (v3); \draw (v3) -- ([shift=(204:.2)]v3); \draw (v3) -- ([shift=(264:.2)]v3); \filldraw[black!20] (v4) -- ([shift=(276:.2)]v4) arc (276:336:.2) -- (v4); \draw (v4) -- ([shift=(274:.2)]v4); \draw (v4) -- ([shift=(336:.2)]v4); \filldraw[black!20] (v5) -- ([shift=(348:.2)]v5) arc (-12:48:.2) -- (v5); \draw (v5) -- ([shift=(348:.2)]v5); \draw (v5) -- ([shift=(48:.2)]v5); \filldraw (v1) circle (1.3pt); \filldraw (v2) circle (1.3pt); \filldraw (v3) circle (1.3pt); \filldraw (v4) circle (1.3pt); \filldraw (v5) circle (1.3pt); \filldraw (vm) circle (1.3pt); \draw (vm) -- (v2); \draw[dashed] (vm) -- (v3); \draw[dashed] (vm) -- (v4); \draw (vm) -- (v5); \draw[dotted] (v2) .. controls ([shift=(126:.5)]vm) and ([shift=(54:.5)]vm) .. (v5); \end{tikzpicture}, \label{eqibp6} \end{align*}
where solid edges 
$\solidedge$ 
have weight $1$, dashed edges 
$\dashededge$
have weight $\frac12$ and dotted edges
$\dottededge$
have weight $-\frac12$.
Again, this is a \emph{local} graph identity that has to be interpreted as in the last section.
The graphical functions, conformal four-point integrals or Feynman periods associated to the full graphs fulfill the identity indicated above.
The graphs on the right hand side of the IBP identity contain an obvious four-vertex-cut. We can therefore apply the twist identity~(Section~\ref{sec:twist}) to them to obtain further identities.
\section{A systematic quest for Feynman periods}\label{sectsyst}
\begin{figure}
\begin{align*} & \begin{tikzpicture}[baseline={([yshift=-0.7ex]0,0)}] \useasboundingbox (-8,{-1.4}) rectangle (6,4); \node (Gt) at (0,2) {$G_1^\star, G_2^\star, \ldots, G_{1\,000\,000}^\star,\ldots$}; \node (Tt) at (-6,2) {Feynman periods}; \draw (Gt) ellipse (3 and 1); \coordinate (vl) at (2.5,1); \draw[->] (vl) to[out=-60,in=-30,looseness=3] node[below right=.0,pos=.5]{ \parbox{1cm}{ \normalsize\raggedleft Twist \\$\Delta$-Y \\IBP}} ++(.7,.7); \end{tikzpicture}\\
 & \begin{tikzpicture}[baseline={([yshift=-0.7ex]0,0)}] \useasboundingbox (-8,{-1}) rectangle (6,1); \draw[<-] (0,1) -- node[right=.0,pos=.5]{ \parbox{4cm}{ \normalsize\raggedleft Internalize vertex $z$}} (0,-1); \end{tikzpicture}\\
& \begin{tikzpicture}[baseline={([yshift=-0.7ex]0,0)}] \useasboundingbox (-8,{-4}) rectangle (6,4); \node (Gt) at (0,2) {$G_1, G_2, \ldots, G_{10\,000},\ldots$}; \node (Tt) at (-6,2) { \parbox{3.5cm}{ \centering graphical functions } }; \draw (Gt) ellipse (3 and 1); \coordinate (vl) at (2.5,1); \draw[->] (vl) to[out=-60,in=-30,looseness=3] node[below=.0,pos=.5]{ \parbox{3cm}{ \normalsize\raggedleft Twist \\$\Delta$-Y \\IBP\\Append edges\\Add ext.~edges\\Products\\Factorization\\Planar duality\\Permutation}} ++(.7,.7); \end{tikzpicture} \end{align*}
\caption{Illustration of the Feynman period computation workflow:
Upstairs, we act with the twist and $\Delta$-Y identities on the 
completed Feynman periods and solve the IBP relations on classes of periods. Downstairs, we compute a broad set of graphical functions by applying identities to a set of known functions.
}
\label{fig:periodhunt}
\end{figure}
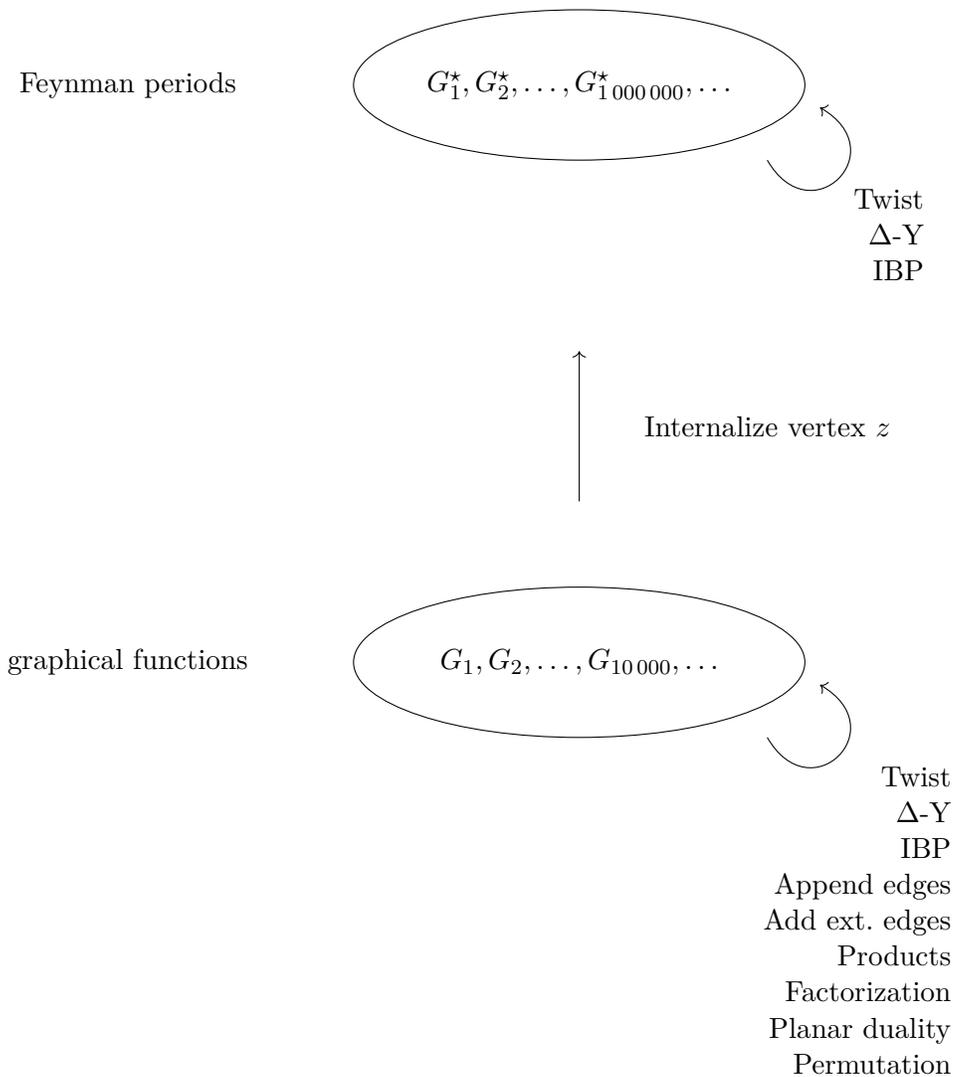

In this section we describe our strategy to compute a given set of Feynman periods systematically using the graphical function method. 
There are two sides to this task: The combinatorial problem of finding a chain of transformation rules for each graph that simplify it successively to a known graphical function. The analytic side of the task is then to \emph{backtrack} the discovered transformation chain and perform the symbolic computation using the analytic version of the transformation rules.

The, computationally more demanding, combinatorial side of the computation was performed using a \texttt{C++} program. 
The computed chains of transformations are then fed into \texttt{HyperlogProcedures} which performs the analytic part of the computation. 

Our approach for the combinatorial side can be separated into two parts again. This separation is illustrated in Figure~\ref{fig:periodhunt}.
One part is a \emph{depth-first} exploration of the space of Feynman periods using a limited set of transformation rules. The transformation rules are applied with the aim of simplifying the associated graph such that it can be related to a known period or it is constructible from a known graphical function.
The other part aims to expand the space of known graphical functions using a \emph{breadth-first search} strategy. For this we go through all graphs which are associated to conformal four-point integrals in the relevant dimension. The combinatorial side of transformation rules is then applied to each graph with the aim of simplifying the graph. This way, we compute the set of all graphical functions below a given edge and vertex number threshold which are accessible using the transformation rules. 

For the whole combinatorial computation we make heavy use of the fact that we can impose an ordering on graphs by using a canonical labeling~\cite{mckay2014practical}.
We will proceed to describe both parts of the computation in detail starting with the breadth-first part:

\paragraph{Graphical functions side: breadth-first approach}

The graphical function or breadth-first part (downstairs in Figure~\ref{fig:periodhunt}) of the computation works by grouping graphs into classes such that each graph in a class can be transformed into onto another member of the same class by a chain of transformation rules. To ensure this, we store a chain of transformation rules for each graph, which transforms it into the smallest member of its class. Note that the analytic expressions of the graphs in the same class do not need to be equal as many transformation rules change the analytic expression associated to the graph.

To compute these classes we first generate all graphs with the desired properties using the \texttt{feyngen} program from~\cite{Borinsky:2014xwa} and the \texttt{geng} program from the \texttt{nauty} package~\cite{mckay2014practical}. We work with completed graphical functions, i.e.~conformal four-point integrals, to exploit the full permutation symmetry of the external vertices.  Moreover, we ignore edges between all external vertices as they only amount to a change of the analytic expression by a rational factor. To only have a finite set of graphs to deal with, we limit the total number of edges and vertices. 

At first, each such graph represents its own class. In the next step, we go through the list of graphs and apply the transformation rules: appending edges, twist, $\Delta$-Y, products, factorization, planar duality and the IBP identity in cases where we can solve it explicitly for one graph. If a transformation rule results in a graph that we already encountered, we merge the associated two classes. To do so we concatenate the respective transformation chains in the class such that all chains end up at the unique simplest graph. Eventually, we end up with a list of classes of graphical functions that cannot be merged anymore using any of the applied transformation rules. All graphical functions that are in the same class as the trivial graphical function are obviously solvable by applying the stored list of transformations. The smallest graphs in every other class are natural candidates for an attempted evaluation using \texttt{HyperInt}. If successful, the whole class can be evaluated as well by applying the transformation chains to the one externally computed graphical function.
In our concrete $\phi^3$ theory example, we computed 11 of these smallest inaccessible graphical functions using \texttt{HyperInt} to further expand the space of known graphical functions.
Unfortunately, \texttt{HyperInt} is not (yet) very efficient in 6 dimensions and these computations were partially very time consuming. All other graphical functions were computed by applying the transformation rules to these or the trivial graphical function.

We used this procedure to generate a list of $\approx 10000$ graphical functions of low vertex and edge number together with their analytic expressions. This list serves as a look-up-table for the other combinatorial part: the depth-first Feynman period computation.

\paragraph{Iterated depth-first period computation}
The second part of the computation (upstairs in Figure~\ref{fig:periodhunt}) starts with a list 
of completed Feynman period graphs $G_1^\star, G_2^\star, \ldots$ which are provided as input. In our explicit example, this list 
consists of all subdivergence-free $\phi^3$ theory Feynman periods up to 7 loops. Additionally, we have the precomputed look-up table of graphical functions from the first part available.

Again we aim to group graphs into classes. However, in this second part of the computation, each class only contains completed Feynman period graphs that evaluate to exactly the same number and there is no need to store the transformation chains.

To every graph of the initially provided list and to each new graph that we encounter we apply the following procedure: We convert the graph to a completed graphical function in all possible ways by choosing every possible set of four external vertices. That means, we invert the combination of period completion (see Section~\ref{sec:period_completion}) and vertex internalization (see Section~\ref{sec:internalizevertex}) in all possible ways. For each graphical function that we obtain, we strip off external edges, try to remove appended edges or factorize the graph into products with the aim of reaching a known graphical function that has been computed in the breadth-first part. If we are successful, we found an explicit way to compute  the respective Feynman period by applying the transformation rules~\ref{sec:addedges}--\ref{sec:internalizevertex}.

We hence start with the reduced list of unknown completed Feynman period graphs. At first each such graph represents one individual class. We go through this list and apply a limited set of transformation rules to each entry, the twist and the $\Delta$-Y identity. If a transformation results in a graph that has been encountered before, the two associated classes are merged. If a new graph is found, it is added to the class of the original graph and the transformation rules are again applied to the new graph recursively. Moreover, we also try to reduce each new graph to a known graphical function with the procedure described in the last paragraph. If successful, the whole class of the original graph can be computed.  
Even though many classes can be computed using the reduction to a known graphical function, the total number of period completed graphs which are encountered can reach into the millions as the chains of transformation rules that connect different classes can be quite long. 

After we have exploited all possible twist and $\Delta$-Y transforms that do not produce too complicated graphs (i.e.~graphs with not too many edges), we apply the IBP identity. Each application of the IBP identity to each graph in every class gives a new linear relation of the different classes. 
The application of the IBP identity can also produce large amounts of graphs that have not been encountered before. We store these graphs separately. We try to solve the system of all IBP identities that only contain graphs that have been encountered before using a sparse linear algebra solver. The solution of this system may result in explicit expressions for the Feynman periods of some classes. 

Afterwards we repeat the process using the period graphs that are still unknown together with the newly encountered graphs from the IBP identities as input.

Eventually, we either produced integration paths for all initial Feynman periods, exhausted all identities within the defined limits or either the resulting IBP system or the number of graphs that have to be checked for reductions becomes too big to be solved in reasonable time.

In summary, we use the depth-first strategy (upstairs in Figure~\ref{fig:periodhunt}) for the 
Feynman period computation and a breadth-first strategy for the graphical function computation (downstairs in Figure~\ref{fig:periodhunt}). Both parts of the computation are connected by the vertex internalization transformation in Section~\ref{sec:internalizevertex}. 

We applied this strategy with the set of all
$D=6$, $\phi^3$ periods up to $7$ loops as input. The results of this computation are given in Appendix~\ref{app}.

As there is usually a large number of different ways to compute the same period, we were able to check our results in highly nontrivial ways. We additionally used the implementation~\cite{tropical-feynman-quadrature} of the tropical numerical integration algorithm described in~\cite{borinsky2020tropical} to check our results via direct numerical evaluation.

\paragraph{Implementation}
All transformation rules are implemented in the \texttt{Maple} package \texttt{Hyp\-er\-log\-Procedures}~\cite{Shlog}. For higher efficiency, we have implemented the combinatorial part of the transformation rules in \texttt{C++}.

In this \texttt{C++} implementation we used \texttt{nauty}~\cite{mckay2014practical} to put graphs into a canonically labeled form. To handle the efficient search for combinatorially transformed graphs and large sparse equation systems, we used the \texttt{flat\_hash\_map} class form the \texttt{Google-Abseil} library~\cite{abseil}. To handle large rational numbers we used the \texttt{GMP} library~\cite{gmp}. For planar duality computations we used \emph{the edge addition planarity suite} \texttt{C} library \cite{planar,planar1}.

\section{Conclusion}
We introduced graphical functions and conformal four-point integrals with the associated transformation rules as a general tool to perform Feynman period computations in even integer dimensions $ \geq 4$. We gave detailed explanations and examples for the transformation rules that lie at the heart of the method. 

Conformal four-point integrals that can be computed directly using the method include the generalized ladder graphs (see Section~\ref{sec:ladder}) in arbitrary even dimensions $\geq 4$. The associated wheel periods~(Section~\ref{sec:wheels}) certify the existence of nontrivial classes in the cohomology of $\mathcal M_g$. This surprising and rather mysterious connection of Feynman integrals with the cohomology of the moduli space of curves remains to be explored.

Brown recently generalized his construction that relates Feynman periods to the homology of a graph complex associated to  the cohomology of $\mathcal M_g$ \cite{Brown:2021umn} to Feynman graphs with kinematics \cite{Brown:2022oix}. It would be interesting to study this generalization directly in the context of graphical functions and conformal four-point integrals. 

We illustrated the workflow for the systematic evaluation of a family of periods in $\phi^3$ theory in six-dimensional spacetime. All Feynman periods up to $6$ loops have been computed analytically. The five loop periods were a necessary ingredient for the $\beta$ function computation in~\cite{Borinsky:2021jdb}. 
At $5$ loops we found the first counterexample to the $\phi^3$ theory variant of the conjecture that the Hepp bound is a perfect period invariant (see Section~\ref{sec:hepp}).
At $7$ loops 561 of the 607 finite Feynman periods could be computed analytically. The remaining 46 periods have been computed numerically up to a limited accuracy using the tropical integration algorithm from~\cite{borinsky2020tropical,tropical-feynman-quadrature}. A list of $\phi^3$ periods up to 7 loops is given in Appendix~\ref{app}.  
 A machine readable list is provided with the ancillary files to the \texttt{arXiv} version of this article.

Our results suggest that $\phi^3$ theory has the same number content as (or a subset of the number content of) $\phi^4$ theory.
The coaction conjecture in $\phi^4$ theory extends to $\phi^3$ theory.
In particular, the space of $\phi^3$ periods is conjectured to be a co-module with respect to the motivic Galois coaction. The existent data is even consistent with a loop order filtration
in the sense that the non-trivial part of the coaction consists of $\phi^3$ periods of strictly lower loop order. This refinement of the coaction conjecture is false in $\phi^4$ theory where
one has to include all ($\phi^4$ and non-$\phi^4$) periods to obtain a loop order filtration. Note that it is possible that the loop order filtration in $\phi^3$ theory will fail due to
yet unknown Feynman periods at higher loop orders. Another difference between $\phi^3$ and $\phi^4$ theories is \emph{weight mixing}.
In a $\phi^3$ period of $n$ loops all weights $\leq 2n-3$ can mix whereas in $\phi^4$ weight mixing is heavily constrained.

We included illustrations of the graphs in the $7$ loop period list in Appendix~\ref{app} to facilitate the search for an explicit \emph{diagrammatic coaction formula}. Such a formula would likely be extremely helpful for the proof of Conjecture~\ref{con:coaction} and related conjectures.

Most of the presented transformation rules (Sections~\ref{feynpergf} and \ref{sec:trafo}) can be applied immediately in the case of dimensionally regularized Feynman periods, graphical functions and conformal four-point integrals. In the presence of subdivergences subtleties regarding the regularization of the associated integrals have to be dealt with. Such subtleties will be discussed in detail in~\cite{7loops}. The evaluation of dimensionally regularized Feynman periods is computationally more challenging than integer dimensional Feynman periods, as the difficulty of computing coefficients in the $\varepsilon$ expansion is increasing with the order. A solution could be the systematic subtraction of the subdivergences using BPHZ-type Hopf algebra methods~(see,~e.g.,~\cite{Connes:1999yr,Brown:2015fyf,borinsky2018graphs,Beekveldt:2020kzk}). The systematic graphical reduction described in Section~\ref{sectsyst} only operates on Feynman periods that are single numbers and not Laurent expansions in $\varepsilon$. An extension of this combinatorial computation to dimensionally regularized Feynman periods is feasible, but more involved bookkeeping is needed. The equivalence classes of Feynman periods would have to represent Laurent expansions instead of numbers. Such an extension would likely enable the computation of all necessary periods for the $6$-loop $\beta$ function in six-dimensional $\varphi^3$ theory.

Based on the available data, the computation of the $\phi^3$ theory minimal subtraction renormalization group $\beta$-function to $6$ loops seems within reach.
The $6$ loop anomalous dimension (the $\gamma$-function) has already been calculated \cite{7loops}. The computation of renormalization group functions to $7$ loops and beyond
is likely to be more challenging.

Our computation of periods at 7 loops was partially obstructed by the limitations of the \texttt{Maple} implementation \texttt{HyperlogProcedures}~\cite{Shlog}. We were able to compute combinatorial reduction paths for 568 of the 607 Feynman periods at 7 loops. We skipped the evaluation of seven of these reduction paths because they were too large to be fed into a \texttt{Maple} program.
The remaining 39 Feynman periods had too complicated topologies and we were unable to find a reduction chain to compute them. For these cases a systematic application of \texttt{HyperInt} to compute selected necessary graphical functions is a promising way to go forward. In fact, due to the analogy to $\phi^4$ theory in $D=4$ where all 7 loops Feynman periods can be expressed analytically in terms of MZVs and their extensions by third roots of unity \cite{Schnetz:2008mp,PanzerSchnetz:2017coact,Schnetz:2016fhy},
there is reason to believe that the same holds true in $\phi^3$ theory in $D=6$ at 7 loops.
A systematic implementation of the $\Delta$-Y reduction methods developed in \cite{Jeffries:2021aog} for the period reduction might also make further $\phi^3$ periods accessible.
In summery there seems ample headroom for future optimizations which may lead to a complete list of $\phi^3$ periods at loop order 7.

Computations in $\phi^4$ theory in $D=4$ within the minimal subtraction scheme indicate that the contribution from the subdivergence-free logarithmically divergent graphs to the coefficients of the $\beta$ function becomes dominant with higher loop order. In other words, graphs with subdivergences seem to be giving a more and more insignificant contribution to the $\beta$ function (see,~e.g.~\cite{Kompaniets:2017yct,Dunne:2021lie}). Our results indicate that the analogous statement does not hold in $\phi^3$ theory in $D=6$. The $\beta$ function as computed in \cite{Borinsky:2021jdb} seems to also contain significant contributions from graphs with subdivergences at higher loop order. These phenomena are likely to be related to questions on \emph{renormalons} in the respective theories. In comparison to $\phi^4$ theory, much more is known about renormalons in $\phi^3$ theory \cite{Borinsky:2021hnd,Borinsky:2022knn}. It would therefore be worthwhile to repeat the thorough study of \cite{Dunne:2021lie} using the available data on $\phi^3$ theory and compare it with renormalon results.

The transformation rules can also be applied to nontrivial conformal four-point integrals for which analytic expressions have been obtained by other means. Especially resourceful examples are the fishnet graphs which cannot be obtained by any (known) transformation rules. Still, they provide highly nontrivial starting points for the transformation rules and from each fishnet graph a new infinite family of conformal four-point integrals can be obtained. 
In fact, these graphs with their associated relatively simple analytic expressions indicate the existence of further unknown transformation rules with which the fishnet graphs themselves can be constructed from simpler conformal four-point integrals. The discovery of such an additional transformation rule would significantly extend the utility of our methods.

The graphical function method would also greatly benefit from the development of \emph{single-valued elliptic hyperlogarithms}. The theory of generalized single-valued hyperlogarithms is one of the backbones of the method \cite{Schnetz:2021ebf}. Due to the limitations of this function space, graphs such as the one depicted in \eqref{eq:elliptic}, which are conjectured to yield an elliptic function, are inaccessible within the present framework. 
Using bootstrap techniques a $D=2$ analogue for \eqref{eq:clawD} was recently found~\cite[Section~7.2]{Corcoran:2021gda} which features a single-valued elliptic function. 
 Such computations might serve as guiding examples for the development of a theory of single-valued elliptic hyperlogarithms. Single-valued elliptic polylogarithms have already been studied in~\cite{Zagier1990}.

General Feynman integral computations are known to feature more involved  geometries (e.g.~Calabi-Yau manifolds) that go beyond the elliptic case~(see, e.g.~\cite{Bourjaily:2018ycu,Bonisch:2020qmm}). On the level of Feynman periods, the situation is less clear. Up to this date no Feynman period is known analytically that cannot be expressed in terms of MZVs or their extensions.
Only very few periods are \emph{proved not to evaluate} to such numbers \cite{Brown:2010bw}: From 8 loops on some such Feynman periods are known in $\phi^4$ theory. Most of these numbers can be associated to periods of $K3$ surfaces. From the $c_2$-invariant \cite{SFq},
(see \cite{Schnetz:2019cab} for more recent review) it is expected that most Feynman periods at high loop orders are neither MZVs or their extensions.
On the other hand, periods of elliptic curves are not expected to exist in Feynman periods \cite{Schnetz:2019cab}.
An extension of the graphical function method to the elliptic case might help to explain why elliptic periods are absent in QFT.

\section*{Acknowledgements}
MB is grateful to 
Samuel Abreu, 
David Broadhurst,
Gerald Dunne, 
John Gracey,
Franz Herzog,
Shannon Jeffries,
Albrecht Klemm,
Andrew McLeod,
Erik Panzer,
Jos Vermaseren and 
Karen Yeats 
for valuable discussions.

Both authors thank the organizers of the workshop `Higher structures emerging from renormalisation' at Erwin Schrödinger Institute in Vienna
 and Dirk Kreimer for continued motivation, support and discussions.

Parts of the computations presented in this article were performed on high performance computers at Nikhef while MB was employed there and supported by the NWO Vidi grant 680-47-551  ``Decoding Singularities of Feynman graphs''.
MB was supported by Dr.\ Max R\"ossler, the Walter Haefner Foundation and the ETH Z\"urich Foundation.

\providecommand{\href}[2]{#2}\begingroup\raggedright\endgroup

\appendix
\section{List of finite \texorpdfstring{$\phi^3$}{phi3} periods in \texorpdfstring{$D=6$}{D=6}}\label{app}

What follows is a list of known $D=6$, $\phi^3$ periods up to $7$ loops. We consider all three-point $\phi^3$ theory graphs without subdivergences up to $7$ loops. Due to the conformal period completion invariance (see Section~\ref{sec:period_completion}), it is sufficient to only deal with the associated period completed graphs. The list depicts the period completed graph (first column), a name for the period, $P_{\ell,n}$, (second column) where $\ell$ is the loop number of an associated uncompleted graph and $n$ is an auxiliary numbering of the graphs which is consistent with the numbering in the machine readable list of periods in the ancillary files to this article. 
The analytic expression for the respective period (third column) is given in terms of the $Q$-numbers defined in Table~\ref{tabQ} and \ref{tabN}.
We used the program \texttt{graphviz}~\cite{Gansner00anopen} to create the graph drawings in the first column.

For Feynman periods at 7 loops that were not accessible within the computational constrains of Section~\ref{sectsyst} we only give a numerical value of the associated period that has been obtained using the implementation~\cite{tropical-feynman-quadrature} of the Monte Carlo algorithm described in~\cite{borinsky2020tropical}. The value is given with its statistical error of one standard deviation.

A machine readable list is included in the ancillary material to the \texttt{arXiv} version of this article. For each period, this list includes the period completed graph, the analytic expression in terms of the motivic $f$-alphabet, an expression in terms of multiple-$\zeta$-values, the numerical value of the analytic expression up to 100 digits (or a limited accuracy numerical value obtained with~\cite{tropical-feynman-quadrature} if no analytic expression is available), the statistical standard deviation for the numerical value (only relevant if the analytic value is not available), a normalized version of the Hepp bound 
$\widetilde{H}_{\ell,n} = \frac{8}{3} 2^{-3\ell} H_{\ell,n}$
that is invariant under the $\Delta$-Y transformation and the value of the usual Hepp bound $H_{\ell,n}$ as defined in \cite[eq.~(1.5)]{Panzer:2019yxl}. 
We only performed a systematic exploration of the set of $\phi^3$ periods up to $7$ loops, but still obtained many periods beyond that threshold using the methods described in this article. The included machine readable list also contains these additional $\phi^3$ periods up to $9$ loops.

\begin{longtable}{ c  l  l }
    \begin{tikzpicture}[x=1.5em,y=1.5em,baseline={([yshift=-.7ex]current bounding box.center)}] \coordinate (v0) at (-1.000000,-0.943630); \coordinate (v1) at (1.000000,-1.000000); \coordinate (v2) at (0.054346,-0.206302); \coordinate (v3) at (0.018898,1.000000); \draw[preaction={draw, white, line width=.9pt, -},line width=.3pt] (v0) -- (v1); \draw[preaction={draw, white, line width=.9pt, -},line width=.3pt] (v0) -- (v2); \draw[preaction={draw, white, line width=.9pt, -},line width=.3pt] (v0) -- (v3); \draw[preaction={draw, white, line width=.9pt, -},line width=.3pt] (v1) -- (v2); \draw[preaction={draw, white, line width=.9pt, -},line width=.3pt] (v1) -- (v3); \draw[preaction={draw, white, line width=.9pt, -},line width=.3pt] (v2) -- (v3); \draw[line width=.3pt] (v0) -- ($(v0)!1pt!(v1)$); \draw[line width=.3pt] (v1) -- ($(v1)!1pt!(v0)$); \draw[line width=.3pt] (v0) -- ($(v0)!1pt!(v2)$); \draw[line width=.3pt] (v2) -- ($(v2)!1pt!(v0)$); \draw[line width=.3pt] (v0) -- ($(v0)!1pt!(v3)$); \draw[line width=.3pt] (v3) -- ($(v3)!1pt!(v0)$); \draw[line width=.3pt] (v1) -- ($(v1)!1pt!(v2)$); \draw[line width=.3pt] (v2) -- ($(v2)!1pt!(v1)$); \draw[line width=.3pt] (v1) -- ($(v1)!1pt!(v3)$); \draw[line width=.3pt] (v3) -- ($(v3)!1pt!(v1)$); \draw[line width=.3pt] (v2) -- ($(v2)!1pt!(v3)$); \draw[line width=.3pt] (v3) -- ($(v3)!1pt!(v2)$); \filldraw (v0) circle (.3pt); \filldraw (v1) circle (.3pt); \filldraw (v2) circle (.3pt); \filldraw (v3) circle (.3pt); \end{tikzpicture} & 
    $P_{1,1}$ &
    ${\begin{aligned}&{\scriptstyle 1}\end{aligned}}$
    \\
\end{longtable}

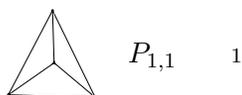
\end{document}